# Nonlinear rotational spectroscopy reveals many-body interactions in water molecules

Yaqing Zhang, Jiaojian Shi, Xian Li, Stephen L. Coy, Robert W. Field, and Keith A. Nelson*

Department of Chemistry, Massachusetts Institute of Technology, 77 Massachusetts Avenue, Cambridge, MA 02139, USA.

*Corresponding author. Email: kanelson@mit.edu.

**Abstract:**

Because of their central importance in chemistry and biology, water molecules have been the subject of decades of intense spectroscopic investigations. Rotational spectroscopy of water vapor has yielded detailed information about the structure and dynamics of isolated water molecules as well as water dimers and clusters. Nonlinear rotational spectroscopy in the terahertz regime has been developed recently to investigate the rotational dynamics of linear and symmetric-top molecules whose rotational energy levels are regularly spaced, but it has not previously been applied to water or other lower-symmetry molecules with irregularly spaced levels. We report the use of recently developed two-dimensional terahertz rotational spectroscopy to observe high-order rotational coherences and correlations between rotational transitions that could not be observed previously. The results include two-quantum (2Q) peaks at frequencies that are shifted slightly from the sums of distinct rotational transitions on two molecules, which directly reveal the presence of previously unseen metastable water complexes with lifetimes of 100 ps or longer.

Several such peaks observed at distinct 2Q frequencies indicate that the complexes have multiple preferred bimolecular geometries. Our results demonstrate sensitivity of rotational correlations measured in 2D THz spectroscopy to molecular interactions and complexation in the gas phase.

**Significance statement**

Since water vapor is everywhere around us and is crucial to life, the stable complexes that water molecules form with each other and with various environmental constituents have been studied extensively. But transient, metastable complexes are more elusive. A recently developed spectroscopic method, *two-dimensional rotational spectroscopy*, directly measures *correlations* between the rotational transitions that appear in a conventional spectrum. Measurements of water vapor showed that rotations of one water molecule can change the rotational frequencies of another molecule – clear evidence of previously unseen complexes between the water molecules involved. The exquisite sensitivity of the method to intermolecular interactions has revealed new behaviour in one of the most extensively studied molecular species and promises new insights about many others.

## Main Text:

## Introduction

Water has attracted extensive spectroscopic interest because of its critical implications for theoretical and applied sciences (1, 2). Water shows anomalous properties because of complicated fluxional hydrogen-bond networks and has been investigated by Raman and infrared spectroscopy (3, 4), sum frequency spectroscopy (5), optical Kerr effect spectroscopy (6), vibration-rotation-tunneling spectroscopy (7) and recently by two-dimensional (2D) infrared spectroscopy (8, 9) and 2D Raman-terahertz (THz) spectroscopy (10). In the gas phase, water is of utmost importance for atmospheric science, astrophysics, combustion research, and fundamental chemistry and physics (1, 11, 12). Although the pure rotational spectrum of water vapor has been well known for decades (13, 14), nonlinear THz spectroscopy of water rotational dynamics has not been previously reported. Nonlinear rotational spectroscopy in the microwave spectral range is well established (15), but because of the small moments of inertia of water, most of its rotational transition frequencies lie in the terahertz frequency range (Fig. 1). Nonlinear THz rotational spectroscopy was first reported only recently (16, 17), and two-dimensional (2D) THz rotational spectroscopy (10, 18, 19) more recently still. As in 2D spectroscopy of vibrational, electronic, and other degrees of freedom (9, 20-25), 2D rotational spectroscopy can reveal correlations between rotational states, many-body effects, and distinct multiple-field interactions that cannot be observed by linear spectroscopy (26). The large dipole moment of water, manifest in strong atmospheric absorption in the THz window (27-31), and the existence of water dimers and larger clusters with complex structures and dynamics (7, 32, 33), suggest that 2D spectroscopy of water could generate important insights that have thus far been elusive (34-37).

2D THz rotational spectroscopy has not been extended previously to water or any asymmetric-top molecules, although such molecules, whose rotational spectra are complex because all three of their moments of inertia are unequal, are the majority of naturally occurring molecular species. Unlike a linear or symmetric top molecule, in which the spectroscopic transitions between successive total rotational angular momentum levels denoted by the quantum number $J$ are spaced by even-integer multiples of a common factor, the rotational constant $B$, the asymmetric top nature of water molecules leads to irregularly-spaced rotational energy levels. These levels are described approximately by quantum numbers $J$, $K_a$, and $K_c$ that indicate the total angular momentum and its symmetry-axis projections (1, 2). The spectrum of water vapor consists of many transitions, with $\Delta J = -1, 0, +1$ (P, Q, and R branches) all allowed and, for each, changes $\Delta K_a = \pm 1, \Delta K_c = \pm 1$ (1, 2). The large centrifugal distortion of water and the distinct rotational states occupied by its spin isomers (even-symmetry for para, odd-symmetry for ortho) further complicate its rotational spectrum (1, 2, 38). A typical THz time-domain free-induction decay (FID) signal from water vapor at ambient conditions, induced by a weak single-cycle THz pulse, is shown in Fig. 1**e**. The irregular oscillations arise from more than 15 transitions (Fig. 1**d**) that contribute significantly to the water vapor absorption spectrum in the 0.1-2 THz region.

## Results

The 2D THz experimental setup is illustrated schematically in Fig. 1f and described further below (Materials and Methods). Briefly, a pair of THz pulses with variable time delays were focused on the water vapor sample in a gas cell at about $60^{\circ}\mathrm{C}$ and 150 Torr pressure. The induced THz signal was detected in an electro-optic crystal by a variably delayed optical readout pulse, the time-dependent polarization rotation of which reveals the temporal profile of the signal field. The signal was collected at a 1 kHz repetition rate, with a total data acquisition time of 2-3 days as both time delays are swept. Fourier transformation (FT) of the signal as a function of both time delays yields the 2D spectra (Figs. 2 -4). In order to resolve closely spaced peaks near 1.0 THz, separate measurements with long FID detection times (up to 300 ps) were recorded (Figs. 3**a-c**).

Figures 2 shows the experimental 2D THz rotational spectra. The spectra mainly consist of non-rephasing (NR) (Figs. 2**a-c**) and photon echo or rephasing (R) (Figs. 2**d-f**) parts. Many nonlinear spectroscopic peaks appear along the $f_{pump} = \pm f_{probe}$ diagonals. These peaks can be understood in terms of a two-level system model and the Feynman diagram in Fig. 2**g**. The field interaction from the first pulse induces a coherence between the rotational states $|0\rangle$ and $|1\rangle$, described by the density matrix element $|1\rangle\langle 0|$; the successive two interactions from the second pulse can convert the coherence first into a population state, $|1\rangle\langle 1|$, and then into the coherence, $|1\rangle\langle 0|$, which radiates at the same frequency as the first. For example, the rotational states $|1_{01}\rangle$ and $|1_{10}\rangle$ denoted as $|0\rangle$ and $|1\rangle$, respectively, give rise to the non-rephasing diagonal peak at $(f_{probe} = 0.753 \text{ THz}, f_{pump} = 0.753 \text{ THz})$, indicated as Peak (1) in Fig. 2**a**. The third field interaction may alternatively convert the population $|1\rangle\langle 1|$ into a new coherent superposition, $|2\rangle\langle 1|$, which radiates at a new frequency, leading to off-diagonal or cross peaks, such as the one at $(f_{probe} = 1.23 \text{ THz}, f_{pump} = 0.753 \text{ THz})$ (Fig. 2**b**, Feynman diagram (2) in Fig. 2**i**) that results from the three levels, $|2_{02}\rangle$, $|2_{11}\rangle$, and $|2_{20}\rangle$. Similar sequences, in which the first field interaction produces the coherence $|0\rangle\langle 1|$, yield rephasing signals observed as $(f_{probe} = 0.753 \text{ THz}, f_{pump} = -0.753 \text{ THz})$ (Peak (1′) in Fig. 2**d**) or $(f_{probe} = 1.23 \text{ THz}, f_{pump} = -0.753 \text{ THz})$ (Peak (2′) in Fig. 2**e**), respectively. An example of the time-dependent nonlinear signal field $E_{NL}(t)$, recorded with a specified time delay $\tau$ between the incident THz pulses, is shown in Fig. S4 of the supplementary information. Fourier transformation of signals with respect to temporal variables $t$ and $\tau$ yields the 2D spectrum as functions of $f_{probe}$ and $f_{pump}$ respectively.

Some diagonal peaks between 0.5 and 0.8 THz, which stem from coherences of an extremely small amount of adventitious gaseous acetonitrile residue from previous experiments (Figs. 2**a,b,** and **d**), appear but do not significantly affect the 2D spectra of the water molecules except for the resonance associated with the $|J = 30\rangle \rightarrow |J' = 29\rangle$ transition at 0.552 THz in acetonitrile which interferes with that of the $|1_{01}\rangle \rightarrow |1_{10}\rangle$ transition in water, leading to slightly split double-peaks at $(f_{probe} = 0.558 \text{ THz}, f_{pump} = 0.558 \text{ THz})$ (Fig. 2**a**). The 2D THz spectra show no cross-peaks at mixed frequencies that belong to both water and acetonitrile molecules nor between para- and

ortho- water molecules, indicating that distinct constituent species and different nuclear spin isomers of the same molecular source generate independent spectra, similar to the behavior of alkali-metal atoms in 2D electronic spectra (39-41). We note that far weaker peaks than those from acetonitrile, due to water dimers, also appear. Their analysis requires careful subtraction of the acetonitrile peaks and consideration of signal levels far weaker than those of interest in this paper, and will be presented in a separate publication.

Sum-frequency signals (SFs) and difference-frequency signals (DFs) appear, which arise from ladder-type (Fig. 2**h**) and V-type (Fig. 3**g**) transitions, respectively. The SFs (Figs. 2**c** and **f**) are two-quantum (2Q) signals in which two successive field interactions from the first THz pulse generate a coherent superposition of three rotational states $|0\rangle, |1\rangle$, and $|2\rangle$, which includes a 2Q coherence $|2\rangle\langle 0|$. The 2Q coherence does not radiate, but its phase-dependent interaction with the THz field in the second pulse produces a radiating 1Q coherence $|2\rangle\langle 1|$ or $|1\rangle\langle 0|$ whose amplitude and/or phase oscillate at the $|2\rangle\langle 0|$ coherence frequency as the inter-pulse time delay $\tau$ is varied. The 2Q frequency is thereby revealed in the FT with respect to the inter-pulse delay, $\tau$, along with the radiating signal coherence frequency in the FT with respect to the detection time, $t$. An example of this type of coherence is the peak at ($f_{probe} = 1.23$ THz, $f_{pump} = 1.98$ THz) in Fig. 2**c** (labelled as Peak (3)). DFs can be observed specifically in a so-called V-type system consisting of the three ortho water states, $|0\rangle = |1_{01}\rangle$, $|1\rangle = |1_{10}\rangle$, and $|2\rangle = |2_{12}\rangle$ (Figs. 1**d** and 3**g**). The excited rotational states $|1\rangle$ and $|2\rangle$ share the same ground state $|0\rangle$, and the transition between the two excited states is forbidden by the selection rule ($\Delta K_a = \pm 1$, $\Delta K_c = \pm 1$). The DF signals come from the coherence $|1\rangle\langle 2|$ induced by two interactions from the first THz pulse, applied in succession starting from the initial population state, $|0\rangle\langle 0|$ (diagram (i) in Fig. 3**h**). This is also a 2Q coherence, and although (just like the SF 2Q coherence $|2\rangle\langle 0|$) it is not radiative and therefore cannot be observed in a conventional linear spectrum, the $|1\rangle \leftrightarrow \langle 2|$ coherence becomes observable in 2D spectroscopy through the third field interaction, from the second THz pulse, which projects onto either of the coherences $|1\rangle\langle 0|$ or $|0\rangle\langle 2|$ that radiate at frequencies of $f_{probe} =$ 0.558 THz and 1.670 THz respectively.

Based on the energy levels of individual water molecules, no 2Q resonant peak is anticipated near ($f_{probe} = 0.753$ THz, $f_{pump} = 1.506$ THz) along the line $f_{pump} = 2f_{probe}$, since the $|2_{02}\rangle \rightarrow |2_{11}\rangle$ transition at 0.753 THz is not followed by a second transition at the same frequency. Surprisingly, a strong 2Q signal appears at $f_{probe} = 0.753$ THz as shown in Fig. 4**a**. In addition, 2Q off-diagonal features, that consist of discrete peaks, extend downward from both of the 2Q diagonal peaks at $f_{probe} = 0.558$ THz and $f_{probe} = 0.753$ THz. These features can be explained by the collective resonances induced by many-body interactions in water molecules, similar to effects observed in 2D electronic spectra of atomic vapors (40, 41) and quantum wells (24). For example, two molecules with identical two-level systems can create a system with a ground state, a degenerate pair of singly excited states, and a doubly excited state (Fig. 4**b**). Dipole-dipole interactions between the water molecules could split and shift the singly-excited-state energies and create correlations between the successive transitions, yielding 2Q diagonal and off-diagonal peaks as shown in Fig. 4**a**. The three-quantum (3Q) signal can be understood similarly but does not play a major role in this work because only a diagonal 3Q peak was observed (Fig. 4**a**; see also Figs. S13 and S22).

**Discussion**

Owing to the small shifts (~20 GHz) measured in our 2Q off-diagonal frequencies, in contrast to the completely different rotational spectrum of the known hydrogen-bonded water dimer (33, 42, 43), we attribute the 2Q and 3Q features to the transient formation of weakly bound metastable water complexes (32, 43, 44). We assume these are predominantly molecule-pairs, which we refer to below as “complexes” to avoid confusion with the H-bonded dimer. A lower limit of ~100 ps for the lifetime of the complexes can be estimated from the linewidths of the 2Q off-diagonal peaks along the pump axis. This value is equal to our maximum time delay between the two THz pump pulses and is therefore instrumentally limited. Note that at the experimental pressure of 150 Torr, the mean free time between molecular collisions is about 1.1 ns, so weakly interacting complexes can persist for longer than 100 ps without being destroyed by collisions with other molecules. 2D spectroscopy permits direct access to the metastable complexes, which may represent precursor or intermediate states on paths toward forming H-bonded dimers.

None of the one-quantum (1Q) signals in our 2D spectra (e.g. peaks in Figs. 2**a**, **b**, **d**, and **e**,) show any off-diagonal peaks or continua arising from many-body interactions. The doubly excited states revealed in the off-diagonal 2Q peaks are different from the rotational states of individual water molecules. In a molecule-pair complex, the rotational excitation of one molecule influences the rotation of the other in a way that modifies the 2Q energy. The many-body interactions that form the complex are not strong enough to prevent independent rotations of the individual molecules (in contrast to the H-bonded dimer, which rotates as a single unit), but are strong enough to cause measurable shifts in the energy when both molecules are rotationally excited. In addition, the off-diagonal 2Q signals in Fig. 4**a** (and also in supplementary information Fig. S16) show multiple closely spaced peaks. We believe these must be associated with multiple preferred intermolecular complex geometries. The fine structure in these spectral features highlights the sensitivity of rotational correlations to intermolecular interactions and the metastable complexes that arise from them.

Another possible source for 2Q and 3Q signals is the radiative interaction (17) in which the emitted THz field that results from the interaction of the incident THz field with one molecule can subsequently serve as the excitation field for another molecule. However, this radiative interaction does not generate molecular excited states different from those that exist in individual water molecules, and thus makes no contribution to the 2Q off-diagonal peaks. A simulation of the 2D spectra based on the rotational Hamiltonian for non-interacting symmetric-top molecules and including the known water transition dipoles and centrifugal distortion (described in detail in the supplementary information) reproduces the main features well, as shown in Fig. 5. To confirm the distinct sources of the 2Q off-diagonal and diagonal peaks, we analyzed the observed ratio between their heights at two sample temperatures (Fig. S21). The ratio of the 2Q off-diagonal peak heights to the 2Q diagonal peaks increases by a factor of about 4.9 when the temperature is increased from room temperature ($21^{\circ}$C) to 60ºC, which increases the water vapor pressure from 18 Torr to 150 Torr, i.e. by a factor of 8.3. We show further in the supplementary information (Figs. S15-S16) that the ratios of 2Q off-diagonal peaks to a strong 1Q peak ($f_{probe} = 0.558$ THz, $f_{pump} = 0.558$ THz) that clearly arises from individual water molecules change by comparable amounts with temperature, while the ratios of the 2Q diagonal peaks to 1Q peaks do not change significantly (Fig. S21). These results clearly show that the 2Q off-diagonal peaks arise uniquely from molecule-pair complexes, the concentrations of which depend on the square of the water vapor pressure. We note that the increase with temperature in 2Q off-diagonal peaks relative to that of the others is less than a factor of 8.3, that is, the complex concentration increases by somewhat less than the square of the water vapor pressure, because the higher temperature shifts the chemical equilibrium toward separated water molecules. A straightforward calculation including the appropriate Boltzmann factors (presented in the supplementary information) indicates molecule-pair complex binding energies of roughly 10 kJ/mol, with a considerable range of values (8.4-17 kJ/mol) deduced from the different 2Q peaks. The average value is smaller than the binding energy of the H-bonded dimer (33, 42).

**Conclusions:** Nonlinear 2D rotational spectra of water vapor have shown rephasing (photon echo), non-rephasing, and multiple-quantum signals that arise from different field-induced pathways between rotational states. The sensitivity of 2D spectra to correlations between transitions has enabled observations of such correlations between rotations of distinct molecules. Intermolecular rotational correlations have not been observed previously. In the case of water, energy shifts of 2-quantum rotational coherences and the nonlinear dependence of 2Q spectral peak heights on water vapor pressure clearly indicate the presence of previously unseen metastable molecule-pair complexes with multiple preferred geometries and allow an estimate of the binding energies of these complexes. Our work paves the way for systematic study of intricate intermolecular interactions and energy transfer mechanisms that give rise to correlations between ordinarily unrelated rotational transitions. Many experimental refinements are possible, including the introduction of a third THz pulse (45, 46), which would enable measurements at variable population times and could yield the lifetimes of transient complexes, distinct from the dephasing times derived from spectral linewidths. Considerable reduction in data acquisition time may be possible by incorporating single-shot spectroscopy methods to eliminate the need for one of the variable delays (47-49) or through high-resolution multidimensional spectroscopy with frequency combs (39), which have been generated and used for linear spectroscopy of water vapor at THz frequencies (50). Multidimensional THz rotational spectroscopy holds promise for many applications including measurement of enantiomeric excess (15), molecular cluster identification and characterization, and coherent manipulation of molecular orientation.

## Materials and Methods

### Sample preparation

A capped storage vessel (25ml, Chemglass Life Sciences) with a sidearm served as a liquid water container. About 5 ml liquid water was pipetted into the storage vessel at room temperature. The sidearm of the vessel was connected to a stainless steel transfer line (approximately 50 cm long), the other end of which was linked to a gas cell with a 1.25 cm path length. Three valves were equipped to adjust the gas flow. Liquid water in the storage vessel was frozen into ice using liquid nitrogen, following a vacuum pumping process to eliminate air in the storage vessel, the transfer line, and the gas cell before ice melted.

The storage vessel was then sealed by the cap and immersed in a warm water bath heated by a hotplate. The temperature was stabilized at approximate 60°C to avoid water condensation. A temperature-controllable silicone heating tape (Cole-Parmer) was wrapped around the system in order to maintain a stable temperature (copper wires were also used at local turning points for the same purpose). By slightly unscrewing the cap of the storage vessel, water vapor was allowed to slowly diffuse into the gas cell via the transfer line. A few minutes was allowed for the gas pressure to equilibrate (approximately 150 Torr at 60°C). The gas cell was then sealed and moved into the experimental setup while the heating tape was still attached and turned on until the experiment was completed.

### THz time-domain spectroscopy (THz-TDS)

A typical THz-TDS system (51) was used to record the temporal profile of water vapor from ambient air at room temperature (Fig. S1). A mode locked Ti: Sapphire oscillator (78 MHz rep rate, 150 mW average power, 800 nm central wavelength, < 100 fs pulse duration) was used to pump a DC biased photoconductive antenna (PCA) made of low-temperature-grown gallium arsenide (LT-GaAs). The generated single-cycle THz field (~1 ps, 0.1-5 THz spectrum) was collected by four 90-degree off-axis parabolic mirrors and detected by another PCA. A small portion of the femtosecond laser beam was split and sent into an optical delay line, attenuated, and used to read out the THz field profile. The THz field amplitude was proportional to the time-averaged photocurrent, and detected by a lock-in amplifier. The coherent nature of the detection ensures a high signal-noise ratio on the order of $1.0 \times 10^4$.

**2D THz spectroscopy**

Two Ti:Sapphire amplifier systems (Coherent, 1 kHz repetition rate, 800 nm central wavelength) were used in our experiments. One delivered 35-fs pulses and had an output power of 7 W; the other delivered 100-fs pulses and had 3.5 W output power. Using either system, 95% of the output power was divided into two equal parts with time delays between them controlled by a mechanical delay stage. Both parts were directed into the same lithium niobate (LN) crystal to generate two time-delayed, collinearly propagating THz pulses by the tilted-pulse-front technique. The generated THz pulses (Fig. S2) were collected by four 90-degree off-axis parabolic mirrors, and focused on the gas sample in the gas cell. The remaining 5% laser output was attenuated, routed into a third path to read out the THz signals by EO sampling in a 2 mm ZnTe crystal. Dry air was used to purge the experimental setup to avoid THz absorption by water vapor from ambient air. The two THz electric field amplitudes were 440 kV/cm and 290 kV/cm using the 35-fs laser system, and 300 kV/cm and 350 kV/cm using the 100 fs laser system.

**Differential chopping technique**

Differential chopping was employed to separate nonlinear THz signals from strong linear signals (Figs. S3-S4). Two optical choppers were placed in the two optical paths to modulate the optical pulses used for THz generation. One chopper was operating at 500-Hz frequency, the other at 250 Hz. Each chopper was put in one optical path. Within four sequential laser shots, THz pulses from both paths 1 and 2 denoted as $E_{12}$, only from path 1 denoted as $E_1$, only from path 2 denoted as $E_2$, and none denoted as $E_{BG}$ were produced, respectively. The two choppers and the laser were synced to a data acquisition card (National Instruments) that was used to measure the signal out of EO sampling. Each data point was averaged over 100 laser shots in order to achieve a signal-noise ratio of around 5000. The total acquisition time period for a complete 2D scan was roughly 2.5 days.

**Acknowledgments: Funding:** This research was supported in part by National Science Grant CHE-1665383.

**Author contributions:** K.A.N. supervised the project. K.A.N. and Y.Z. conceived the project. Y.Z., J.S., and X.L. performed the experimental measurements. Y.Z., K.A.N., J.S., S.L.C., and R.W.F., analyzed the data. Y.Z. wrote the paper with input from all authors.

**Competing interests:** The authors have no competing interests.

**Data availability:** All data are available in the main text and the supplementary information, and are available from the corresponding author on reasonable request.

**Supplementary Information:**

Supplementary Text

Figs. S1 to S27

Tables S1 to S7
Supplementary References

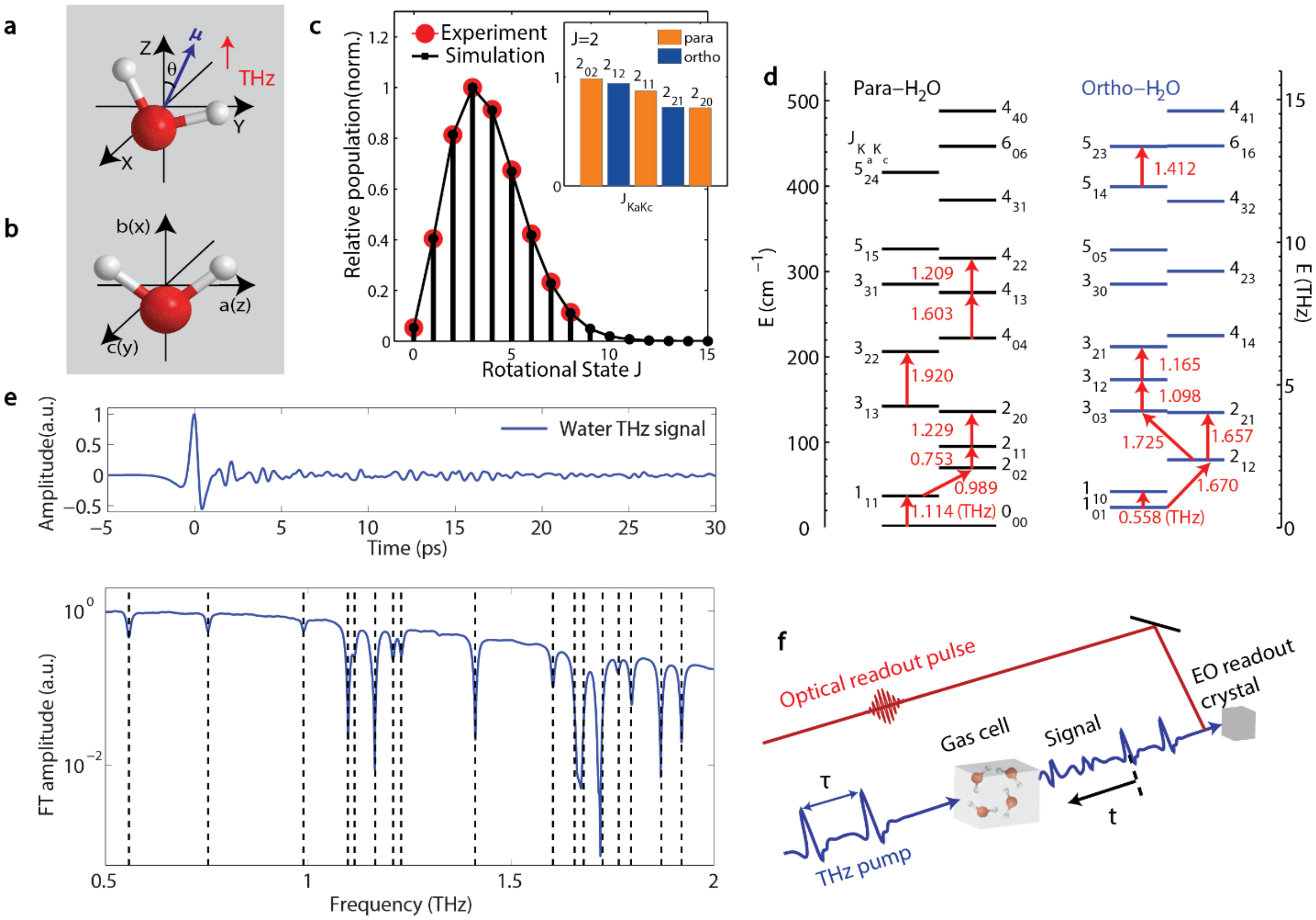

**Fig. 1. Overview of the experiment. a**, Water molecule in the lab frame, showing the dipole moment $\boldsymbol{\mu}$ at an angle $\theta$ from the $Z$ axis (THz polarization direction). **b**, Water molecule in the molecule-fixed frame with the three moments of inertia $I_a$, $I_b$, and $I_c$ along the corresponding axes. **c**, Relative population distribution as a function of the $J$ rotational quantum number. All relevant Ka and Kc components are included in the population distribution. Inset shows the relative population distribution within the state $J$=2. **d**, Rotational energy levels of para-$H_2O$ and ortho-$H_2O$ molecules. Red arrows illustrate rotational transitions and transition frequencies involved in this work. **e**, Measured THz free-induction decay (top) and Fourier transform (bottom) showing rotational transitions (marked by dashed vertical lines) of water vapor in ambient air. **f**, Schematic illustration of the 2D THz experimental setup. Linear THz spectra (example in **e**) are measured with only one THz pump pulse.

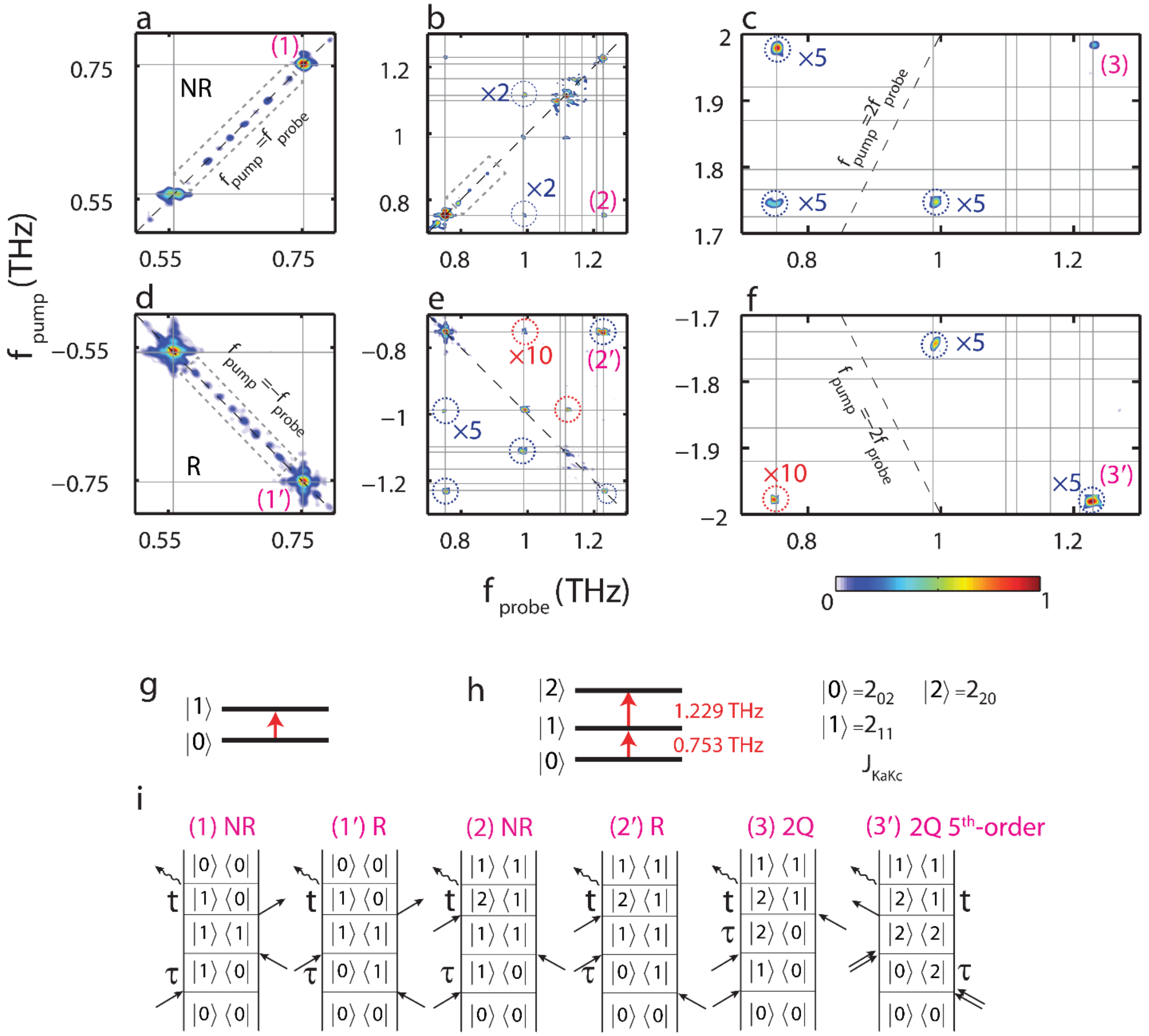

**Fig. 2. 2D rotational spectra of water vapor at 60 °C. a** to **c**, Non-rephasing (NR) 2D rotational spectrum. **d to f**, Rephasing (R) 2D rotational spectrum. **g**, Two-level system diagram. **h**, Three-level system diagram. **k**, V-type system diagram. All 2D spectra are based on the same color bar. Continuous, evenly-spaced features in gray dashed boxes (in **a**, **b**, and **d**) arise from acetonitrile residue in the gas cell, and they do not significantly affect the nonlinear signals of water vapor. Representative peaks are labelled ((1) - (3) and ((1′) - (3′)) and their Feynman diagrams are listed in **i**. The inter-pulse delay $\tau$ and the detection time $t$ are also shown. All other 2D signals shown here can be interpreted in a similar manner.

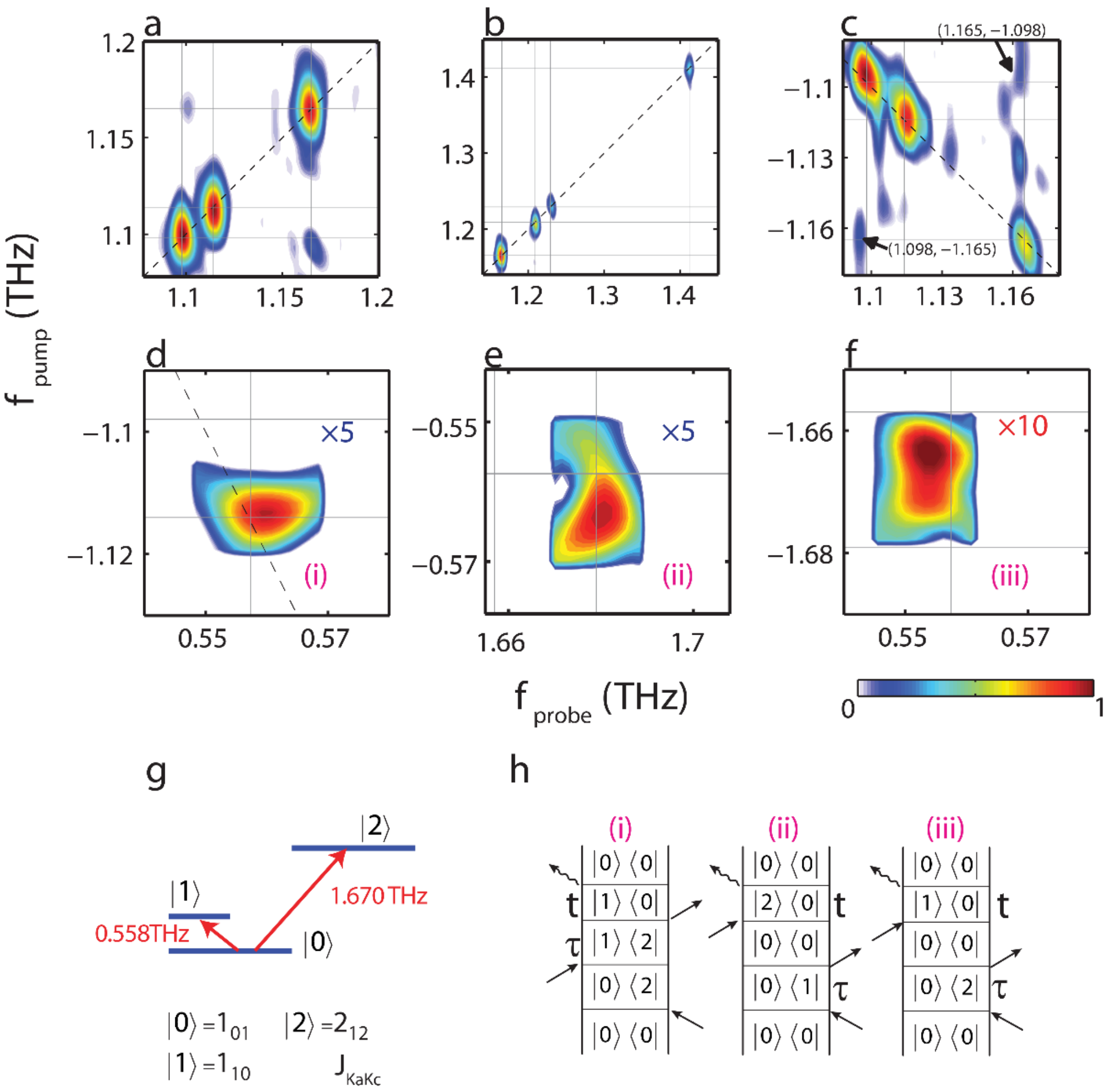

**Fig. 3. Close-up view of 2D rotational spectra of water vapor at 60 ºC. a-c**, Spectra recorded with extended detection time (up to 300 ps) in order to resolve closely spaced peaks. The two cross peaks at ( $f_{probe} = 1.098$ THz, $f_{pump} = -1.165$ THz ) and ( $f_{probe} = 1.165$ THz, $f_{pump} = -1.098$ THz) labelled in **c** arise from rotational coherences among levels $|3_{03}\rangle$, $|3_{12}\rangle$, and $|3_{21}\rangle$. **d-f**, Close-up views of three peaks. **g**, V-type system diagram. All 2D spectra are based on the same color. Representative peaks ((i) - (iii)) and their Feynman diagrams are listed in **h**.

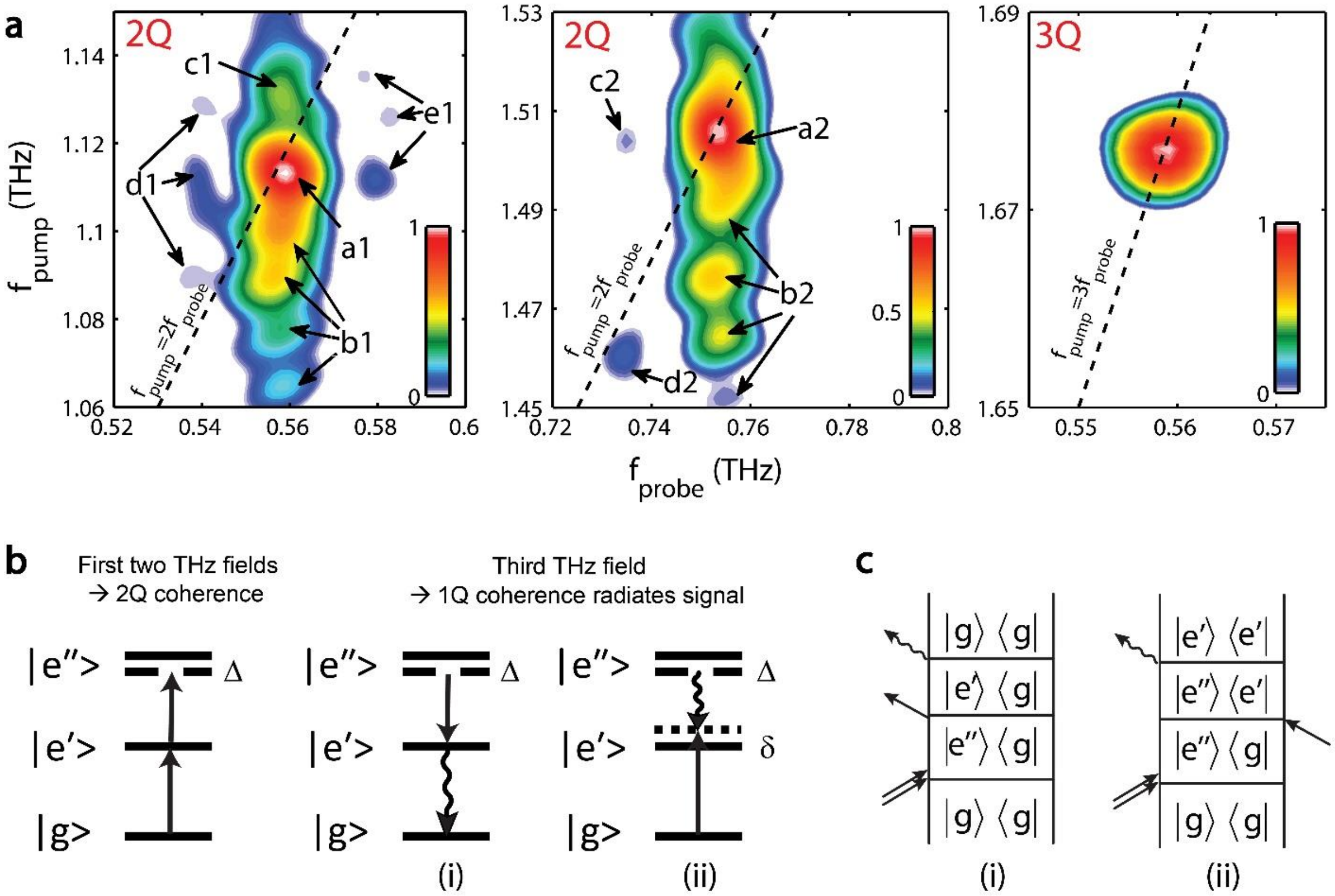

**Fig. 4. Many-body interactions in water vapor.** **a**, 2Q spectra near ( $f_{probe} = 0.558$ THz, $f_{pump} = 1.115$ THz ) (Left) and ( $f_{probe} = 0.753$ THz, $f_{pump} = 1.506$ THz ) (Middle) and 3Q spectrum at ( $f_{probe} = 0.558$ THz, $f_{pump} = 1.673$ THz ) (Right) from water vapor at 60 ºC. (a1 and a2) are 2Q diagonal peaks. (b1, c1, and b2) are 2Q off-diagonal features. (d1, e1, c2, and d2) are side peaks arising from distinct coherence pathways. **b**, Energy ladder diagram of water complexes with many-body interactions. The ground state $|g\rangle$, the singly excited state $|e'\rangle$, and the doubly excited states $|e''\rangle$ are illustrated. Energy level shifts (denoted as Δ and δ, see also Fig. S18) occur due to many-body interactions. Solid lines show the excited levels without interaction-induced shifts. The first two THz fields generate the 2Q coherence, which does not radiate. The third THz field may project |e″> downward (i) to the singly excited level |e′>, resulting in a coherence between |e′> and |g> that radiates at the usual 1Q transition frequency. Alternatively, the third THz field may promote from |g> to |e′>, resulting in a coherence between |e′> and |e″> whose frequency depends on the intermolecular geometry and whose shift from usual 1Q transition frequency may be different from the 2Q frequency shift Δ, **c**, Coherence pathways corresponding to the field interactions (i) and (ii) in **b** and to 2Q peaks which may show different spectral shifts along the pump (2Q) and probe (1Q) axes.

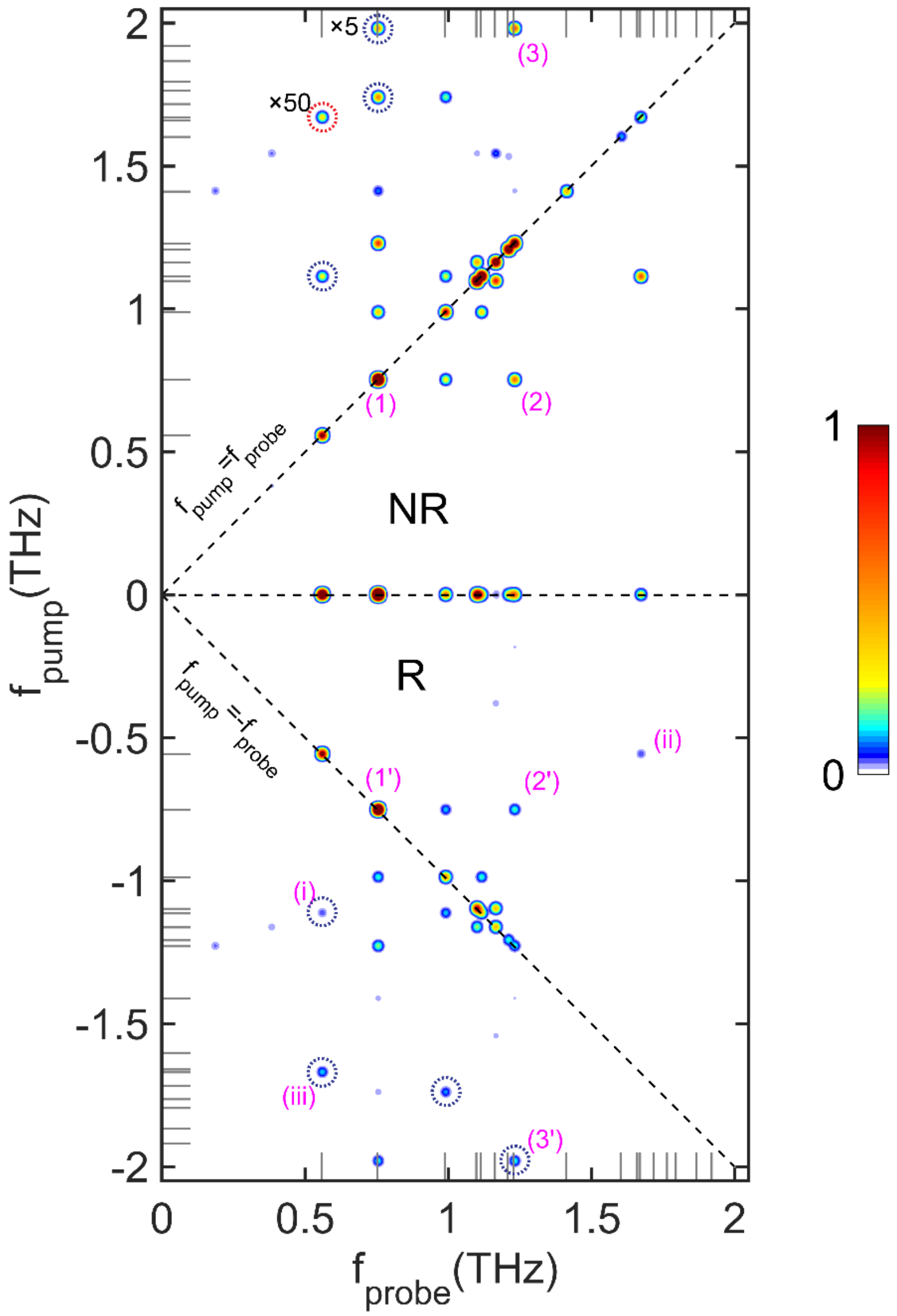

**Fig. 5. Simulated 2D spectrum for water vapor at 60 ºC.** Positive and negative $f_{pump}$ frequencies correspond to non-rephasing (NR) and rephasing (R) or echo signals respectively. No intermolecular interactions are included in the simulation. Several weaker peaks in circles are magnified by factors indicated (blue circle: ×5; red circle: ×50). Peaks with labels ((1)-(3), (1′) – (3′), and (i) – (iii)) correspond to those in the experimental spectra of Figs. 2 and 3. Short gray lines at the edges indicate rotational transition frequencies. Simulations including phenomenologically introduced interactions are presented in the supplementary information (Fig. S15), where weak higher-order nonlinear spectral peaks are also discussed.

# Supplementary Information for

## Nonlinear rotational spectroscopy reveals many-body interactions in water molecules

Yaqing Zhang, Jiaojian Shi, Xian Li, Stephen L. Coy, Robert W. Field, and Keith A. Nelson*

*Corresponding author. Email: kanelson@mit.edu.

**This PDF file includes:**

Supplementary Text

Figs. S1 to S27

Tables S1 to S7
Supplementary References

**Simulation details**

Coordinate system

Before we discuss any simulation details, it is beneficial to clarify the relevant coordinate systems (**Fig. S5**). The excitation THz field is defined in the space-fixed frame $(X, Y, Z)$. For a linearly polarized THz field, the polarization vector is assigned along the $+Z$ axis without any loss of generality.

Hamiltonian

A quantum mechanical rotational system is governed by the rotational Hamiltonian $H(t)$, which consists of a field independent Hamiltonian $H_0$, and a field-molecule interaction term $H_1(t)$,

$$H(t) = H_0 + H_1(t) \tag{S.1}$$

where $H(t)$ is the total rotational Hamiltonian.

The field-free term is

$$H_0 = AJ_a^2 + BJ_b^2 + CJ_c^2 \tag{S.2}$$

where $A$, $B$, and $C$ are three rotational constants inversely proportional to the three components of moment of inertia along $a$-, $b$- and $c$- axes, respectively ($J_a$, $J_b$, and $J_c$ are the three components of the total angular momentum $J$).

If we choose the following coordinate identifications:

$$a \leftrightarrow z, b \leftrightarrow x, c \leftrightarrow y \tag{S.3}$$

Accordingly,

$$H_0 = AJ_z^2 + BJ_x^2 + CJ_y^2 \tag{S.4}$$

The matrix elements, for example in the basis set of the symmetric top

$$\langle JKM | H_0 | J'K'M' \rangle \tag{S.5}$$

have been evaluated in textbooks(1). The other part of the total Hamiltonian is dictated by the field-dipole interaction,

$$H_1(t) = -\boldsymbol{\mu} \cdot \boldsymbol{E}_{THz}(t) \tag{S.6}$$

which can be expressed in the language of the irreducible spherical tensor operator as follows(1),

$$H_1(t) = -\sum_{p=-1}^{1} (-1)^p T_p^{(1)}(\mu) T_{-p}^{(1)}(E_{THz}(t)) \tag{S.7}$$

Evaluation of matrix elements

In the Cartesian coordinate system (space-fixed frame), we define the components of the THz field,

$$E_{THz}(t) = (E_X, E_Y, E_Z) = (0, 0, E_{THz}(t)) \tag{S.8}$$

$$T_0^1(E) = E_Z = E_{THz}(t)$$
$$T_{\pm1}^1(E) = \frac{\mp(E_X \pm iE_Y)}{\sqrt{2}} = 0 \tag{S.9}$$

On the other hand, the tensor components of the permanent dipole vector are given in the molecule-fixed frame in a similar manner,

$$\mu = (\mu_x, \mu_y, \mu_z) = (\mu_b, \mu_c, \mu_a) \tag{S.10}$$

$$T_0^{(1)}(\mu) = \mu_z = 0$$
$$T_{\pm1}^{(1)}(\mu) = \frac{\mp(\mu_x \pm i\mu_y)}{\sqrt{2}} = \frac{\mp\mu_b}{\sqrt{2}} \tag{S.11}$$

In order to evaluate the field-molecule interaction $H_1(t)$, a coordinate transformation is needed. If we transform the dipole moment from molecule-fixed frame into space-fixed frame, that is

$$T_0^{(1)}(\mu)|_{space} = \sum_{q=-1}^{1} D_{0q}^{(1)*} T_q^{(1)}(\mu)|_{molecule} = \frac{\mu}{\sqrt{2}}(D_{0,-1}^{(1)*} - D_{0,1}^{(1)*})$$
$$T_1^{(1)}(\mu)|_{space} = \sum_{q=-1}^{1} D_{1q}^{(1)*} T_q^{(1)}(\mu)|_{molecule} = \frac{\mu}{\sqrt{2}}(D_{1,-1}^{(1)*} - D_{1,1}^{(1)*}) \tag{S.12}$$
$$T_{-1}^{(1)}(\mu)|_{space} = \sum_{q=-1}^{1} D_{-1q}^{(1)*} T_q^{(1)}(\mu)|_{molecule} = \frac{\mu}{\sqrt{2}}(D_{-1,-1}^{(1)*} - D_{-1,1}^{(1)*})$$

then $H_1(t)$ is ready to be evaluated in that both the dipole moment and the excitation THz field are in the same frame (in this case, space-fixed frame).

The matrix element of $H_1(t)$ in the basis set of the symmetric top wavefunction $|JKM\rangle$ is given as

$$\begin{aligned}
&\langle JKM | H_1(t) | J'K'M' \rangle \\
&= -\mu_b E_{THz}(t) \left\langle JKM \left| \frac{1}{\sqrt{2}}(D_{0,-1}^{(1)*} - D_{0,1}^{(1)*}) \right| J'K'M' \right\rangle \\
&= -\frac{\mu_b E_{THz}(t)}{\sqrt{2}} (-1)^{M-K} \left[(2J+1)(2J'+1)\right]^{1/2} \times \\
&\left\{ \begin{pmatrix} J & 1 & J' \\ -K & -1 & K' \end{pmatrix} - \begin{pmatrix} J & 1 & J' \\ -K & 1 & K' \end{pmatrix} \right\} \begin{pmatrix} J & 1 & J' \\ -M & 0 & M' \end{pmatrix}
\end{aligned} \tag{S.13}$$

where $D$ is a rank (1) Wigner rotational matrix, and $\begin{pmatrix} \cdots \\ \cdots \end{pmatrix}$ is a 3-j symbol(1). The nonvanishing condition is

$$\Delta M = 0; \Delta J = 0, \pm1; \Delta K = \pm1 \tag{S.14}$$

The following relation is used in the calculation,

$$\left\langle J\Omega M \mid D_{Pq}^{(k)*}(\omega) \mid J'\Omega' M' \right\rangle = (-1)^{M-\Omega}\left[(2J+1)(2J'+1)\right]^{1/2} \times \begin{pmatrix} J & k & J' \\ -\Omega & q & \Omega' \end{pmatrix}\begin{pmatrix} J & k & J' \\ -M & P & M' \end{pmatrix} \tag{S.15}$$

The Wigner 3-j symbol is evaluated numerically by the Racah formula(1). Alternatively, the matrix element can also be expressed in the basis set of the asymmetric-top wavefunction $| JK_aK_cM >=| J\tau M >$ . Expanding $| J\tau M >$ as a linear combination of the symmetric-top wavefunction, one obtains

$$\begin{aligned} | J'\tau' M' \rangle &= \sum_{K'} a_{K'\tau'}^{J'} | J'K'M' \rangle \\ \langle J\tau M | &= \sum_{K} \langle JKM | (a_{K\tau}^{J})^* \end{aligned} \tag{S.16}$$

Inserting the identity operator, we find,

$$\begin{aligned} &\left\langle J\tau M \mid H_1(t) \mid J'\tau' M' \right\rangle \\ &= \left\langle JKaKcM \mid -\mu_b E_{THz}(t)\cos\theta_b^Z \mid J'Ka'Kc'M' \right\rangle \\ &= \sum_{K''}\sum_{K'''}\left\langle J\tau M \mid J''K''M'' \right\rangle\left\langle J''K''M'' \mid H_1 \mid J'''K'''M''' \right\rangle\left\langle J'''K'''M''' \mid J'\tau'M' \right\rangle \\ &= \sum_{K}\sum_{K'}(a_{K\tau}^{J})^*\left\langle JKM \mid H_1(t) \mid J'K'M' \right\rangle(a_{K'\tau'}^{J'}) \end{aligned} \tag{S.17}$$

Therefore,

$$\begin{aligned} &\left\langle J\tau M \mid H_1(t) \mid J'\tau' M' \right\rangle \\ &= \begin{pmatrix} a_{K_{(1)\tau}}^{J} \\ a_{K_{(2)\tau}}^{J} \\ \vdots \\ a_{K_{(n)\tau}}^{J} \end{pmatrix}^* \begin{pmatrix} \left\langle JK \mid H_1 \mid J'K' \right\rangle_{1,1} & \cdots & \left\langle JK \mid H_1 \mid J'K' \right\rangle_{1,n} \\ \left\langle JK \mid H_1 \mid J'K' \right\rangle_{2,1} & \cdots & \left\langle JK \mid H_1 \mid J'K' \right\rangle_{2,n} \\ \vdots & \ddots & \vdots \\ \left\langle JK \mid H_1 \mid J'K' \right\rangle_{n,1} & \cdots & \left\langle JK \mid H_1 \mid J'K' \right\rangle_{n,n} \end{pmatrix} \begin{pmatrix} a_{K_{(1)\tau}}^{J} \\ a_{K_{(2)\tau}}^{J} \\ \vdots \\ a_{K_{(n)\tau}}^{J} \end{pmatrix} \\ &= (\left\langle JKM \mid J\tau M \right\rangle)^*(H_1(t)_{JKM,J'K'M'})(\left\langle J'K'M' \mid J'\tau'M' \right\rangle) \end{aligned} \tag{S.18}$$

The nonvanishing condition is the same as above. A typical implementation of the density matrix has been illustrated (**Fig. S6**) below.

<u>Time evolution</u>

At the condition of thermal equilibrium, the initial density matrix is

$$\begin{aligned} \rho(0) &= \frac{e^{-\beta H_0}}{Tr(e^{-\beta H_0})} \\ \beta &= \frac{1}{k_B T} \end{aligned} \tag{S.19}$$

or simply by

$$\rho(0) = \sum_i P_i |\psi_i\rangle\langle\psi_i| \tag{S.20}$$

where $P_i$ is the statistical probability of the system in the normalized pure state $|\psi_i\rangle$,

$$\sum_i P_i = 1 \tag{S.21}$$

The system evolution is described by the well-known Liouville-von Neumann equation,

$$\frac{d\rho(t)}{dt} = \frac{-i}{\hbar}\left[\rho(t), H(t)\right] \ . \tag{S.22}$$

the time-dependent density matrix can be calculated by the time propagation operator,

$$\rho(t) = e^{-i\frac{H(t)}{\hbar}t}\rho(0)e^{+i\frac{H(t)}{\hbar}t} \tag{S.23}$$

<u>Orientation calculation and THz field representation</u>

The ensemble averaged orientation is,

$$\langle\cos\theta\rangle(t) = Tr\left[\rho(t)\cos\theta\right] \tag{S.24}$$

And the nonlinear orientation factor(2),

$$\langle\cos\theta(\tau,t)\rangle|_{NL} = \langle\cos\theta(\tau,t)\rangle|_{12} - \langle\cos\theta(\tau,t)\rangle|_{1} - \langle\cos\theta(\tau,t)\rangle|_{2} - \langle\cos\theta(\tau,t)\rangle|_{BG} \tag{S.25}$$

Where subscript 12 denotes both THz fields on, 1 denotes only THz field 1 on, 2 denotes only THz field 2 on, and *BG* denotes both THz fields are off (see **Fig. S4**). Note that the orientation angle $\theta$ here is understood as the angle between THz field polarization (along +$Z$ axis) and the molecular dipole moment (along $b$ axis), so that $\theta = \theta_{b|molecule}^{Z|space}$, which can be further evaluated by the Wigner rotational matrix,

$$\theta = \theta_{b|molecule}^{Z|space} = \frac{1}{\sqrt{2}}(D_{0,-1}^{(1)*} - D_{0,1}^{(1)*}) \tag{S.26}$$

A THz field pulse is characterized by a Gaussian-envelope modulated sine or cosine wave with a center frequency $\omega_0$ ,

$$E_{THz}(t) = E_1 e^{-t^2/\tau^2}\cos(\omega_0 t + \varphi_1) \tag{S.27}$$

Two pulses with a time delay are described by,

$$E_{THz}(t) = E_1 e^{-t^2/\tau^2}\cos(\omega_0 t + \varphi_1) + E_2 e^{-(t+\delta t)^2/\tau^2}\cos(\omega_0(t+\delta t) + \varphi_2) \tag{S.28}$$

For the sake of simplicity, the initial phases are set to zero, $\varphi_1 = \varphi_2 = 0$.

<u>Watson's reduced Hamiltonian</u>

Centrifugal distortion has to be taken into consideration in order to accurately find the positions of rotational transition lines. Non-rigidity causes significant spectral line shifts, on the order of 1% of the line frequencies, and needs to be accounted for to avoid misassignments of the spectral lines. The line positions can be calculated with around 0.1% accuracy (absolute error on the order of 1 GHz) with centrifugal distortion taken into account (**Fig. S7-S10**).

One approach to including centrifugal distortion is Watson's reduced Hamiltonian. This approach is based on a perturbative treatment to high order, and is used extensively although it inherently diverges(3). Watson's reduced Hamiltonian, up to the $6^{th}$ order, is sufficient to address our concern. The total Hamiltonian is written as(4, 5),

$$H = H_r + H_d^{(4)} + H_d^{(6)} \tag{S.29}$$

where the unperturbed ($2^{nd}$ order) term is

$$H_r = \frac{1}{2}(B+C)J^2 + \left[A - \frac{1}{2}(B+C)\right](J_z^2 - b_p J_-^2) \tag{S.30}$$

The $4^{th}$ order term is given as

$$H_d^{(4)} = -\Delta_J J^4 - \Delta_{JK} J^2 J_z^2 - \Delta_K J_z^4 - 2\delta_J J^2 J_-^2 - \delta_K (J_z^2 J_-^2 + J_-^2 J_z^2) \tag{S.31}$$

And the $6^{th}$ order term,

$$\begin{aligned} H_d^{(6)} &= H_J J^6 + H_{JK} J^4 J_z^2 + \\ &H_{KJ} J^2 J_z^4 + H_K J_z^6 + 2h_J J^4 J_-^2 + h_{JK} J^2 (J_z^2 J_-^2 + J_-^2 J_z^2) + h_K (J_z^4 J_-^2 + J_-^2 J_z^4) \end{aligned} \tag{S.32}$$

with

$$\begin{aligned} b_p &= \frac{C-B}{2A-B-C} \\ J_-^2 &= J_x^2 - J_y^2 \end{aligned} \tag{S.33}$$

The rotational constants used in this work are listed (**Table S1**).

In order to implement the calculation of matrix elements of the angular moment operator, we follow the method in textbooks(1) using commutation rules and matrix multiplication. Matrix elements of the angular momentum operator through $6^{th}$ order are evaluated and listed below (**Tables S3 and S4**).

Qualitative simulation of many-body interactions between water molecules

We model water intermolecular interactions qualitatively by treating an individual transition as a two-level quantum system and considering three nearby molecules. The three two-level systems yield together a four-level system via intermolecular interactions such as dipole-dipole interactions among the molecules (**Fig. S13**). The singly excited and doubly excited energy levels are degenerate, each consisting of three energy levels. The small energy shifts induced by many-body interactions are not shown in the figure. Combined three-level and multiple-level pictures can be interpreted in a similar way. A small energy shift of the combined energy levels is needed to break the symmetry of the combined states so that emission signals from distinct pathways do not completely cancel each other(6-8).

In order to understand the many-body interactions among water molecules, a qualitative simulation is developed. The orientation simulation follows the time propagation mechanism described above, with a new Hamiltonian in which the interactions give rise to perturbation terms represented by off-diagonal elements (labelled as $a$ in **Fig. S14**). Diagonalization gives a set of new eigen-energies. Generally, the coupling strength is very small ($a \ll \hbar\omega$) and a multi-level quantum system with an approximately equal energy spacing $\hbar\omega$ between neighboring, non-degenerate energy levels is created by many-body interactions. Note that the coupling elements (the assignments of $a$) are not necessarily the same as shown in **Fig. S14**. Different arrangements of the coupling elements produce similar energy levels.

Up to 6 combined energy levels have been included in the simulation. The transition frequency is specifically chosen as 0.558 THz (the first ortho-type transition $1_{01} \leftrightarrow 1_{10}$) in water. 2Q and 3Q peaks with respect to the 0.753 THz transition can be implemented in a similar way. The transition dipole matrix is set as that of a linear or prolate symmetric-top molecule in which only coherences between distinct J states are considered. A quantitative and accurate simulation of both diagonal and off-diagonal features may require a complete knowledge of the intermolecular interaction and the newly formed Hamiltonian.

## Supplementary Text

Rotational energy levels of water molecules

The population distribution of rotational states of a linear molecule at the condition of thermal equilibrium follows the Boltzmann distribution(5) (**Fig. S11**)

$$Population \propto (2J+1)e^{-E_{JKaKc}/k_B T} \tag{S.34}$$

where the factor $(2J+1)$ is introduced due to the $M$ sublevel degeneracy. As a result, at or not far from ambient temperature, most of the rotational population stays in low $J$ states. The majority of the rotational population is under $J = 10$ at room temperature, and this value is used as a cut-off number to simplify the simulation. We note that (i) the population distribution within a single $J$ state is not uniform (**Fig. S11)**, in contrast to linear or symmetric-top molecules; (ii) any two successive para (ortho) rotational states are always separated by an ortho (para) state, leading to no adjacent rotational states with the same nuclear spin configuration within one specific $J$ quantum state.

The rotational energy levels of a quantum-mechanical rigid rotor are given by

$$H_0 = \frac{J_a^2}{2I_a} + \frac{J_b^2}{2I_b} + \frac{J_c^2}{2I_c} = AJ_a^2 + BJ_b^2 + CJ_c^2 \tag{S.35}$$

where $A$, $B$, and $C$ $(A \geq B \geq C)$ are three rotational constants, each of which is inversely proportional to $I_a, I_b,$ and $I_c$ $(I_a \leq I_b \leq I_c)$, the three components of the moment of inertia. For water, the values are listed with other related rotational constants below (**Table S1**).

For a linear polar molecule (e.g. OCS), $I_a = 0; I_b = I_c$, and $A = \infty; B = C$. The energy levels are solely determined by one rotational constant $B$, and no angular momentum occurs along the molecular axis (the direction of the permanent dipole moment) because a torque cannot exist along the molecular axis. A (prolate) symmetric top(2) (e.g. $CH_3CN$) , satisfying ($I_a < I_b = I_c$, and $A > B$ =C), has a slightly complicated rotational energy level structure described by introducing a new quantum number $K$, giving the projection of angular momentum onto the molecular axis. Fortunately, the selection rule requires that $\Delta J = \pm 1; \Delta K = 0$, thus reducing the energy level structure into multiple blocks distinguished by $K$ values. Within each energy level block, a symmetric top resembles a linear molecule.

When one moves to a small asymmetric-top molecule, especially one with a large asymmetry (e.g. water), the three principle moments of inertia are all unequal, as are the rotational constants($I_a < I_b < I_c$, and $A > B >$C), leading to an extremely irregular rotational energy level structure(4, 5). The transition energy values are no longer multiples of a single fundamental frequency(2). There is no simple pattern describing the various absorption peaks (**Fig. S12**).

<u>Selection rules</u>

The rotational selection rules for the water molecule is given (for linearly polarized THz fields) by(4, 5)

$$\begin{aligned} \Delta M &= 0, \\ \Delta J &= 0, \pm 1, \\ \Delta K_a &= \pm 1, \pm 3, \ldots \\ \Delta K_c &= \pm 1, \pm 3, \ldots \end{aligned} \tag{S.36}$$

Therefore, rotational transitions of water molecules involve not only those within one specific $J$ state (only $K$ quantum number is changing, Q-branch) but also between distinct $J$ states (both $J$ and $K$ are changing, P-/R-branches). In addition, many weak transition lines from $\Delta K = \pm 3$ also contribute to the final spectrum.

<u>Wavefunctions of an asymmetric top</u>

The wavefunctions of an asymmetric top are generally given as linear combinations of symmetric-top wavefunctions,

$$\begin{aligned} &| JK_aK_cM \rangle = | J\tau M \rangle = \sum_{K=-J}^{J} a_K (JK_aK_c) \, | JKM \rangle \\ &\tau = K_a - K_c \end{aligned} \tag{S.37}$$

Two pseudo-quantum numbers $K_a, K_c$, or their difference $\tau$, are introduced to characterize the wavefunction of an asymmetric top. One can choose either $|JK_aK_cM\rangle$ or $|JKM\rangle$ as the basis set to construct the Hamiltonian matrix although the latter is only exact for symmetric tops.

<u>Spin statistics</u>

Water naturally exists as a mixture of two nuclear-spin species, para and ortho, which are determined by the total spin of the two hydrogen nuclei. For total spin $I = 0$, water is para, and $K_a + K_c$ is even; for total spin $I = 1$, water is ortho, and $K_a + K_c$ is odd. No transition between ortho and para is allowed, separating the water energy levels into two parts (**Fig. S12**). Different nuclear spin statistical weights for para- and ortho- transitions strongly affect transition line strengths.

Centrifugal distortion

Because the water molecule is very light, its rotational constants are notably large compared to most others. For example, the rotational constant $B$ of water molecule is 70 times larger than that of $\mathrm{OCS}$ and 47 times larger than $\mathrm{CH_3CN}$. As a consequence, water molecule shows an extraordinarily large centrifugal distortion effect even when the angular momentum quantum number $J$ is very small (e.g. $J < 5$), since the product of the rotational constant and the square of the angular momentum component (e.g. $BJ_b^2$) contributes to the Hamiltonian(4, 5). This non-rigid behavior shifts the absorption and emission spectra and further affects the rotational spectrum.

Mixing and collision

For our nonlinear 2D THz measurements, a gas cell was used. We discovered afterward that a small amount of the previous sample, gaseous acetonitrile, was adsorbed onto the aluminum surface of the gas cell. Although no couplings between water molecule and acetonitrile were found (a similar case has been reported where no coupling occurs between atoms of different rubidium isotopes (7)), a diagonal peak splitting around 0.557 THz is clearly observed. This is formed by the coincidence and interference of two third-order coherent processes: one in the water molecule with the $1_{01} \leftrightarrow 1_{10}$ coherence, and the other in acetonitrile with $J = 29 \leftrightarrow 30$.

At the typical relative humidity of 40% in our lab, there is about 10 Torr water vapor in the ambient air, and one can make full use of the water vapor for the linear THz absorption measurement (**Fig. S1**). Frequent collisions with nitrogen and oxygen molecules in the air broadens the linewidths to roughly twice those of pure water vapor.

Radiative interaction

One possible contribution to the two-quantum peaks is radiative interaction (9). After the main THz pulse interacts with one water molecule, the emitted THz field from this molecule will subsequently excite another water molecule, and the cascaded two-step excitation can also create two-quantum diagonal peaks. Similarly, three-quantum diagonal peaks can be generated involving radiative interactions among three water molecules or many-body interactions between two nearby molecules and radiative interactions with a third molecule. Radiative interactions cannot yield energy shifts away from simple sums of the separate transition energies, so such shifts provide clear signatures of many-body interactions.

2Q and 3Q peaks at 60ºC and 21ºC

All the 2D spectra shown in the main paper were recorded at approximately 60ºC in order to provide sufficient vapor pressure to yield acceptable signal/noise ratios. We recorded spectra at room temperature (21ºC) in order to confirm that the off-diagonal 2Q features in the spectra arose from intermolecular complexes(10) whose concentrations would be expected to decrease far more sharply with temperature than the water vapor pressure itself. As discussed in the main paper, the intensities of the off-diagonal 2Q peaks were observed to decrease much more sharply than the other peaks in the spectra, approximately proportional to the square of the ratio of vapor pressures as would be expected for peaks arising from molecular-pair complexes. The spectra for two groups of 2Q peaks at 60ºC and 21ºC are shown in **Figs. S16** and **S17** respectively. A spectrum showing a diagonal 3Q peak at 60ºC is also shown in **Fig. S16**. This peak was too weak to measure at 21ºC. At both temperatures, the 2Q spectra show multiple off-diagonal features which we believe indicate the presence of multiple intermolecular geometries. Their linewidths yielded an approximate dephasing time of 100 ps, which we take as an instrumentally limited lower limit for the lifetimes of the intermolecular complexes. Due to the limited signal/noise ratios in the room-temperature spectra, some of the features observed at 60ºC are not clearly distinguished, but the main features appear to be preserved with some degree of merging among nearby peaks. The room-temperature 2D spectra are presented with two different levels set for the lowest (white) colorbar scale, with the higher level allowing clearer distinction of the main features and the lower level allowing finer observation of the peak wings and the regions between peaks.

The side peaks (d1, e1, c2, d2, d3, e3, d4, and e4) arise from different coherence pathways. Energy-ladder diagrams and possible Feynman diagrams are illustrated in **Fig. S18**. In **Fig. S18a**, the ground state, the singly excited state, and the doubly excited state are shown as |g>, |e′>, and |e″>, respectively. The solid lines show the energy-level positions without shifts due to many-body interactions. The 2Q pump frequency is that of the coherence |e″><g|. Due to the energy shifts of the different intermolecular complexes, the corresponding peaks can be shifted upward or downward with respect to the 2Q diagonal peak. Peaks can also appear with shifts in the 1Q emitted signal frequency, which depends on the third THz field interaction which can yield the 1Q coherence |e′><g| or |e″><e′|), and which depends on the particular intermolecular geometry. Note that the shift in the 1Q frequency |e″ ><e′ |) may be different from the shift Δ in the 2Q frequency |e″ ><g|) because both the |e″ > and |e′ > energies may be shifted by many-body interactions involving multiple excitations. Similar effects have been observed in 2D electronic spectra of biexcitons(11).

Estimate of molecule-pair binding energies

The ratio of off-diagonal peak intensities at temperatures $T_2$ = 333 K (60ºC) and $T_1$ = 294 K (21ºC) can be used to estimate the binding energies for the intermolecular complexes. The water vapor pressure at 60ºC is greater than that at 21ºC by a factor of $X$ = 8.3. Peaks arising from water monomers should be more intense at 60ºC by a factor of $X$. We select the strong non-rephasing diagonal peak at ($f_{\mathrm{probe}}$ = 0.75 THz, $f_{\mathrm{pump}}$ = 0.75 THz) as an internal standard for the monomer peak intensity $I_{\mathrm{m}}$ at each temperature. Considering only the vapor pressure, the intensity $I_{\mathrm{p}}$ of an off-diagonal 2Q peak arising from a molecule-pair complex would increase by a factor of $X^2$ at the higher temperature, and thus the intensity ratio $I_{\mathrm{p}}/I_{\mathrm{m}}$ would increase by a factor of $X$. However, given the weak binding energies $\varepsilon_P$ of the complexes, the increased thermal energy at 60ºC will shift the equilibrium toward separated monomers, thereby reducing the molecule-pair peak intensities $I_{\mathrm{p}}$. The dependence of the intensity ratio $I_{\mathrm{p}}/I_{\mathrm{m}}$ on temperature can be calculated by considering the equilibrium between water monomers M and molecule-pair complexes C, i.e. 2M⇌C, with equilibrium constant $K_p$ given in terms of the vapor pressures $p_M$ and $p_C$ by

$$K_p = \frac{\left(\frac{p_C}{p^\circ}\right)}{\left(\frac{p_M}{p^\circ}\right)^2} = \frac{p_C p^\circ}{p^2{}_M} \tag{S.38}$$

where $p^\circ = 1\ \text{bar}$. $K_p$ can be calculated from the molecular partition functions $q_M$ and $q_C$ using the expression

$$K_p = e^{-\varepsilon_P/k_BT} \frac{\left(\frac{q_C}{V}\right)\left(\frac{k_BT}{p^\circ}\right)}{\left[\left(\frac{q_M}{V}\right)\left(\frac{k_BT}{p^\circ}\right)\right]^2} = e^{-\varepsilon_P/k_BT}\left(\frac{h^2}{2\pi k_BT}\right)^{3/2}\left(\frac{2}{m_M}\right)^{3/2}\frac{p^\circ}{k_BT}\left(\frac{k_BT}{h\nu}\right)^3 = e^{-\varepsilon_P/k_BT}\frac{p^\circ\left(k_BT\right)^{1/2}}{\left(\pi m_M\right)^{3/2}\nu^3} \tag{S.39}$$

where $m_M$ is the water monomer mass and $\nu$ is the frequency of any of three vibrational modes in which the two molecules in a complex move relative to each other. The three frequencies would not really be equal to each other, but for our purposes their values will not matter except that they are assumed to be low enough that the modes are thermally excited so the vibrational partition function for each takes the simple high-temperature limiting form $q_{vib} = k_BT/h\nu$. We have simplified the treatment greatly by assuming that the two molecules in a complex have the same internal molecular vibrational modes and the same independent rotations as the separated molecules, so all the partition functions are identical except for the masses in the translational partition functions and partition function for the three intermolecular vibrational modes that account for the translational degrees of freedom that are lost when two monomers come together to form a molecule-pair complex. We have not accounted explicitly for zero-point vibrational energies of these modes since we are seeking only a rough estimate for the binding energy $E_B$.

Combining the expressions for $K_p$ gives

$$p_C = p_M^2 e^{-\varepsilon_p/k_BT}\frac{(k_BT)^{1/2}}{(\pi m_M)^{3/2}\nu^3} \tag{S.40}$$

The ratios of spectral features arising from complexes and monomers (i.e. 2Q off-diagonal peaks and any others in the spectrum) are proportional to the ratio of vapor pressures $p_C/p_M$. We measure these ratios at two temperatures $T_1$ = 294 K and $T_2$ = 333 K, then take their ratio. The water monomer vapor pressure $p_M$ increases by a factor of $X$ = 8.3 going from $T_1$ to $T_2$.

$$\frac{\frac{p_C}{p_M}(T_2)}{\frac{p_C}{p_M}(T_1)} = \frac{p_M(T_2)}{p_M(T_1)} \frac{e^{-\varepsilon_p/k_B T_2}}{e^{-\varepsilon_p/k_B T_1}} \left(\frac{T_2}{T_1}\right)^{1/2} = X \frac{e^{-\varepsilon_p/k_B T_2}}{e^{-\varepsilon_p/k_B T_1}} (1.06) \tag{S.41}$$

We believe the peaks with different 2Q frequency values represent different molecule-pair complex geometries, and in principle we could evaluate the each of the peak intensities at both temperatures and determine the binding energy for each distinct geometry. Given the signal/noise limitations in the spectra and the associated uncertainties in peak intensities $I_p$, we have instead calculated an average value of the left-hand side of eq. (S.41) by averaging the intensities of peaks b1 ($b1_3$ – $b1_6$) and b3 ($b3_1$ – $b3_3$) in the spectra shown in **Figs. S16** and **S17** respectively. The resulting value is 5.5, which as expected is somewhat less than $X$ = 8.3. From this we estimate the molar binding energy $\varepsilon_P \approx 10\ \mathrm{kJ/mol}$, smaller than the binding energy of the well-known hydrogen-bonded water dimer (12-14). The somewhat wide range of intensity ratios (using the peaks ($b1_6$ and $b3_3$) gives a ratio of 5.9, while using peaks ($b1_3$ and $b3_1$) gives a ratio of 3.9 may be due to some combination of experimental uncertainties and real variation among the binding energies associated with different complex geometries.

Field dependences and Feynman diagrams for selected spectral peaks

Feynman diagrams of many spectral peaks of water molecules can be identified and assigned based on previous experimental results and analyses(2, 15), especially the diagonal peaks and off-diagonal peaks not far away from the diagonal. Other off-diagonal peaks arise from either 2Q or 3Q coherences (See **Figs. 2** and **3**). The 2D water rotational spectra from both simulations and experimental data reveal the complexity of the 2Q and 3Q signals, because they can occur due to three distinct origins: (i) many-body interactions; (ii) radiative interactions; and (iii) nonlinear responses of individual water molecules. Since the simulation conditions are similar to those in the experiments, the much smaller 2Q and 3Q signal amplitudes (blue and black signals in **Fig. S19b**) than the experimental signals (red signals in **Fig. S19b**) reveal that weak nonlinear responses (**Fig. S22b**) from non-interacting water molecules are not contributing significantly to 2Q and 3Q features. (i) and (ii) together are estimated to have a dominant contribution (>90%) to the overall multiple-quantum signals, while (iii) has a small contribution (<10%). Signals from (iii) are overlapped and merged with the strong 2Q/3Q signals in the NR spectra, but some nonlinear responses from (iii) can be observed in the R spectra, offering another approach to estimating the contributions of distinct origins of the 2Q/3Q signals and giving the same results. In addition, both NR and R contributions from (iii) can be captured in the simulation (without many-body interactions) (**Fig. 5**).

Simulated field dependences of several selected spectral peaks are shown (**Fig. S19a**). As expected, peaks (P2-P6) show a third-order dependence on THz field strength in the moderate THz field regime. P1, overlapped with the 2Q signal, is fifth order on the THz field amplitude, because no corresponding third-order responses can be assigned to it. In particular, one cannot connect the coherences |2><1| and |1><0| by one single interaction between THz fields and water molecules in the V-type energy ladder diagram (**Fig. S20**). However, its counterpart in the R spectra (i.e. P3 in **Fig. S19**) is third order since one THz excitation process can connect |1><2| and |1><0| on the bra side of the Feynman diagram.

As a consequence, only two peaks in our measurement belong to fifth-order signals, namely, (a) 3Q signal (**Fig. 3a**) and (b) the fifth-order signal discussed above which is overlapped with the 2Q signal (**Fig. S22b** and P1 in **Fig. S19**). The rest of the 2D spectral peaks are all third-order signals.

Note that different excitation pathways in the Feynman diagram (for example, the two distinct excitation pathways $|0\rangle\langle 0| \rightarrow |1\rangle\langle 0| \rightarrow |1\rangle\langle 2|$ (**b** (1) in **Fig. S20**) and $|0\rangle\langle 0| \rightarrow |0\rangle\langle 2| \rightarrow |1\rangle\langle 2|$ (**b** (2) in **Fig. S20**) could be distinguished through the use of three pulses to temporally separate each of the field interactions temporally, revealing the first coherence frequency from the first interaction in addition to the second and third field interactions, the variable delays of which reveal the 2Q coherence(11, 16).

<u>Overview of 2D THz spectra of water vapor</u>

We show complete non-rephasing (NR) and rephasing (R) 2D THz spectra in **Figs. S24-27**. Simulated spectra are shown in **Figs. S24** and **S26**, and **Figs. S25** and **S27** illustrate experimental spectra.

Main features in the experiments are reproduced in the simulation. Simulated NR, R, and pump-probe (PP) features match well with those in the experiments. Some peaks are magnified by appropriate factors, for the sake of clarity. In the experimental data, spectral features from acetonitrile residue are indicated. In the simulation data, weak spectral signals involving weak water transitions (e.g. see **Figs. S12** and **S23**) appear (labelled as W1-7 in **Fig. S24** and WR1-5 in **Fig. S26**). Additionally, in order to clearly present spectral peaks within (0.9 – 1.3 THz), enlarged spectra are illustrated in **Figs. S24b** and **S26b**. Peak labels in the experimental spectra (**Figs. S25&S27**) follow those in the simulation (**Figs. S24&S26**).

Minor discrepancies between experimental (**Figs. S25&S27**) and simulated (**Figs. S24&S26**) spectra occur, due to that: (i) signal-to-noise ratio limitations in the measurement; (ii) slight difference between experimental and simulated THz field profiles; (iii) slight difference between rotational transition line intensities in water molecules between experiments and simulations. Note that these discrepancies are trivial, and do not influence present results in this work. Peak positions and their relations with water rotational states are summarized in **Table. S6&S7**. For a qualitative analysis including many-body interactions, see **Fig. S13-S15** and related descriptions.

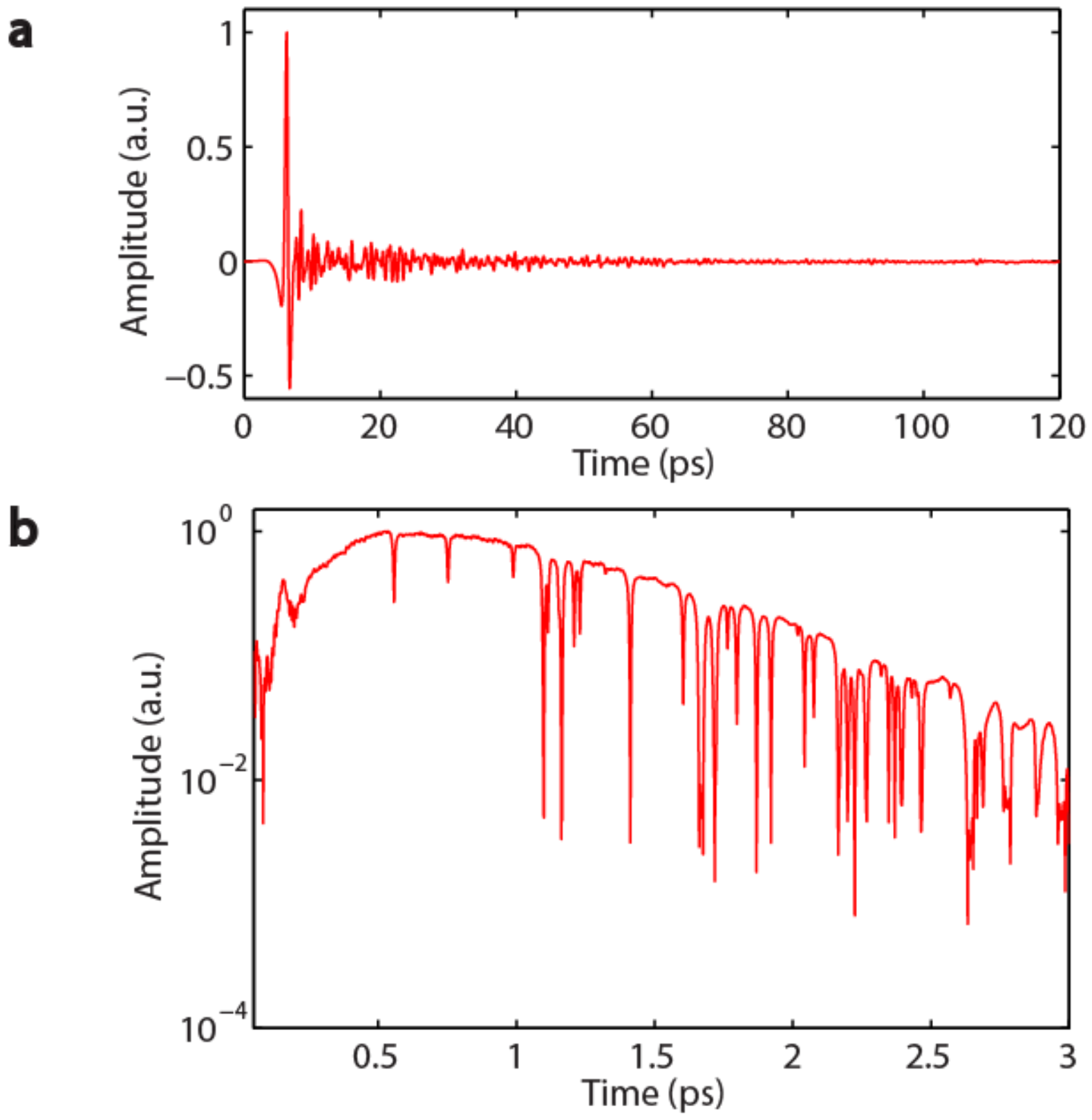

**Fig. S1. Measured THz spectrum of water vapor in ambient air**. **a**, The time domain signal is obtained by a typical THz time-domain spectroscopy (THz-TDS) system in which photoconductive antennae (PCA) are used for THz generation and detection as described in the text. The trailing oscillation after the main THz peak in the top panel stems from THz emission (i.e. the rotational free-induction decay) from water vapor in the ambient air. **b**, Fourier transform of the time domain signal (including both the main peak and the trailing oscillation) reveals the rotational absorption peaks.

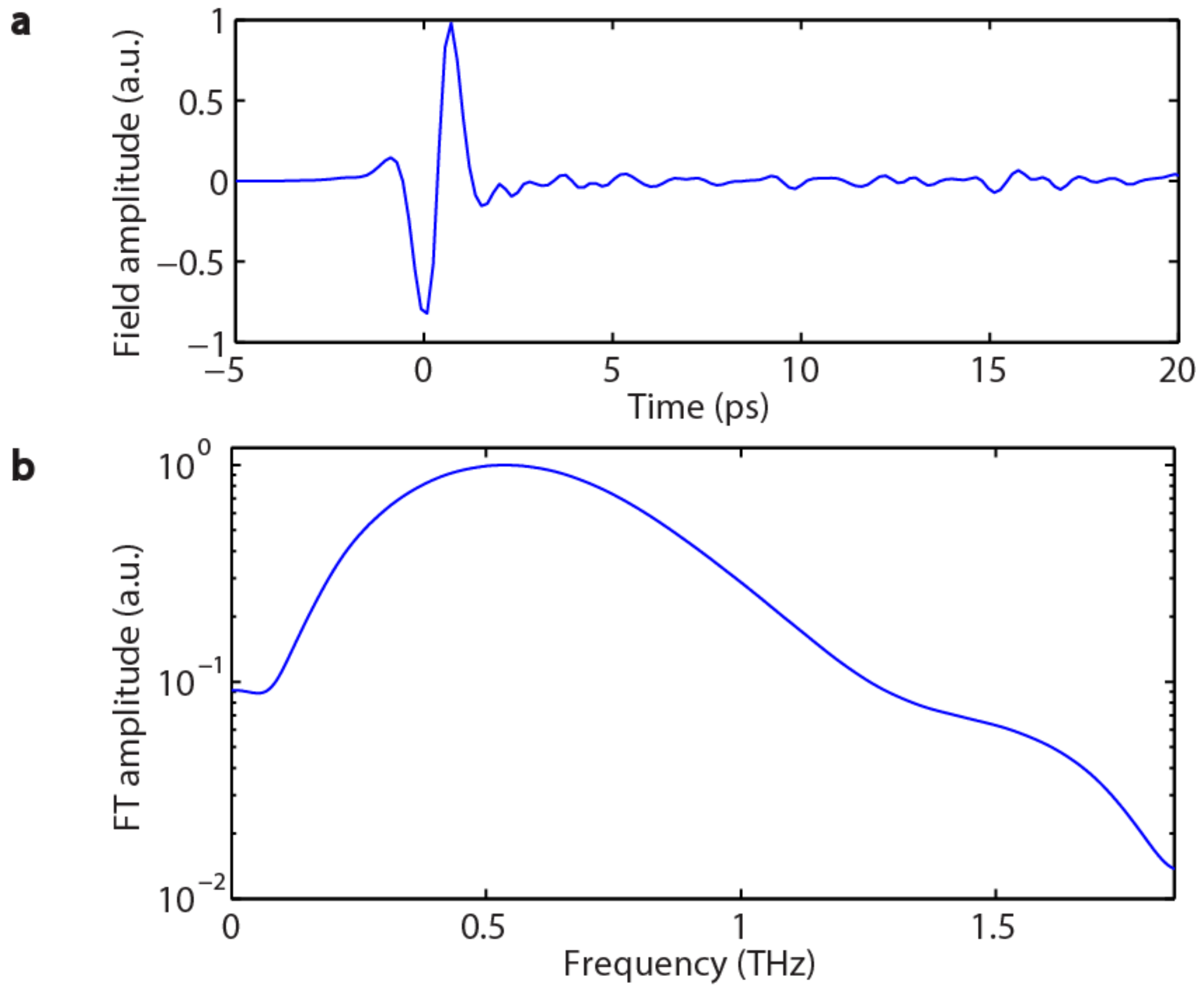

**Fig. S2. Typical profile of a single-cycle THz electric field**. The single-cycle THz field profile shown in this figure was generated in a lithium niobate crystal by the tilted-pulse-front technique. **a**, The time domain signal in the top panel includes the THz main peak around 0 ps and the trailing oscillation from water vapor emission. Double reflections from the gas cell window and optics in the setup appear beyond 20 ps and are excluded. **b**, FT spectrum from top panel, in which only the main THz peak is taken into account (zero-padding was added after the first large THz cycle.

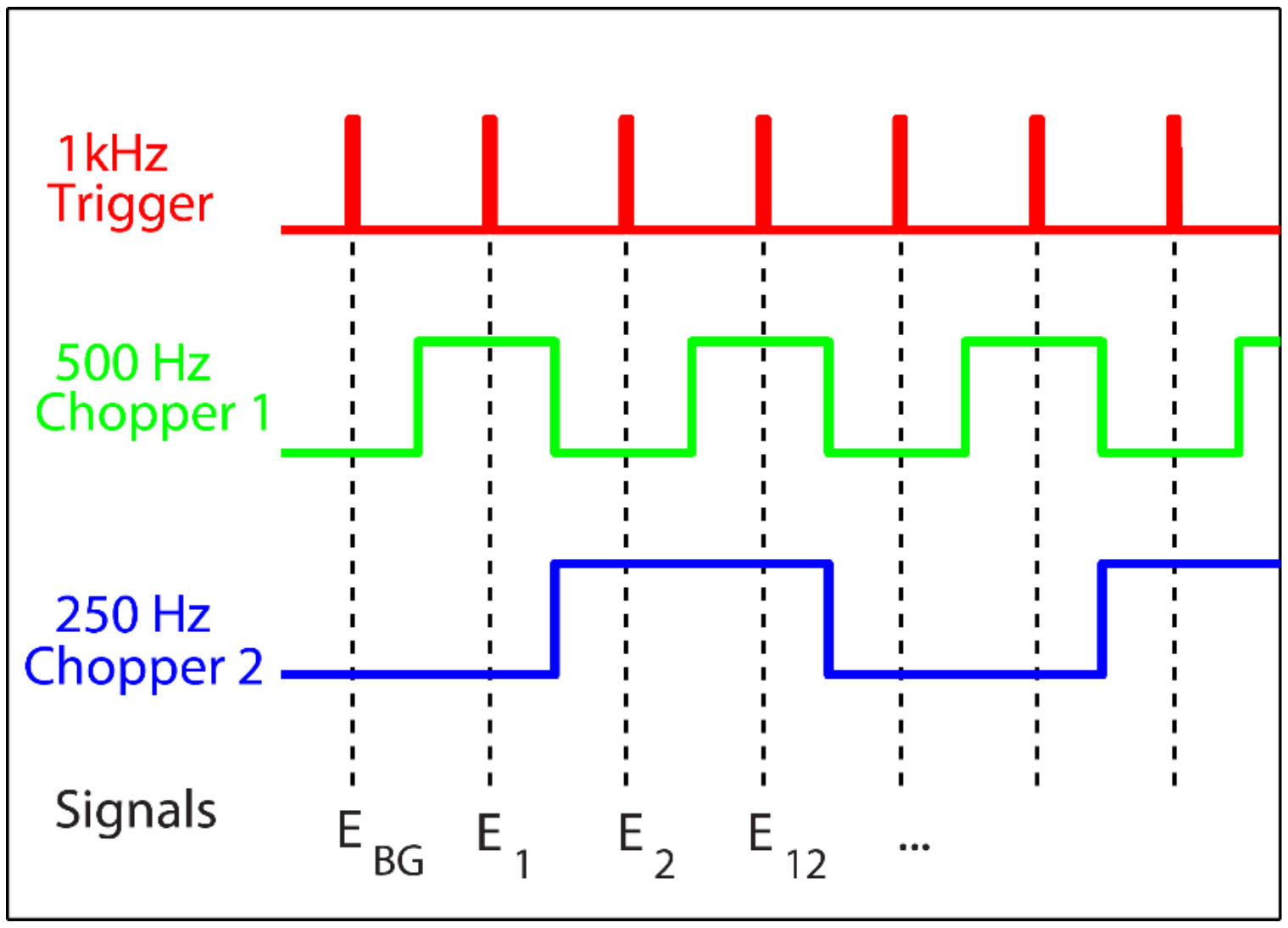

**Fig. S3. Mechanism of differential chopping detection**. The laser is run at a 1 kHz repetition rate (red). Chopper 1 (green) is operated at a 500 Hz frequency while the other chopper 2 (blue) is at 250 Hz. The signal sequence shown here consists of a background signal ( $E_{BG}$ ), an “only THz 1 on” signal ( $E_1$ ), an “only THz 2 on” signal ( $E_2$ ), and a “both THz 1 and 2 on” signal ( $E_{12}$ ).The nonlinear signal can be obtained by subtraction: $E_{NL} = E_{12} - E_1 - E_2$ .

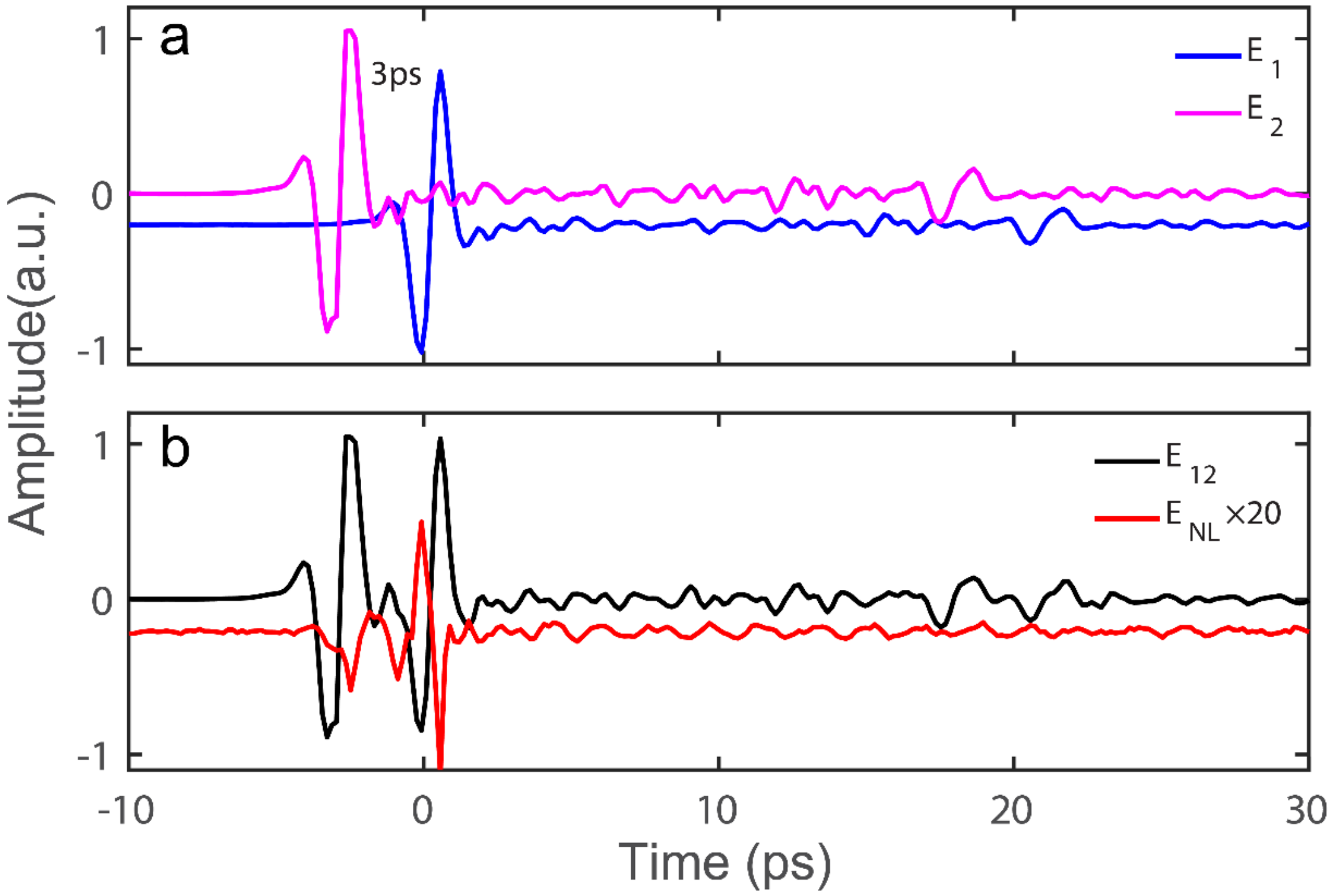

**Fig. S4. Typical linear and nonlinear signals in the experiment**. **a**, THz time-domain signals with only THz 1 (blue) and only THz 2 (magenta) on, respectively. **b**, THz signals when both THz 1 and 2 are on. The nonlinear THz signal (red) is obtained by subtraction of both $E_1$ and $E_2$ from $E_{12}$, namely $E_{NL} = E_{12} - E_1 - E_2$. The background noise can also be subtracted when necessary. The nonlinear signal is magnified by a factor of 20 for the sake of clarity. $E_1$ and $E_{\mathrm{NL}}$ are vertically displaced by the same value of 0.2 with respect to $E_2$ and $E_{12}$, respectively. Time delay between the two THz pulses is 3 ps.

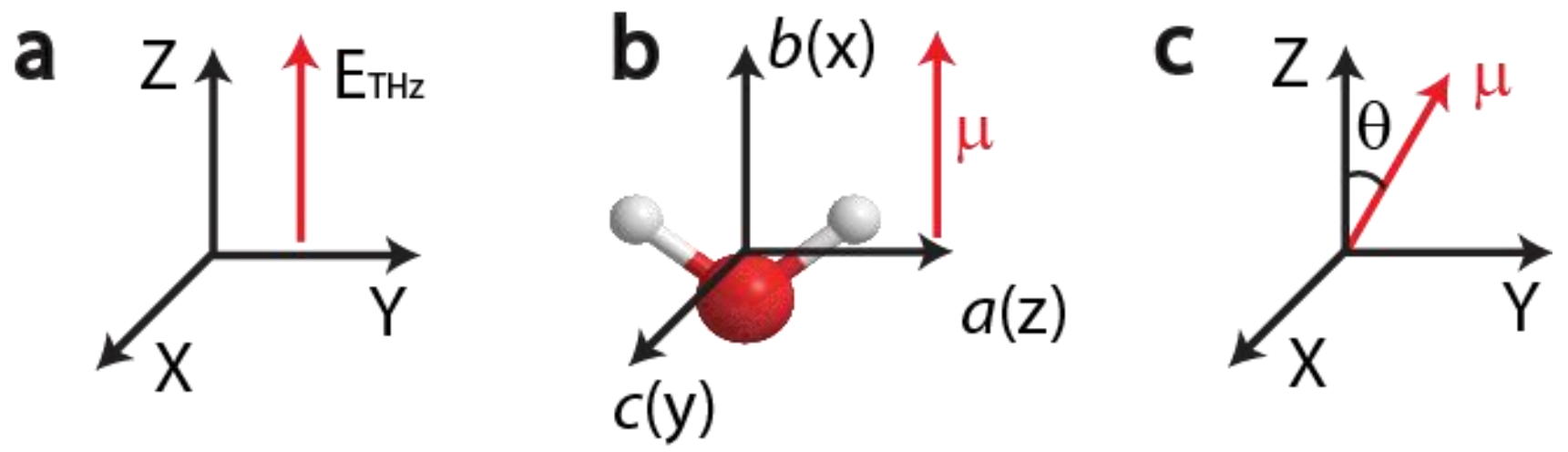

**Fig. S5. Coordinate system assignment**. **a**, THz field is polarized along the *Z* direction in the space-fixed frame $(X,Y,Z)$. **b**, Ball-stick model of a water molecule with its principle axes (*a*, *b*, *c*). The permanent dipole moment of water is along its principle b-axis. **c**, Orientation angle is identified as the intersection angle between the THz field polarization vector (along Z-axis in the space-fixed frame) and the dipole moment vector (along *b*-axis in the molecule-fixed frame). The identification of the principle axes (*a*, *b*, *c*) with molecule-fixed coordinates (x, y, z) is indicated in **b**.

| | | | Col# | 1 | 2 | 3 | 4 | 5 | 6 | 7 | 8 | 9 | ⋯ |
|---|---|---|---|---|---|---|---|---|---|---|---|---|---|
| | | | *J* | 0 | 1 | 1 | 1 | 2 | 2 | 2 | 2 | 2 | ⋯ |
| | | | *K* | 0 | -1 | 0 | 1 | -2 | -1 | 0 | 1 | 2 | ⋯ |
| Row# | *J* | *K* | | | | | | | | | | | |
| 1 | 0 | 0 | | * | | | | | | | | | |
| 2 | 1 | -1 | | | * | | X | | | | | | |
| 3 | 1 | 0 | | | | * | | | | | | | |
| 4 | 1 | 1 | | | X | | * | | | | | | |
| 5 | 2 | -2 | | | | | | * | | X | | | |
| 6 | 2 | -1 | | | | | | | * | | X | | |
| 7 | 2 | 0 | | | | | | X | | * | | X | |
| 8 | 2 | 1 | | | | | | | X | | * | | |
| 9 | 2 | 2 | | | | | | | | X | | * | |
| ⋮ | ⋮ | ⋮ | | | | | | | | | | | ⋱ |

**Fig. S6. Density matrix format of the rotational Hamiltonian**. Each column and row holds one specific rotational state denoted as $|JK>$ ($M$ quantum number is omitted for simplicity). Diagonal matrix elements (red star) are the population states. Off-diagonal matrix elements (green cross) are the coherence states. Matrix elements belonging to the same J quantum number are encompassed by a blue square. When implemented in the eigen-states of an asymmetric top $|JK_aK_c>$ or $|J\tau>$, the density matrix basically remains the same structure except that $K$ in the figure is replaced by $\tau = K_a - K_c$. The $M$ quantum number is omitted for simplicity.

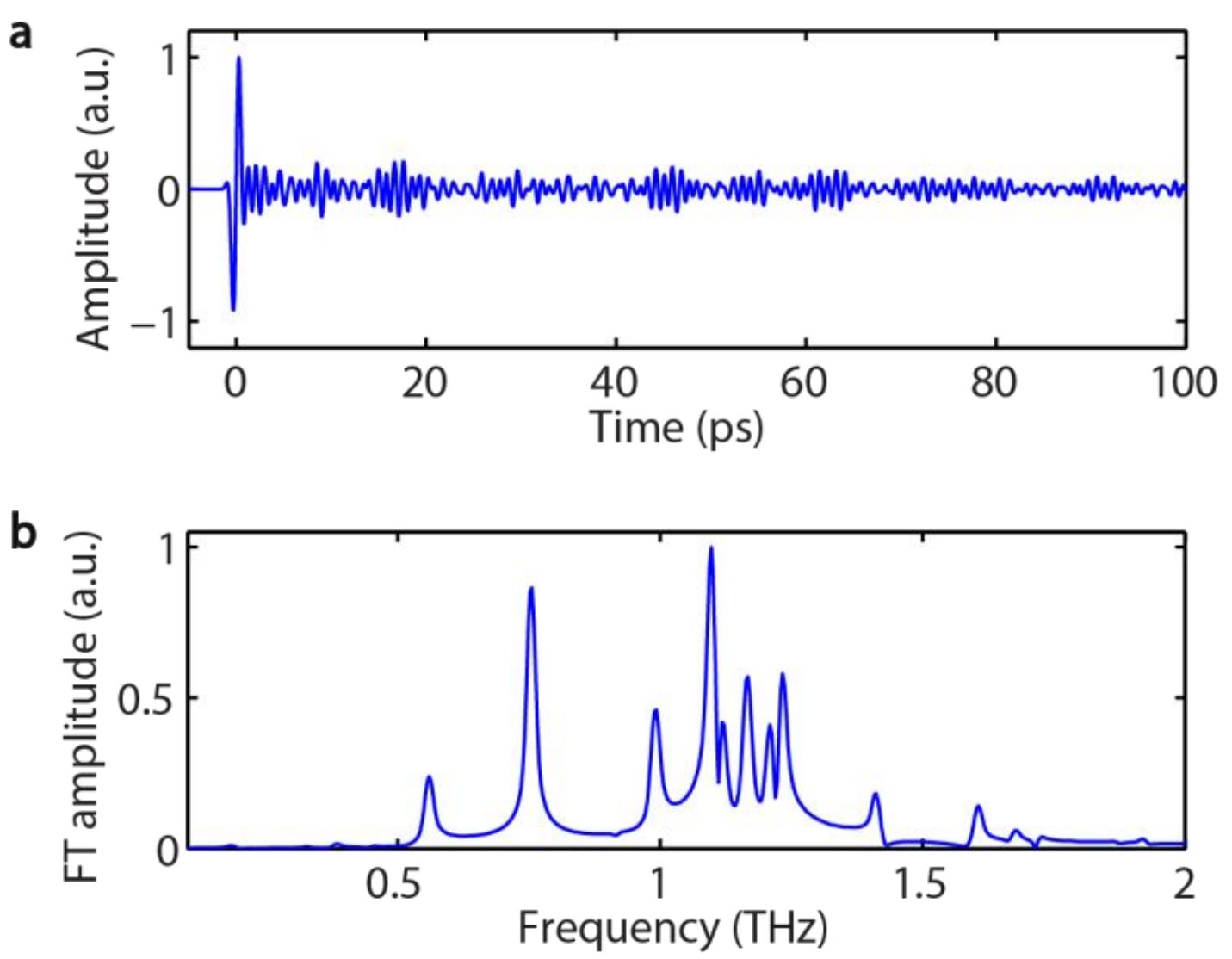

**Fig. S7. Simulation of linear THz spectroscopy of water molecules**. **a**, Simulated THz free-induction decay from water molecules. The maximum *J* quantum number was set at 15. *M* states considered here are those with $|M| \leq 4$. Temperature *T* = 100 K. The THz field strength was set at 300 kV/cm. **b**, FT spectrum from the time domain signal.

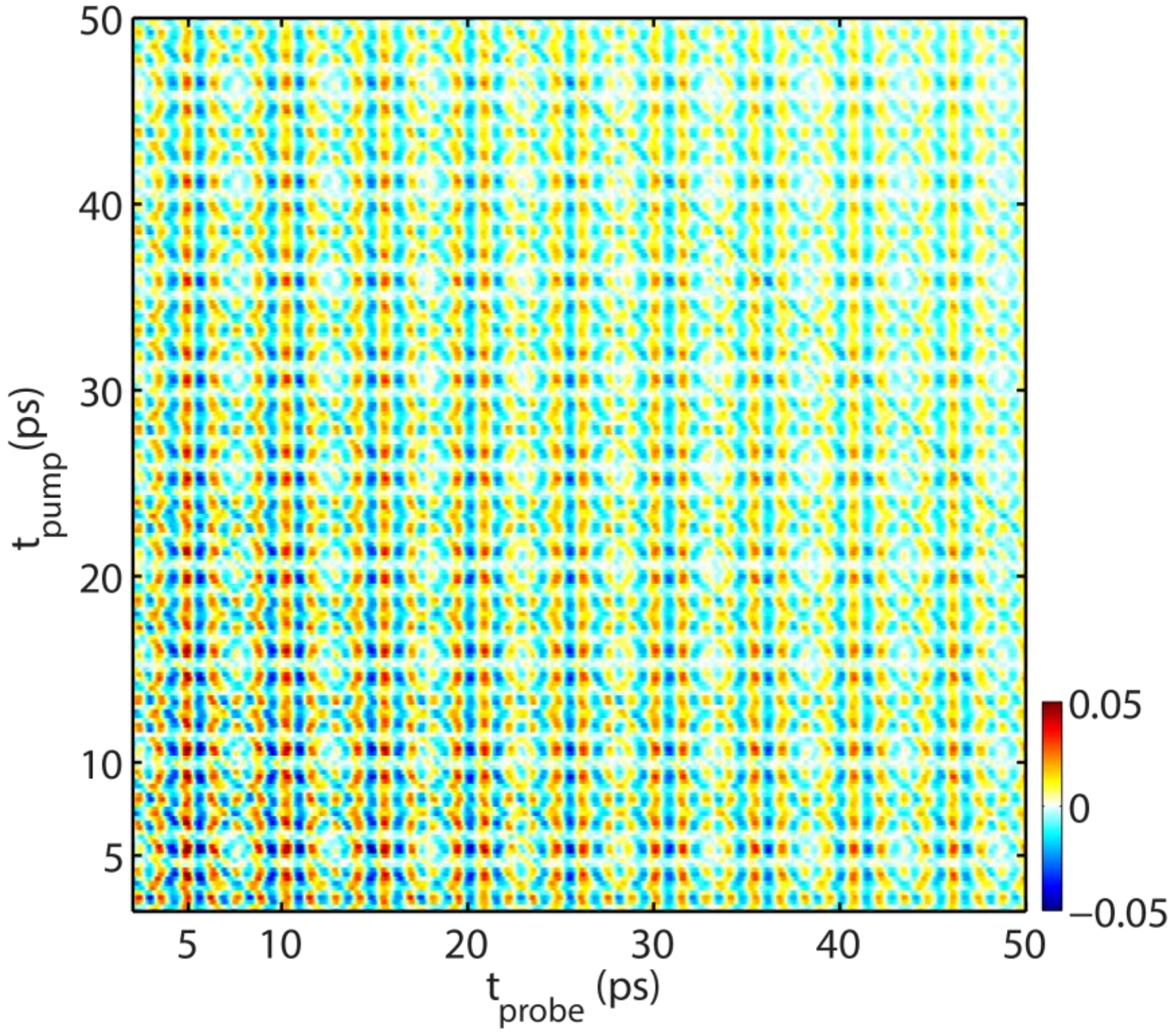

**Fig. S8. Simulated 2D THz signals in the time domain.** The maximum $J$ quantum number was set at 15. $M$ states considered were those with $|M| \leq 4$. Temperature was set at 100 K. Both THz field strengths were set at 300 kV/cm. The step size for the inter-pulse pump axis was 0.25 ps, and that for the probe axis was 0.1 ps. The similar features along both time scales correspond to diagonal peaks in the 2D spectrum. The pronounced periodicity of 1.3 ps (along both axes) corresponds to the strongest diagonal peaks in the spectra, at 0.75 THz (apparent in the measured spectra of **Fig. 2** and the simulated spectra of **Fig. 4**). The features that start at $t_{probe} \approx 0$ at any time along the $t_{pump}$ axis, i.e. signals that begin immediately after the two pulses, correspond to non-rephasing signals. The signals (clearest for the 1.3 ps periodic feature) that are at their approximate maxima along the diagonal correspond to rephasing (photon echo) signals, i.e. they reach their maxima (neglecting dephasing) at a time delay $t_{probe}$ equal to the delay $t_{pump}$ between the THz pump pulses.

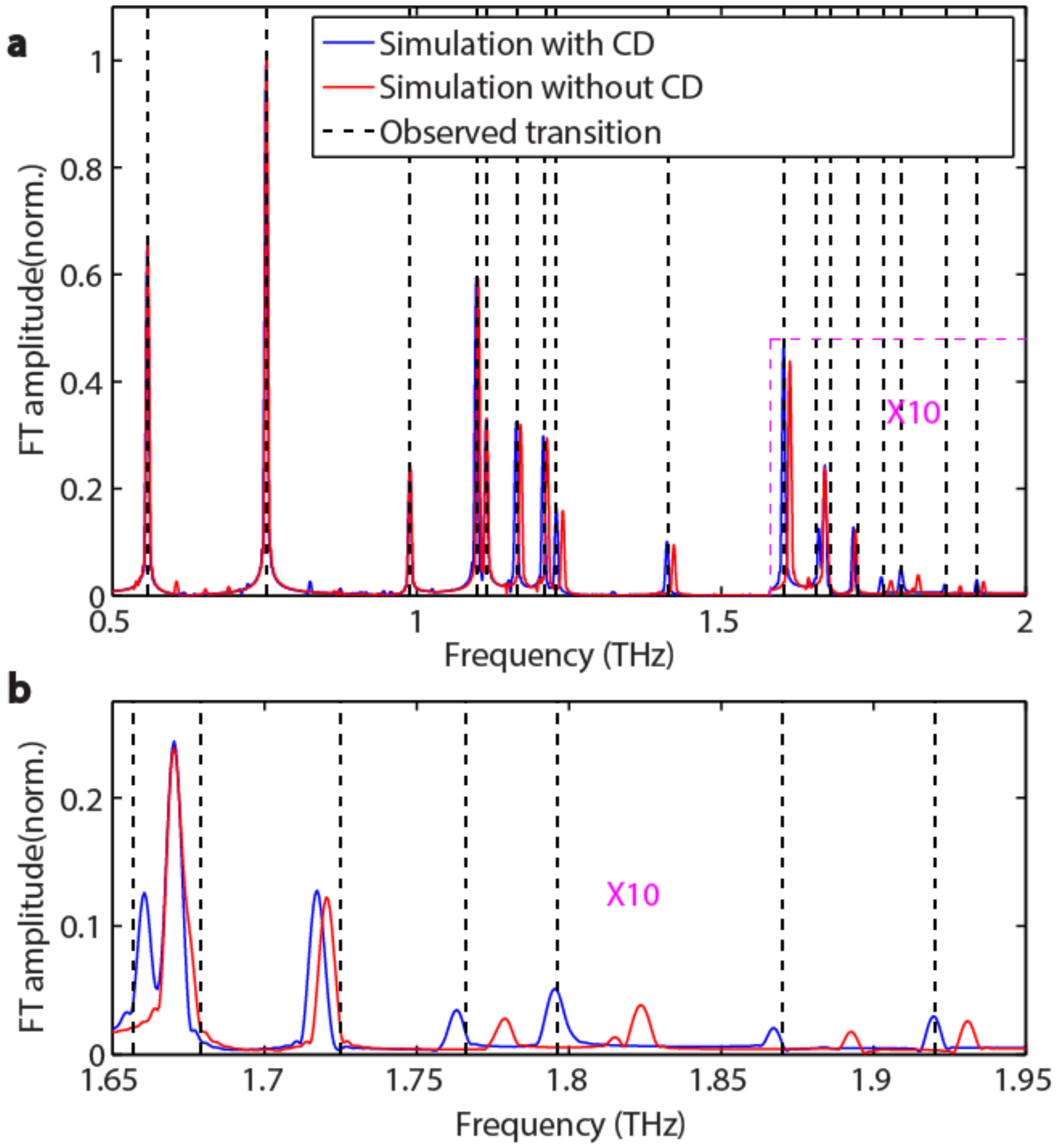

**Fig. S9. Centrifugal distortion effect on transition line positions**. **a**, Comparison between calculated line positions with and without centrifugal distortion (CD). **b**, Enlargement of the top panel in the region (1.65-1.95 THz). Deviations as large as 10 GHz are observed between experimentally observed transition frequencies and those calculated when no centrifugal distortion is taken into account (red line). Watson's reduced Hamiltonian has been employed to account for centrifugal distortion so that the deviation is reduced to around 1 GHz (blue line).

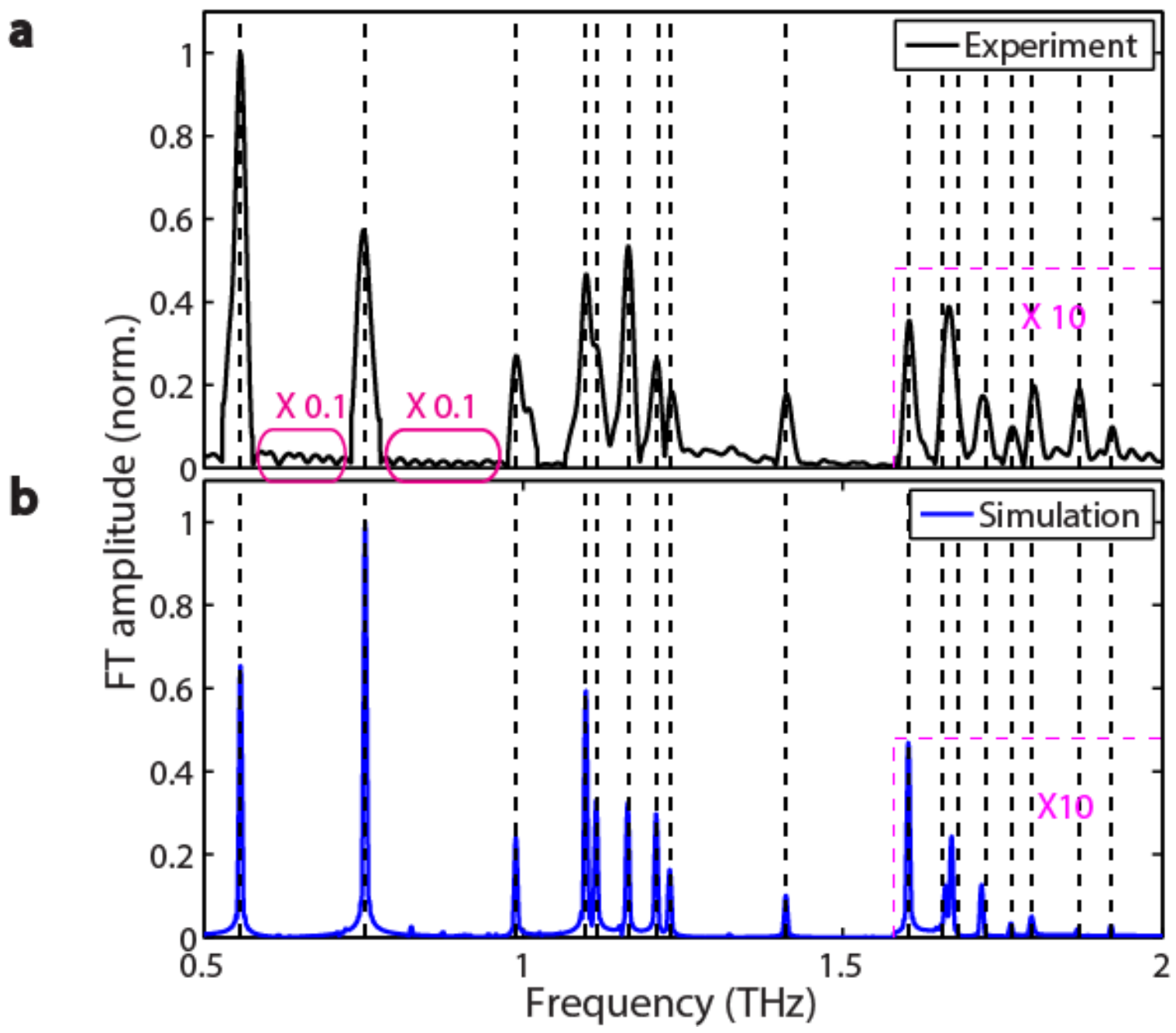

**Fig. S10. Experimental and simulated water THz spectra**. **a**, Measured THz spectrum of water vapor at 60ºC pumped by a THz pulse generated in a lithium niobate crystal. **b**, Simulated water spectrum using Watson's reduced Hamiltonian. The transitions above 1.5 THz are enlarged by a factor of 10 in both panels. Dashed black lines indicate the positions of observed rotational transition lines of water vapor in ambient air by a photoconductive antenna detection. Large oscillations of **a** in the frequency intervals (0.558 THz, 0.753 THz) and (0.753 THz, 0.989 THz) are from double reflections in the setup and reduced by a factor of 0.1, causing a slight distortion at the edges of peaks at 0.558 THz, 0.753 THz ,and 0.989 THz.

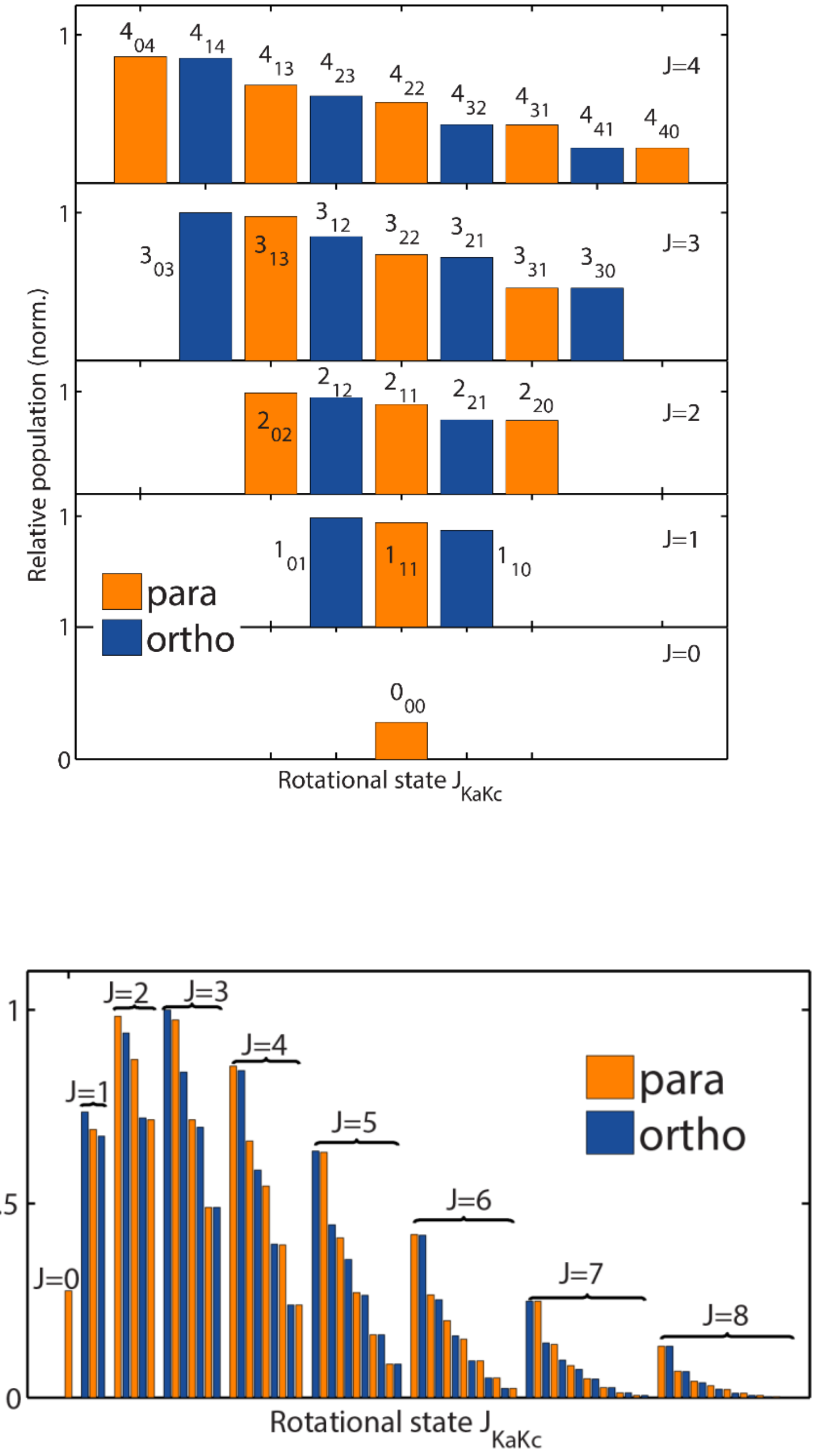

**Fig. S11. Calculated population distribution of water molecules at room temperature**. (Top) Relative population in each *J* (0-4) state is indicated. Orange and blue denote para and ortho water rotational states, respectively. The rotational states $J_{KaKc}$ are indicated. (Bottom) Relative population in each state up to *J* = 8. The largest populations occur at *J* = 2 and *J* = 3 when treating all rotational states sharing the same J quantum number as a whole.

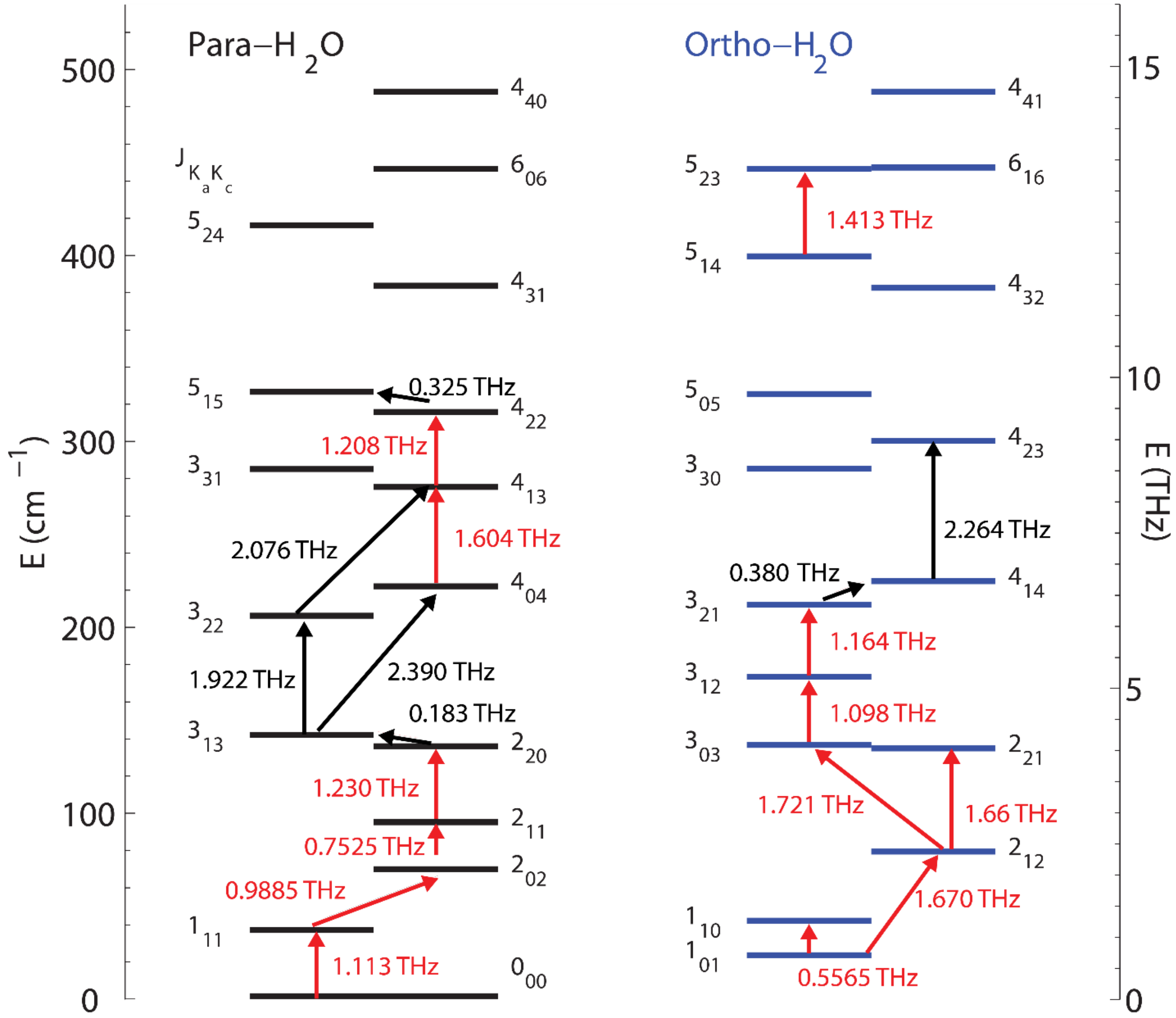

**Fig. S12. Energy-level diagram of water rotational states**. Rotational energy levels are divided into two parts: para-type (left side, in black) where the total nuclear spin of the two hydrogen atoms is 0, and ortho-type (right side, in blue) where the total nuclear spin is 1. Red arrows indicate allowed transitions between 0.5 THz and 1.8 THz observed in this work. Black arrows indicate other possible transitions either lower than 0.5 THz or higher than 1.8 THz. Two transitions below 2 THz, with frequencies of 1.798 THz ($6_{15} \leftrightarrow 6_{24}$ and $7_{25} \leftrightarrow 7_{34}$) and 1.871 THz ($5_{23} \leftrightarrow 5_{32}$)) are associated with higher-energy rotational states than those shown here.

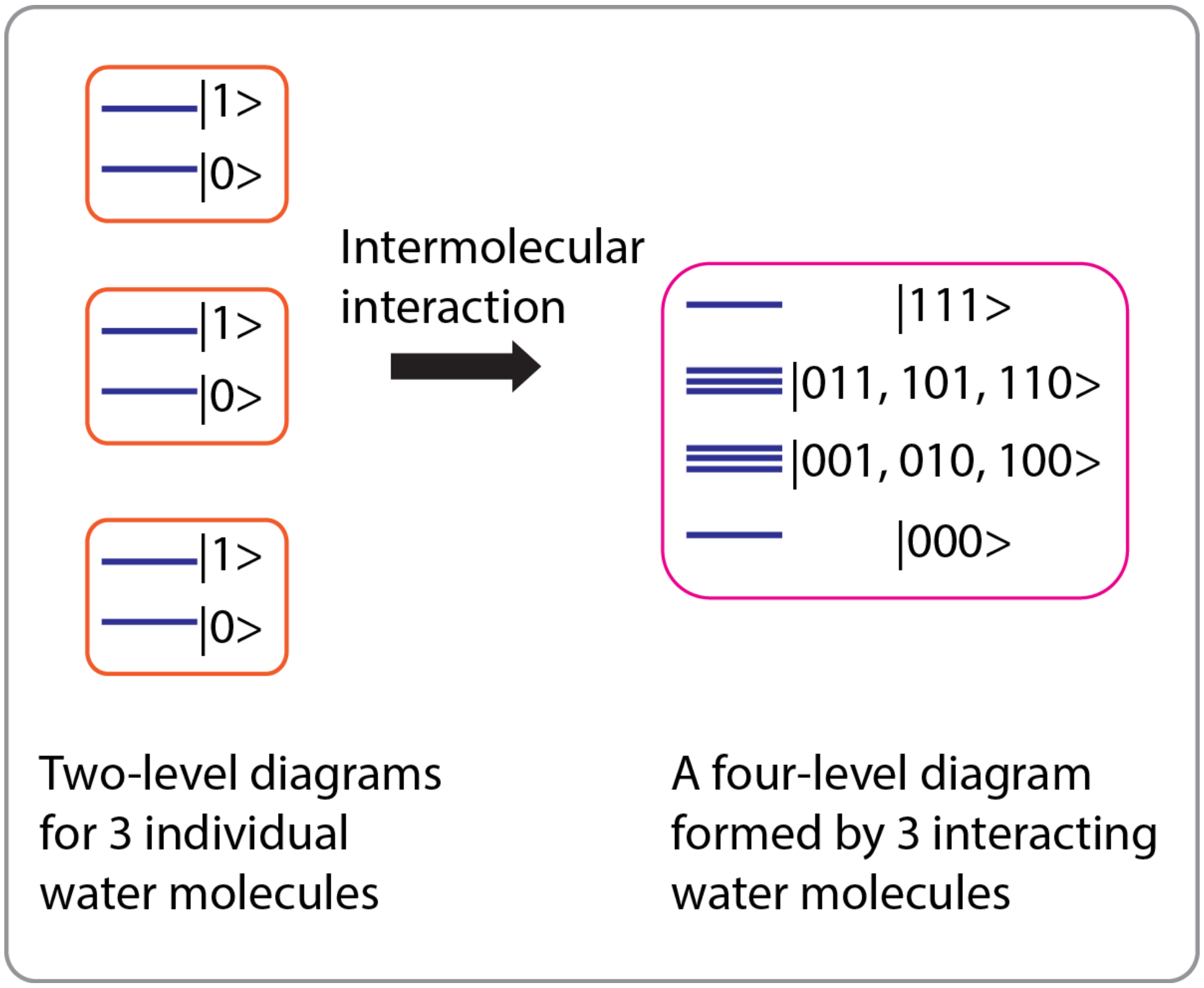

**Fig. S13. Diagram of the formation of a four-level system in water molecules**. Three two-level quantum systems of three individual water molecules generate a four-level quantum system via intermolecular interactions such as dipole-dipole interactions among water molecules. The singly excited and doubly excited energy levels are three-fold degenerate in the absence of interactions but non-degenerate when interactions are taken into account.

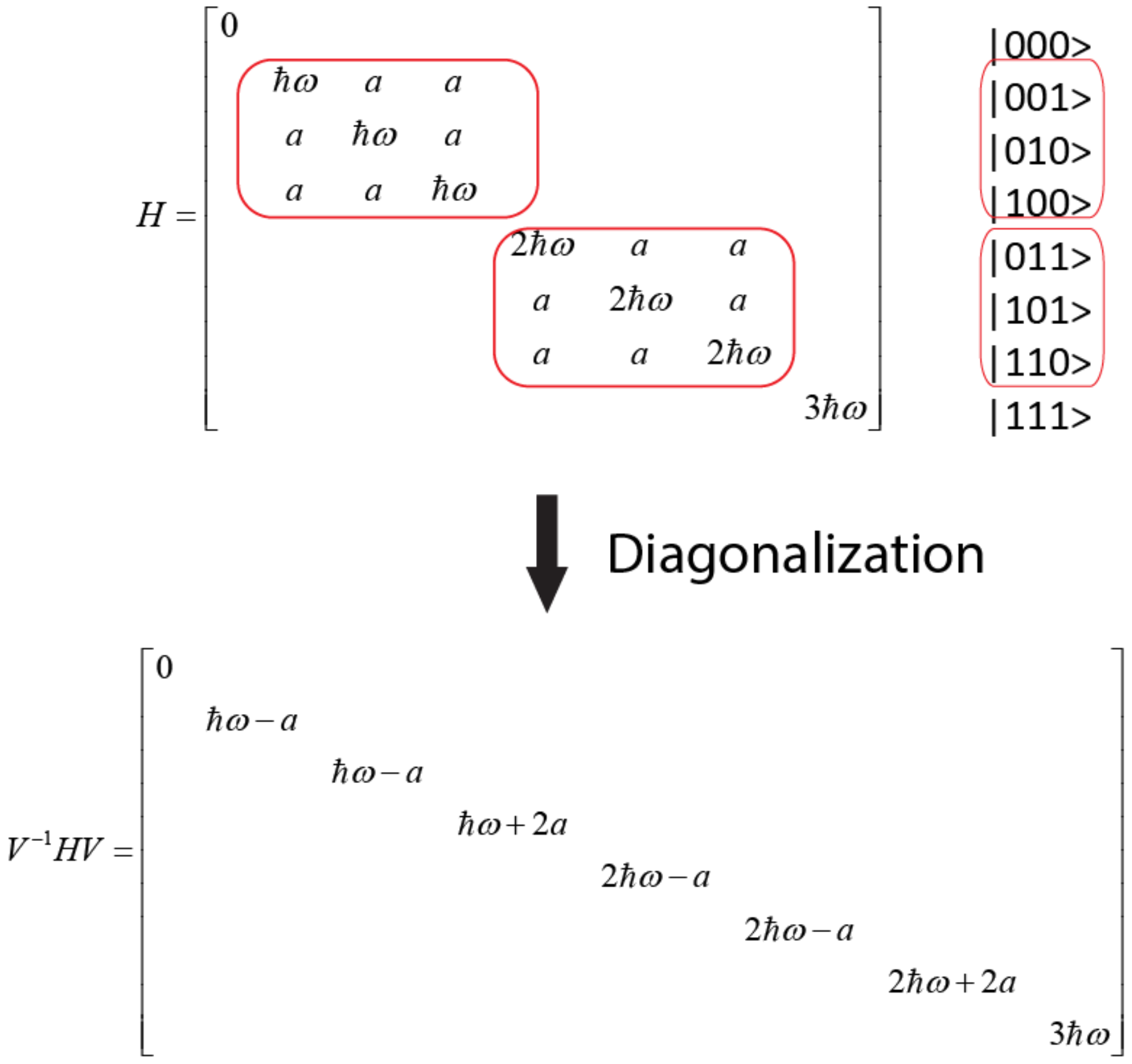

**Fig. S14. Diagram of Hamiltonian under many-body interactions**. (Top) The rotational Hamiltonian under many-body interactions is subject to the impact of perturbation terms, acting as off-diagonal elements (labelled as *a*). (Bottom) Hamiltonian diagonalization gives eigen-energy levels. Note that the coupling elements are not necessarily the same as shown in this figure. Different coupling elements will produce similar splittings among the levels that are three-fold degenerate in the absence of interactions.

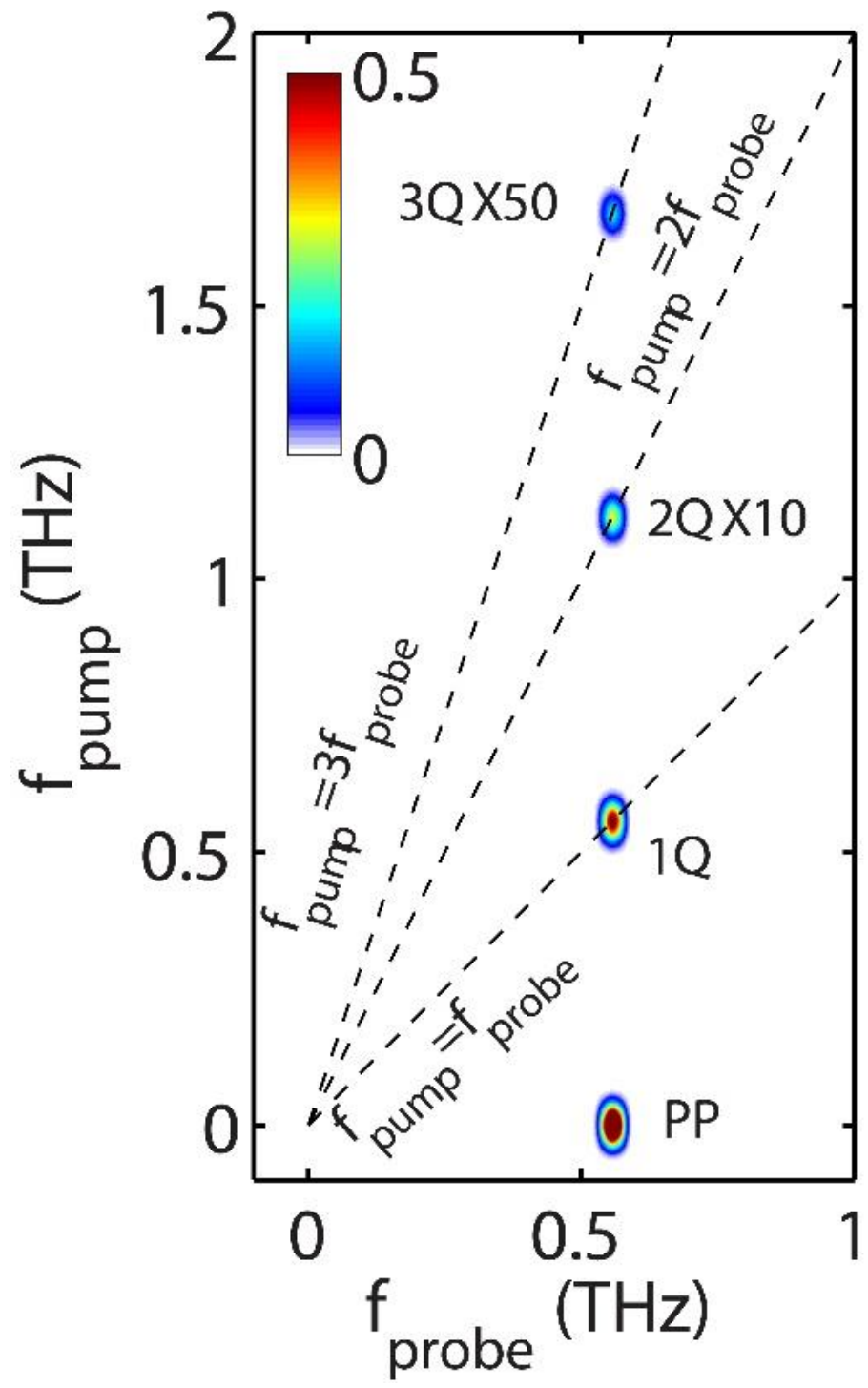

**Fig. S15.** Simulated 2Q and 3Q spectrum with many-body interactions**.** The weak interactions in the simulation are sufficient to allow 2Q and 3Q signals to appear but not enough to induce significant energy shifts, leading to only diagonal peaks. The pump-probe (PP), 1Q, 2Q (magnified by 10), and 3Q (magnified by 50) signals are illustrated. See **Fig. 3** for observed 2Q and 3Q features in the experiment.

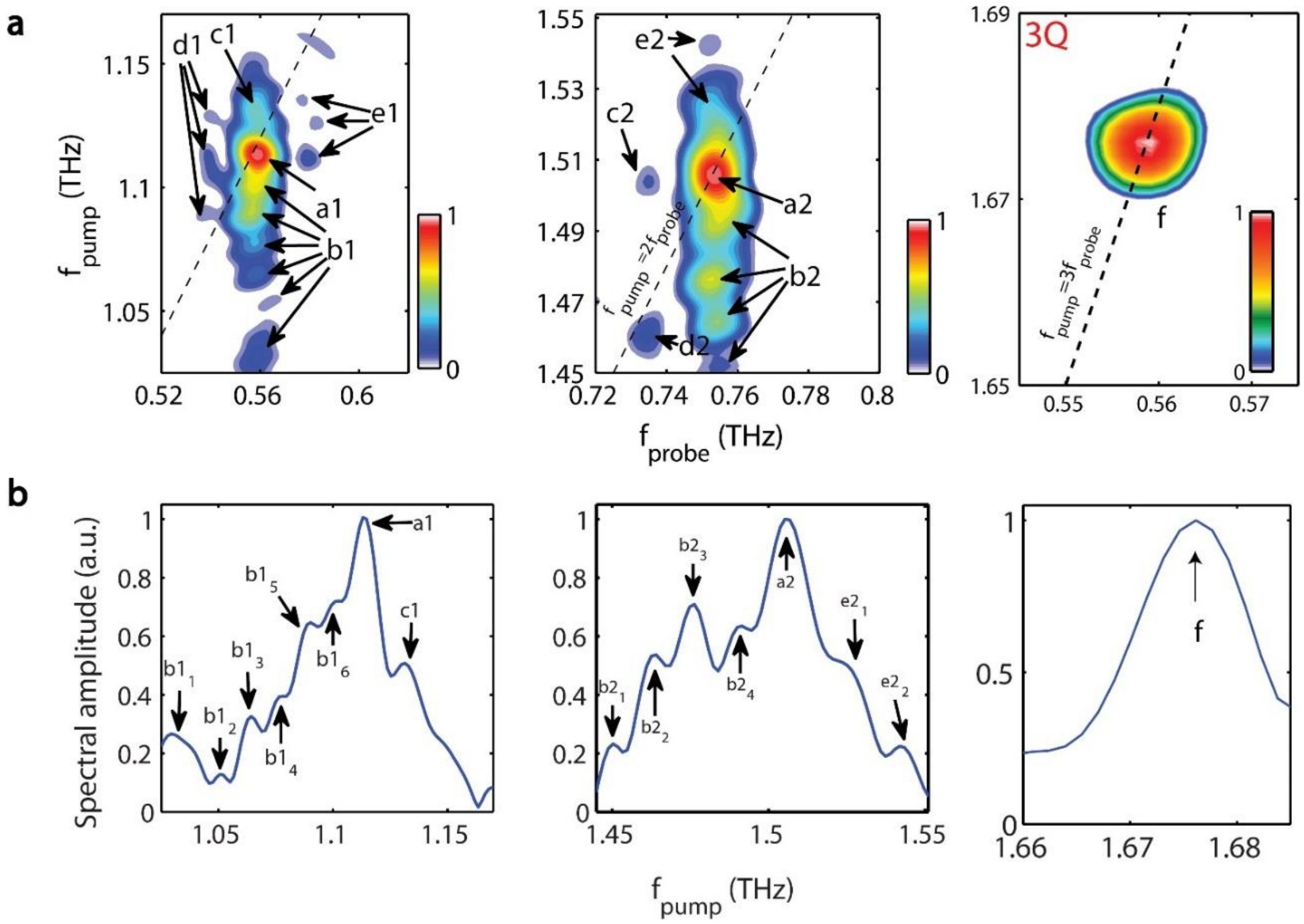

**Fig. S16. Off-diagonal 2Q peak linewidths indicate multiple complex geometries and lifetimes.** The 2Q diagonal, off-diagonal peaks and 3Q peaks from Fig. 3a of the main paper are replotted in **a**, and their spectral slices along the $f_{pump}$-axis are shown in **b**. Peak positions (uncertainties ±2 GHz) and selected peak intensity ratios are indicated in the table below. The estimated linewidths for the main peaks (a1, a2, and f) are 18 GHz, 28 GHz, and 14 GHz, giving dephasing times of 56 ps, 36 ps, and 71 ps, respectively. The 2Q off-diagonal peaks ($b1_{1\text{-}6}$, $b2_{1\text{-}4}$) have an averaged linewidth of ~10 GHz, yielding a dephasing time of ~100 ps which we believe is an instrumentally limited lower limit to the lifetimes of the metastable water complexes. No 3Q off-diagonal features are observed in the present measurements. The averaged intensities of the off-diagonal signals were used to compare with diagonal signals in Figs. 4d and 4e of the main paper. The off-diagonal 2Q spectral fine structure indicates multiple metastable complex geometries.

| 2Q: $f_{pump}$ ($f_{probe}$ = 0.558 THz) | 2Q: $f_{pump}$ ($f_{probe}$ = 0.753 THz) | 3Q: $f_{pump}$ ($f_{probe}$ = 0.558 THz) | Peak intensities (vs 1Q peak at (0.753 THz, 0.753 THz)) |
|---|---|---|---|
| $b1_1$: 1.029 THz | $b2_1$: 1.450 THz | f: 1.676 THz | |
| $b1_2$: 1.051 THz | $b2_2$: 1.464 THz | | |
| $b1_3$: 1.064 THz | $b2_3$: 1.476THz | | $b1_3$: 0.23 |
| $b1_4$: 1.078 THz | $b2_4$: 1.492 THz | | $b1_4$: 0.27 |
| $b1_5$: 1.090 THz | a2: 1.505 THz | | $b1_5$: 0.43 |
| $b1_6$: 1.102 THz | $e2_1$: 1.525 THz | | $b1_6$: 0.48 |
| a1: 1.113 THz | $e2_2$: 1.542 THz | | |
| c1: 1.131 THz | | | |

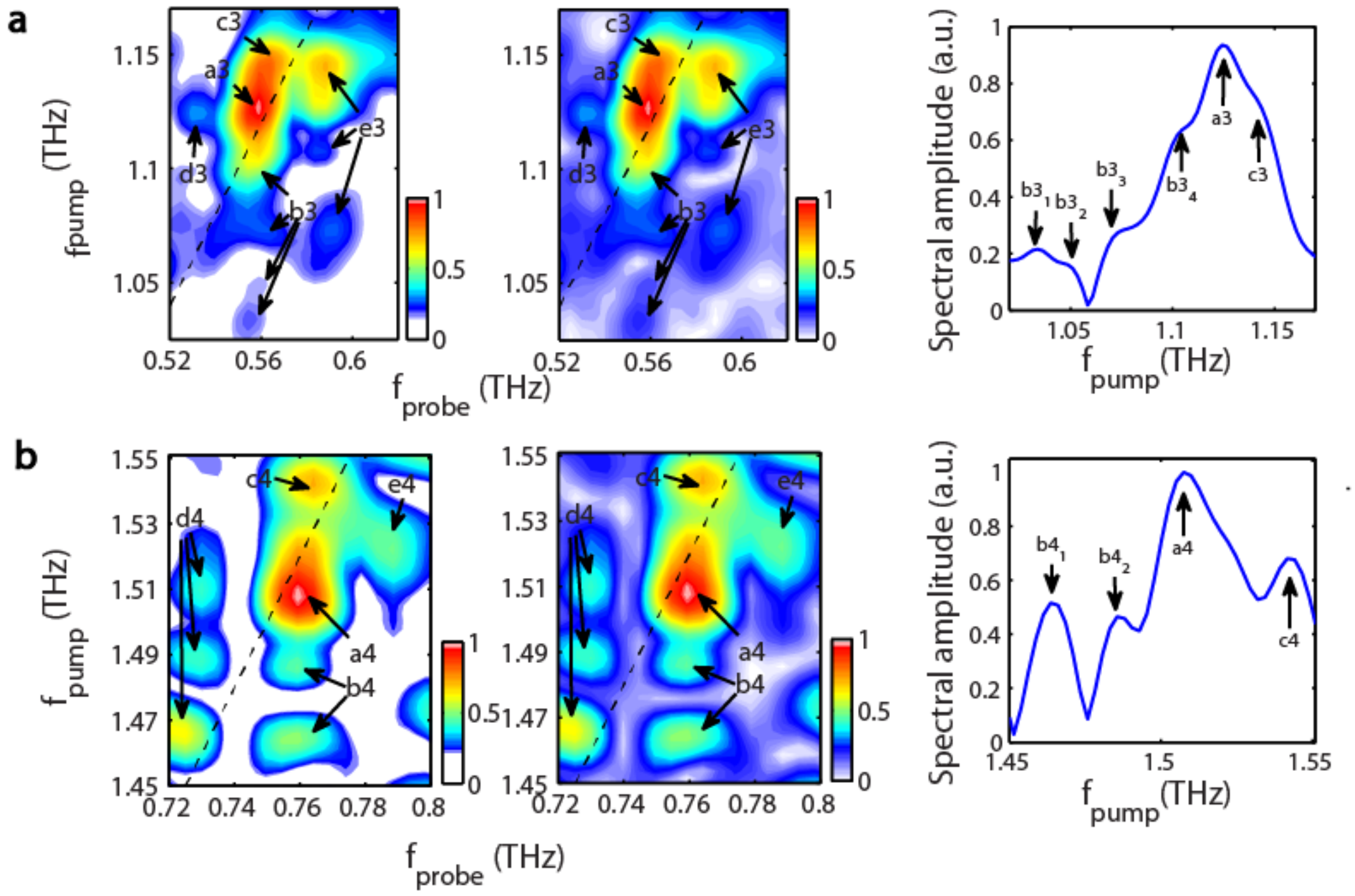

**Fig. S17. Typical 2Q spectra and slices of water vapor at room temperature**. The 2Q peaks and spectral slice near $f_{probe}$=0.558 THz are shown in **a** (Left: Background with intensity < 12% of the 2Q diagonal peak intensity is made white in color; Middle: Raw data with continuous colorbar scale; Right: Slice along the $f_{pump}$ axis ($f_{probe}$=0.558 THz)). Similar features near $f_{probe}$=0.753 THz are shown in **b**. (Left: Background with intensity < 20% of the 2Q diagonal peak intensity is made white in color.) Peak positions (uncertainties ±2 GHz) are indicated in the table below. No 3Q features are observed in the present measurements at room temperature. Comparison between the room-temperature peak positions with those at 60°C (Fig. S15) shows the following features. **(i)** Some peaks remain the same: ($b3_2$,$b1_2$); ($b3_4$,$b1_6$); ($b4_1$,$b2_2$); (c4,$e2_2$). **(ii)** Some peak positions ($b3_1$,$b1_1$); (a4,a2) show slight shifts, probably within uncertainties. **(iii)** Some peaks at room temperature appear to be the spectrally merged averages of peaks that appear distinct at 60°C: (a3,a1-c1); ($b3_3$,$b1_3$-$b1_4$); ($b4_2$,$b2_3$-$b2_4$). **(iv)** Some peaks ($b1_5$, $b2_1$, $e2_1$) in the 60°C spectra are not visible at room temperature where the S/N ratio is lower. Some features (e.g. c3) are subtle and not comparable directly with counterparts at 60°C.

| 2Q: $f_{pump}$ ($f_{probe}$ = 0.558 THz) | 2Q: $f_{pump}$ ($f_{probe}$ = 0.753 THz) | Peak intensities (vs1Q peak at (0.753 THz, 0.753 THz)) |
|---|---|---|
| $b3_1$: 1.034 THz | $b4_1$: 1.464 THz | $b3_1$: 0.059 |
| $b3_2$: 1.051 THz | $b4_2$: 1.486 THz | $b3_2$: 0.052 |
| $b3_3$: 1.073 THz | a4: 1.508 THz | $b3_3$: 0.082 |
| $b3_4$: 1.102 THz | c4: 1.542 THz | |
| a3: 1.124 THz | | |
| c3: 1.144THz | | |

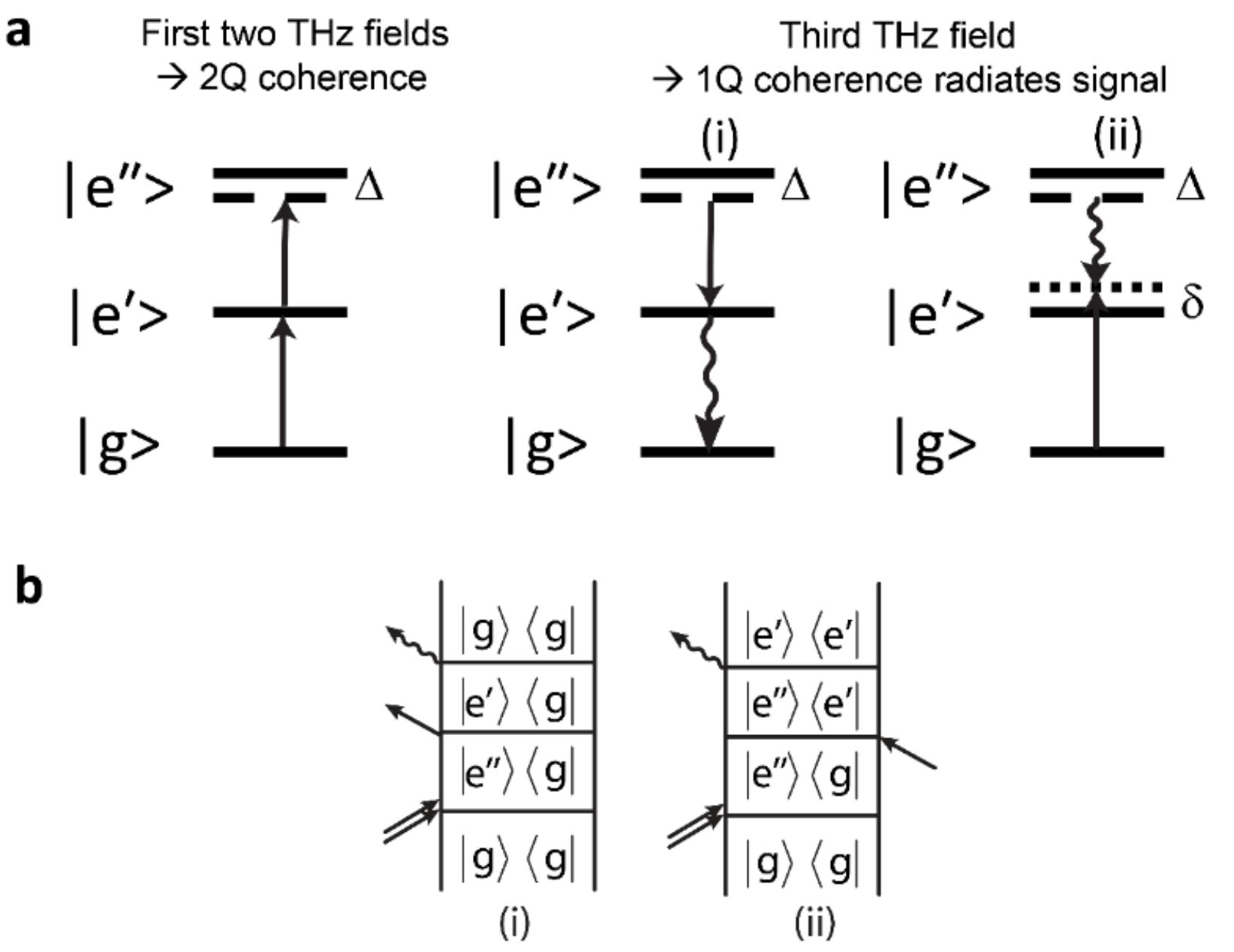

**Fig. S18. Energy-ladder and Feynman diagrams of side peaks in Figs. S15 and S16**. **a**, The ground state, the singly-excited state, and the doubly-excited state due to many-body interactions are shown as |g>, |e′>, and |e″>, respectively. The solid lines show the energy-level positions without many-body interactions. The first two THz fields produce the 2Q coherence, whose frequency depends on the particular intermolecular geometry. The third THz field may project |e″> downward (i) to the singly excited level |e′>, resulting in a coherence between |e′> and |g> that radiates at the usual 1Q transition frequency. Alternatively, the third THz field may promote from |g> to |e′>, resulting in a coherence between |e′> and |e″> whose frequency depends on the intermolecular geometry and whose shift from usual 1Q transition frequency may be different from the 2Q frequency shift Δ, similar to results observed in 2D electronic spectra of biexcitons(11). **b**, The 2Q pump frequency of an off-diagonal peak stems from the coherence |e″><g|. Due to the energy shifts, peaks can be shifted upward or downward with respect to the 2Q diagonal peak. Peaks can also emerge with shifts in the 1Q signal frequency, because the emitted frequency depends on the specific pathway (i) or (ii) determined by the third THz field interaction, yielding a final emitting coherence |e′><g| or |e″><e′| respectively, the latter at a frequency that also depends on the particular intermolecular geometry. The observed off-diagonal peaks and their positions (uncertainties ±2 GHz) are indicated in the table below.

| Side peaks | $f_{probe}$ **(THz)** | $f_{pump}$ **(THz)** |
|---|---|---|
| Peaks at 60°C: | | |
| $d1_1$; $d1_2$; $d1_3$ | 0.538; 0.539; 0.541 | 1.090; 1.112; 1.128 |
| $e1_1$; $e1_2$; $e1_3$ | 0.579; 0.582; 0.578 | 1.113; 1.127; 1.136 |
| c2; d2 | 0.735; 0.735 | 1.504; 1.461 |
| Peaks at 21°C: | | |
| d3; e4 | 0.531; 0.788 | 1.124; 1.522 |
| $e3_1$; $e3_2$; $e3_3$ | 0.590; 0.584; 0.587 | 1.072; 1.107; 1.144 |
| $d4_1$; $d4_2$; $d4_3$ | 0.725; 0.730; 0.730 | 1.466; 1.488; 1.510 |

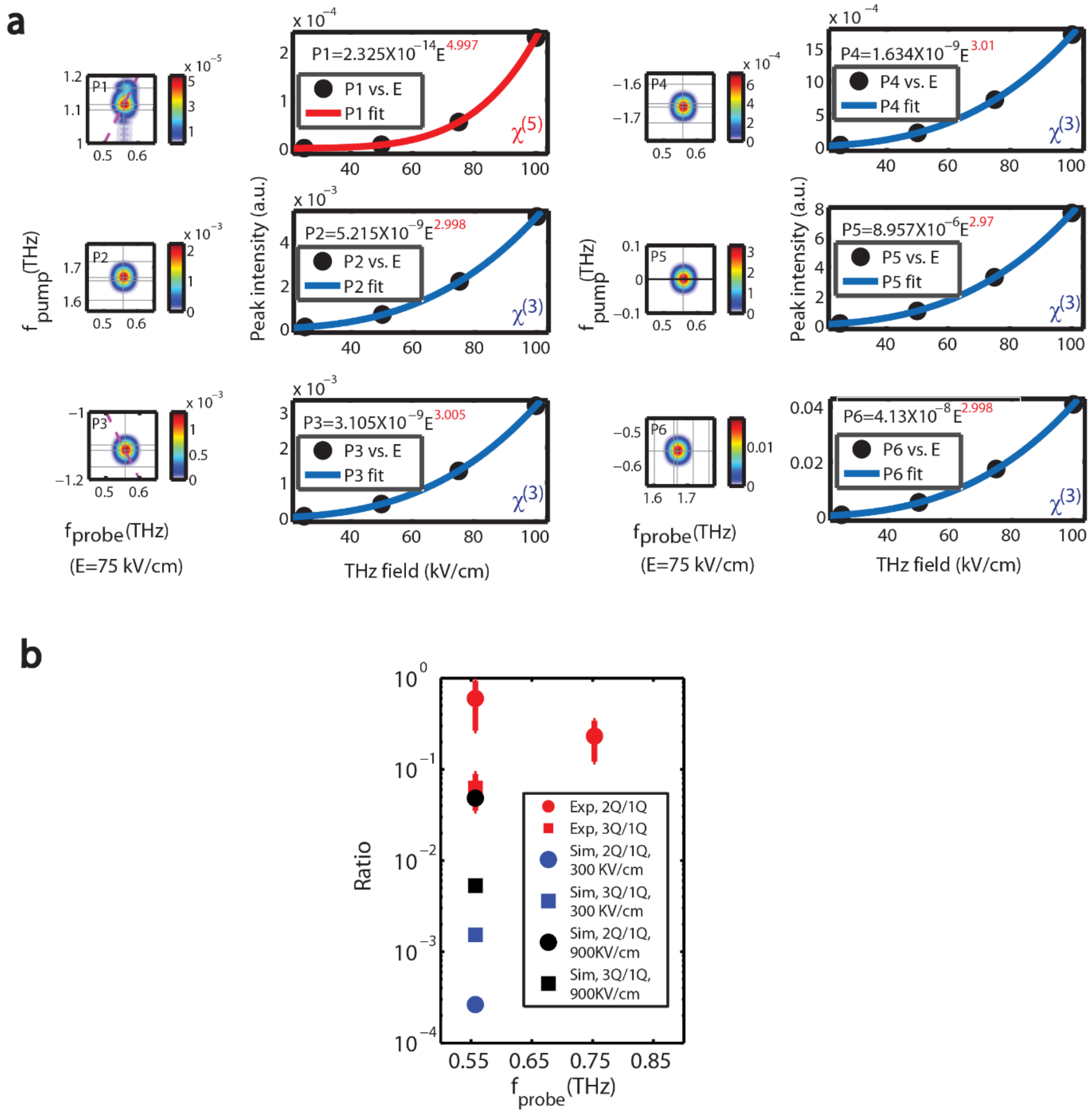

**Fig. S19. Simulated field dependence of selected spectral peaks and peak ratios between experimental and simulation data**. **a**, Peaks (P1-P6) are selected to show their peak intensities and field dependence fittings. In the weak to moderate THz field range, P2-P6 show third-order responses, whereas P1 is fifth order. P1 – P6 correspond to peaks labelled as NR11, NR8, R4, R6, PP1, and R13 in **Figs.S24** and **S26**. **b**, Ratios of 2Q and 3Q diagonal peaks to the corresponding 1Q diagonal peaks. Red circles and squares indicate experimental data. Points in blue and black are simulated results without many-body interactions when THz fields are 300 kV/cm and 900 kV/cm. Since the simulation conditions are similar to those in the experiments except for the absence of intermolecular interactions, the much smaller 2Q and 3Q signal amplitudes (blue and black) in the simulation than the experimental signals (red) indicate that weak nonlinear signals from non-interacting water molecules are not contributing significantly to 2Q and 3Q features.

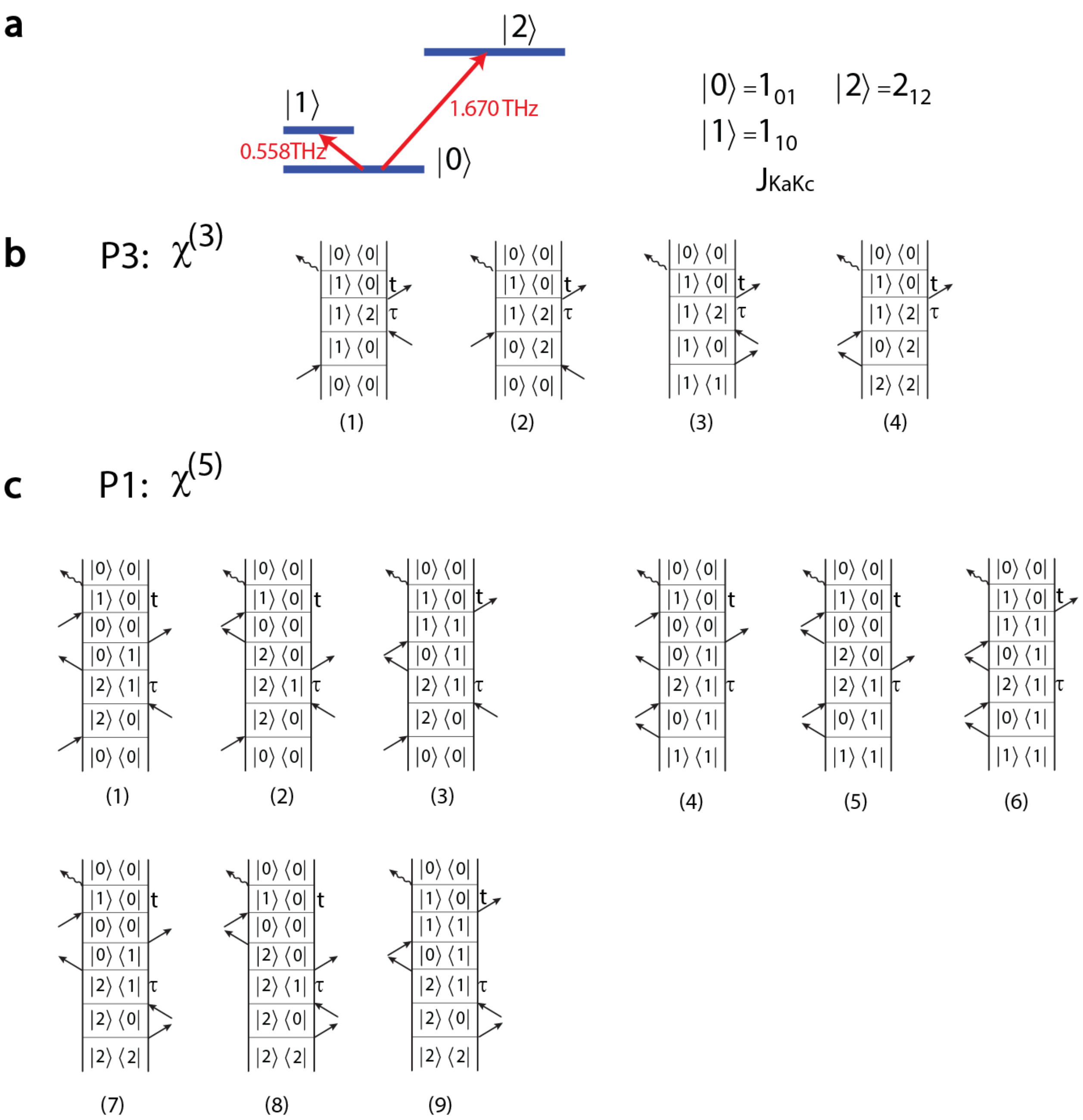

**Fig. S20. Possible Feynman diagrams for selected peaks**. **a**, V-type energy ladder with defined energy level labels. The transition between rotational states |1> and |2> is forbidden. **b**, Four possible Feynman pathways for P3 (as shown in **Fig. S19**). **c**, Nine possible Feynman pathways for P1 (as shown in **Fig. S19**).

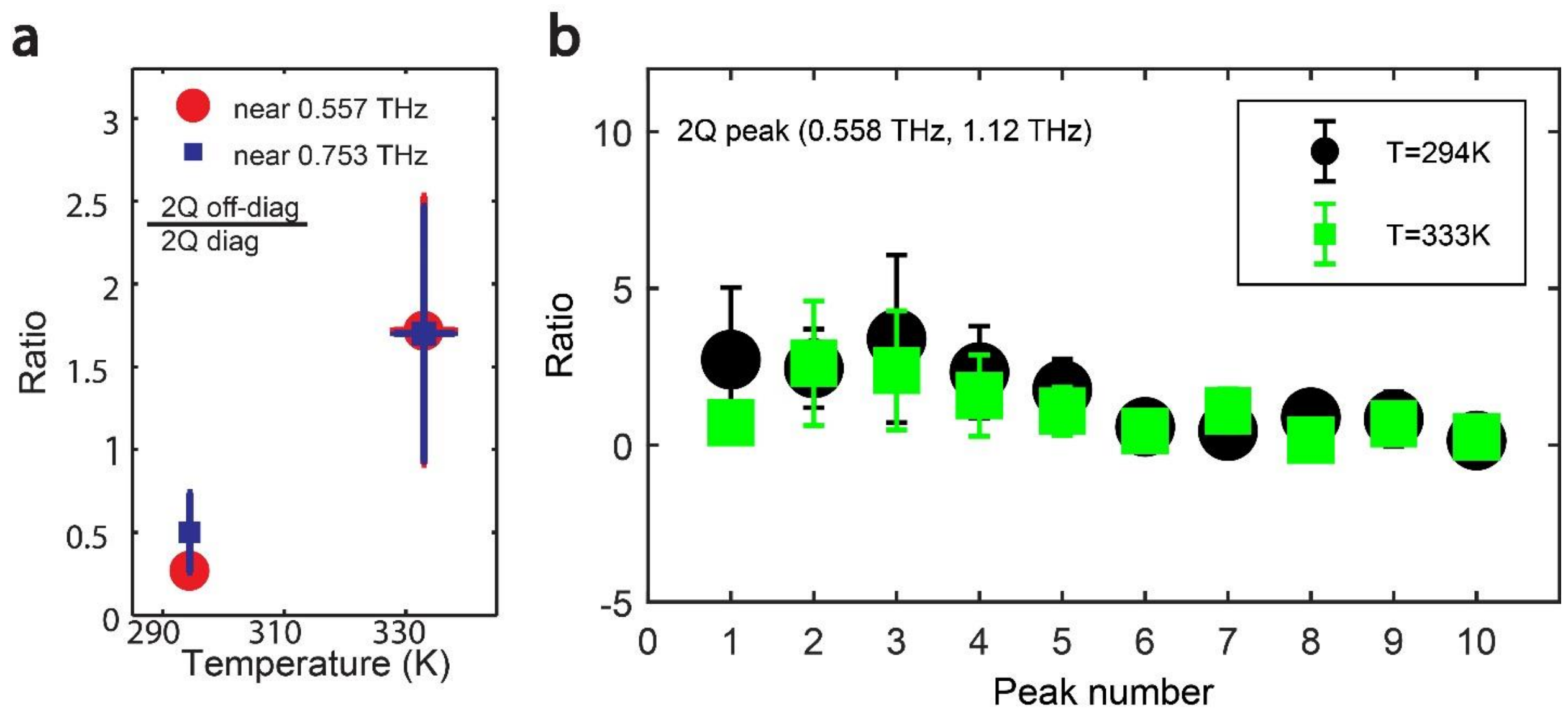

**Fig. S21. Peak ratios in the experiments. a**, Ratio of 2Q off-diagonal peaks to the 2Q diagonal peaks vs temperature. Red circles and blue squares are data near ( $f_{probe} = 0.558$ THz, $f_{pump} = 1.115$ THz ) and ( $f_{probe} = 0.753$ THz, $f_{pump} = 1.506$ THz ), respectively. The strong increase in the ratios with temperature suggests that 2Q off-diagonal features originate from (metastable) water complexes. **b**, Ratio of 1Q diagonal and off-diagonal rephasing and non-rephasing peak intensities to diagonal 2Q peak intensities. Black circles and green squares indicate the ratios with respect to the diagonal 2Q peak (with labels a1 and a3 in Figs. S15-S16) near (0.558 THz, 1.12 THz) at room temperature (21$^{\circ}$C) and at 60 °C, respectively. For example, the data of Peak 1 is the ratio of the diagonal NR peak (0.989 THz, 0.989 THz) to the 2Q diagonal peak near (0.558 THz, 1.12 THz). All these peak intensities increase at the higher temperature due to the increase in water vapor pressure, but their relative intensities (given by the plotted ratios) do not change significantly, as expected for signals that all depend linearly on the water concentration. The ratios with respect to the other diagonal 2Q peak (with labels a2 and a4 in Figs. S15-S16) at (0.753 THz, 1.505 THz) yield the same conclusion. In contrast, Fig. 4d shows that the off-diagonal 2Q peak intensities increase far more than the diagonal 2Q peak intensities (and based on the present figure, far more than the NR and R peak intensities) at the higher temperature, as expected for signals from metastable molecular complexes such as weakly bound molecular pairs whose concentration scales quadratically with the water concentration. The positions of the diagonal and off-diagonal peaks selected for the plot above are as follows: Peak 1 (0.989 THz, 0.989 THz); Peak 2 (1.098 THz, 1.098 THz); Peak 3 (1.165 THz, 1.165 THz); Peak 4 (1.208 THz, 1.208 THz); Peak 5 (1.229 THz, 1.229 THz); Peak 6 (0.753 THz, 1.229 THz); Peak 7 (1.229 THz, 0.753 THz); Peak 8 (0.753 THz, -0.558 THz); Peak 9 (1.229 THz, -0.753 THz); and Peak 10 (0.753 THz, -1.229 THz), where the first (second) coordinate value is the probe (pump) frequency.

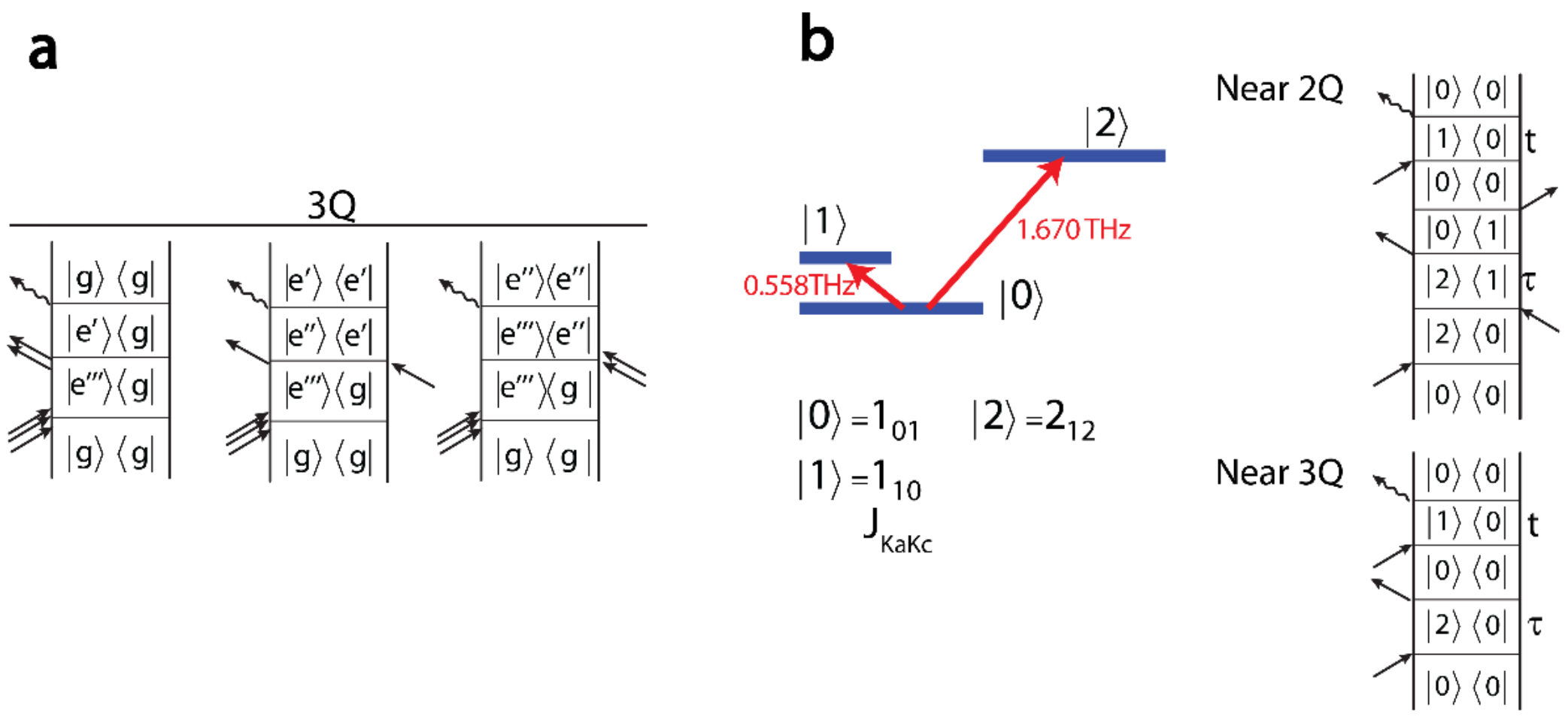

**Fig. S22. Feynman diagrams for 3Q peaks and possible pathways near 2Q and 3Q peaks**. **a**, The labels $|g\rangle$, $|e'\rangle$, $|e''\rangle$, and $|e'''\rangle$ denote the ground state, the singly excited state, the doubly excited state, and the triply excited state, respectively. Different diagrams are possible depending on coherence pathway selections. **b**, For the V-type energy ladder with defined energy level labels. The transition between rotational states |1> and |2> is forbidden. Besides the dominant pathways due to many-body interactions, examples for other possible contributions are illustrated here. See also **Fig. S19** & **20**.

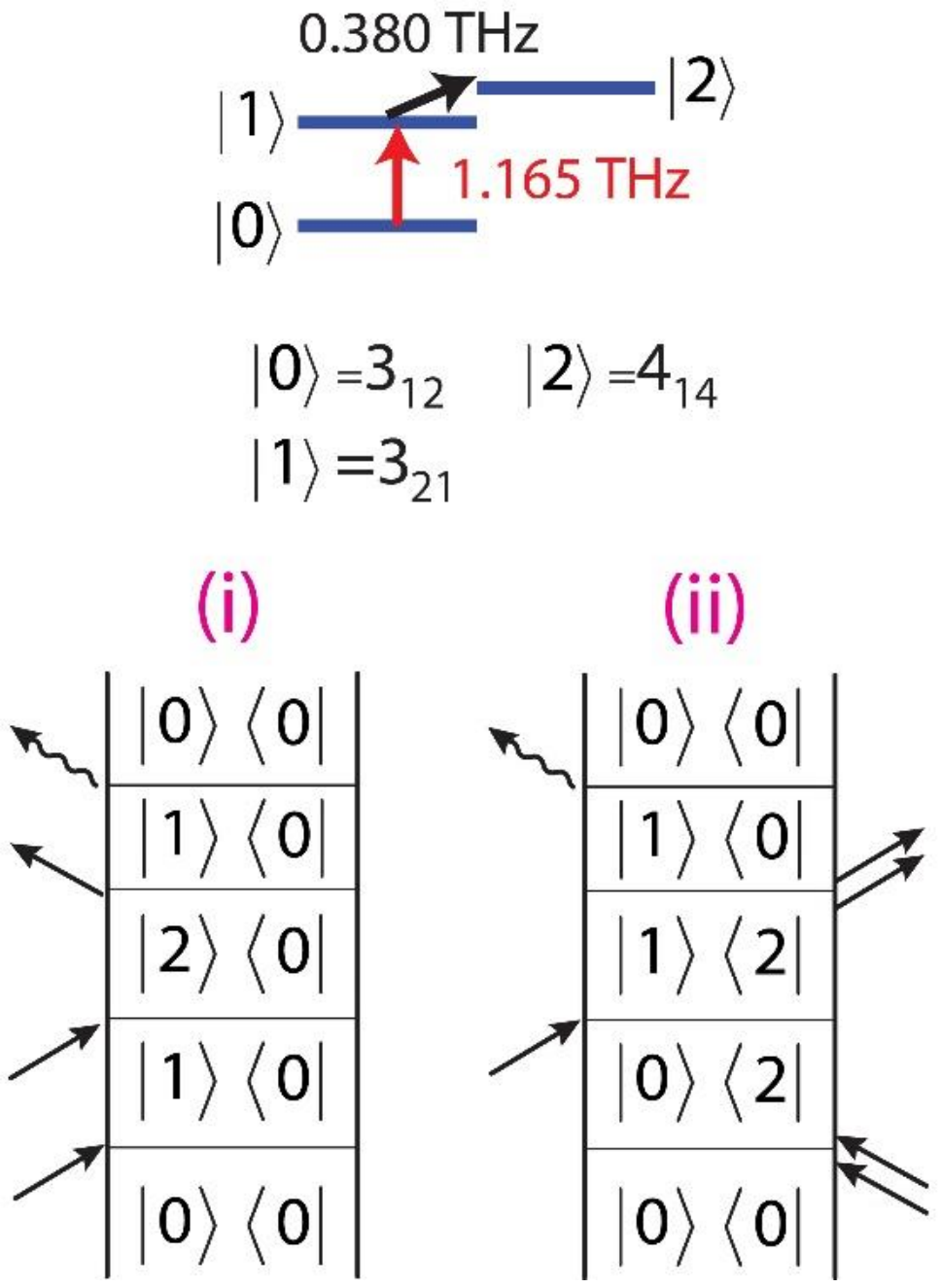

**Fig. S23. Example of rotational couplings involving weak water transitions**. Weak peaks in the simulation (for example, see **Fig. 4**, peak (W2) in **Fig. S24**, and peak (WR2) in **Fig. S26**) are related to the transition $3_{21} \leftrightarrow 4_{14}$ (see **Fig. S12**), the intensity of which is too weak to be observed in the present experiment. Two possible coherence pathways are shown.

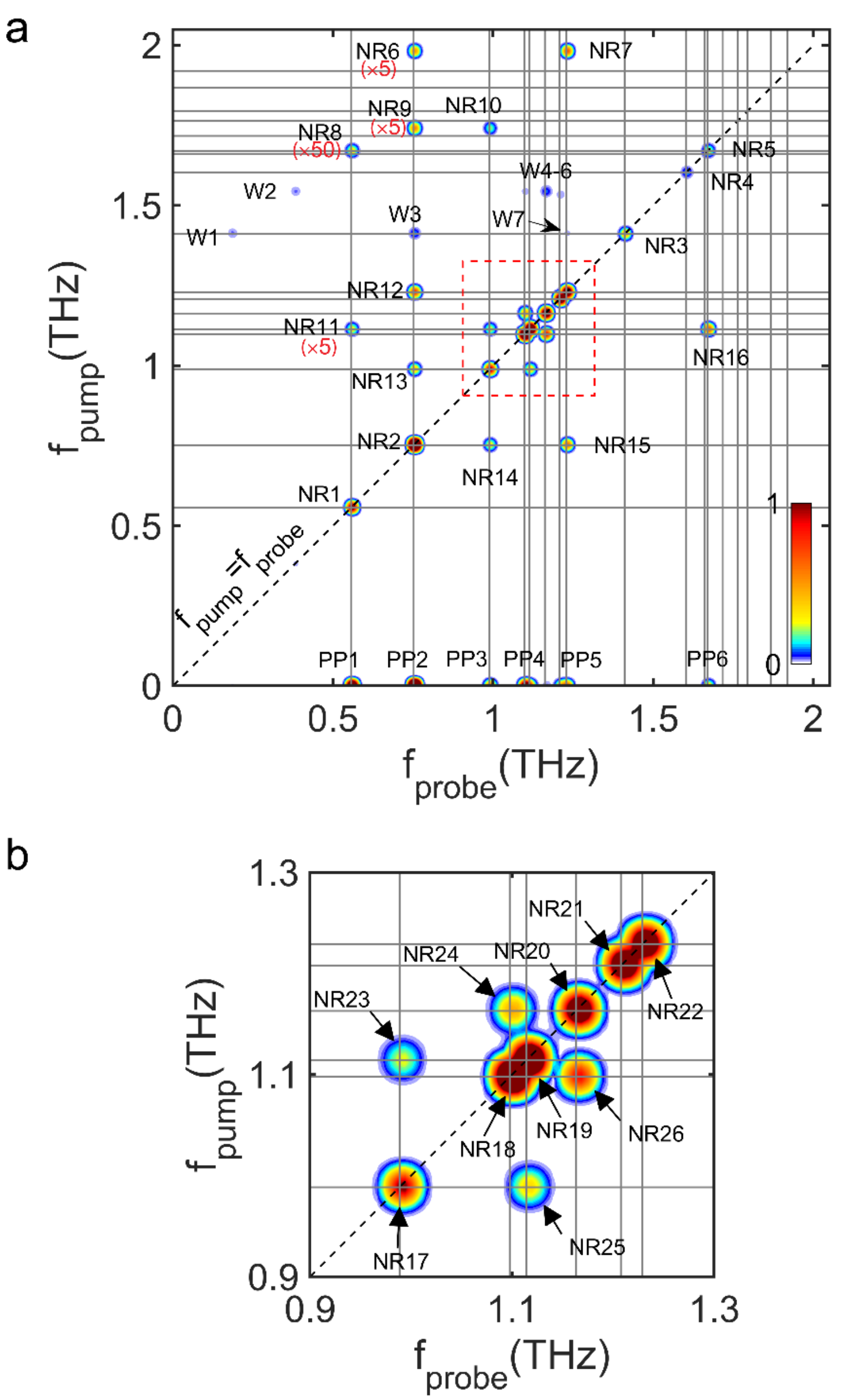

**Fig. S24. The complete simulated non-rephasing (NR) spectrum. a**, Diagonal/off-diagonal NR peaks and PP signals are shown and labelled. Some features are magnified by appropriate factors (indicated in red brackets) for clarity. For example, NR6(×5) means a non-rephasing peak magnified by a factor of 5. Weak peaks (W1-7) are nonlinear signals stemming from weak water transitions. For example, Peak W2 ($f_{pump}$=1.55 THz, $f_{probe}$=0.,38THz) involves the weak water transition of 0.38 THz. See **Fig. S23** for more details. **b**, Enlarged spectrum near 1.1 THz, which is indicated as a red dashed box in **a**. See also **Fig.2** for the measured spectrum.

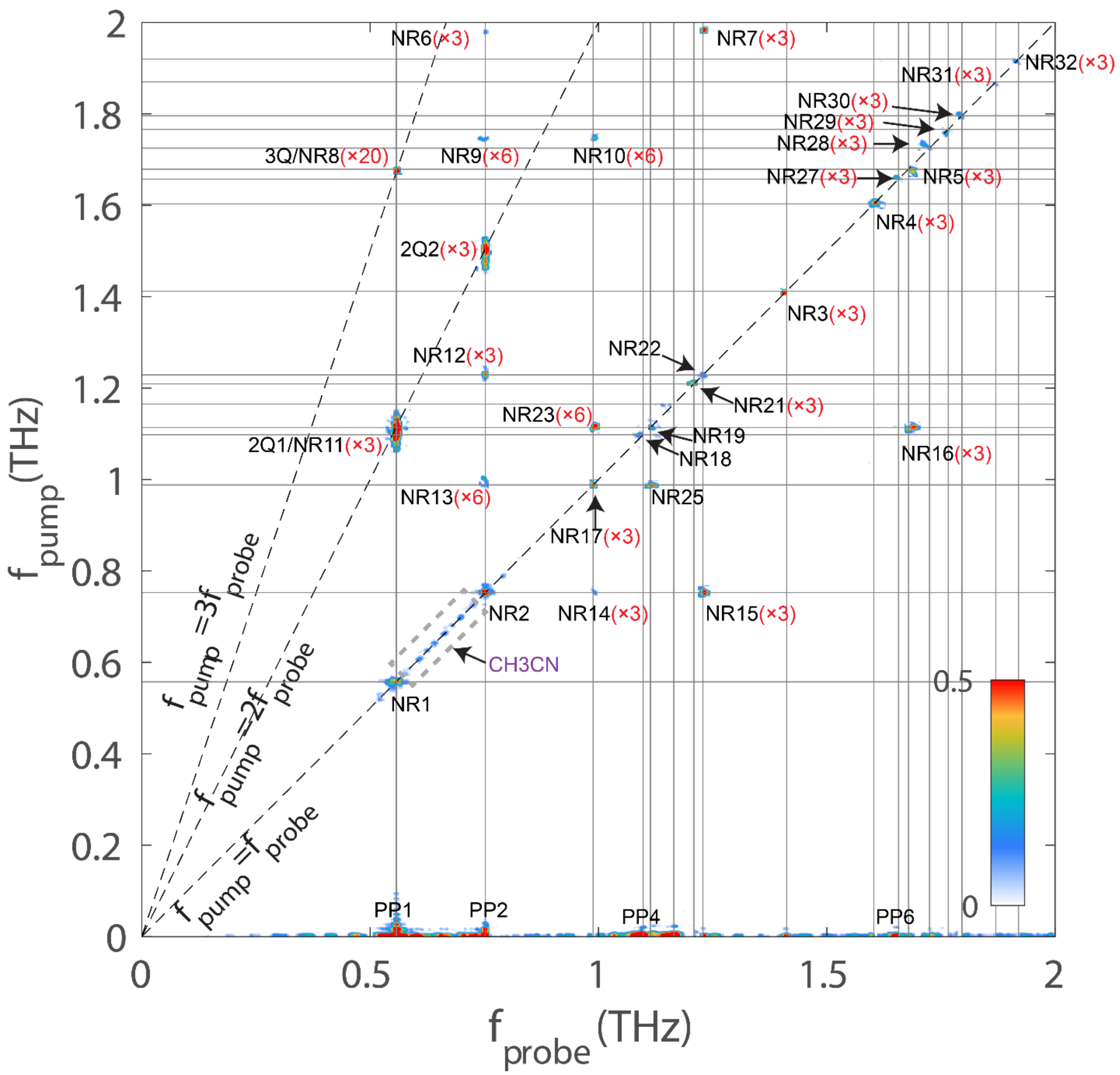

**Fig. S25. The complete experimental non-rephasing (NR) spectrum.** NR, 2Q, 3Q and PP features are shown and labelled. Features from acetonitrile residue are indicated in a gray dashed box. Some features are magnified by appropriate factors (indicated in red brackets), for clarity. For example, NR15(×3) means a non-rephasing peak magnified by a factor of 3. Peak labels follow those in the simulation (**Fig. S24**). See also **Fig.2** for separated spectra.

a

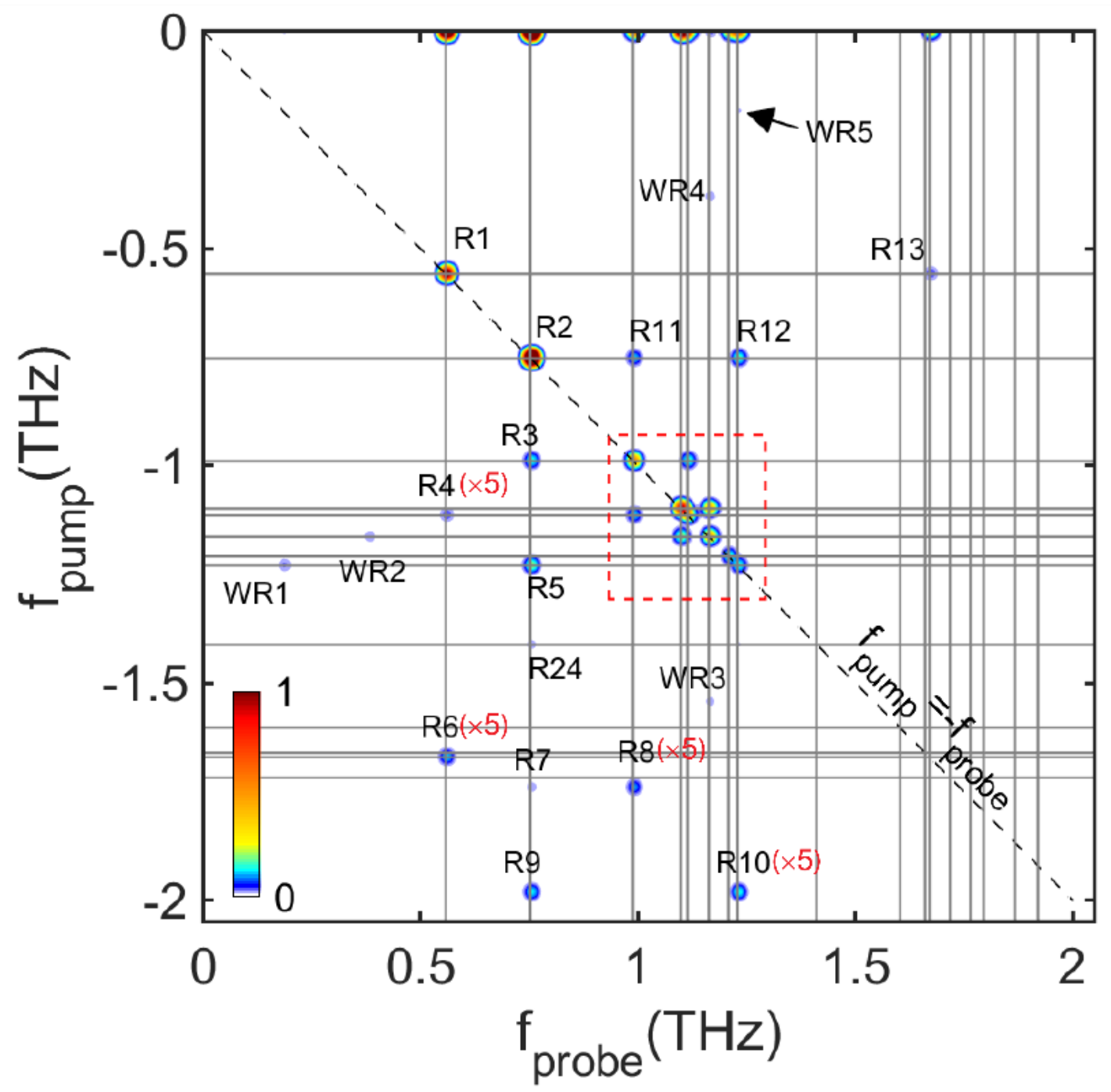

b

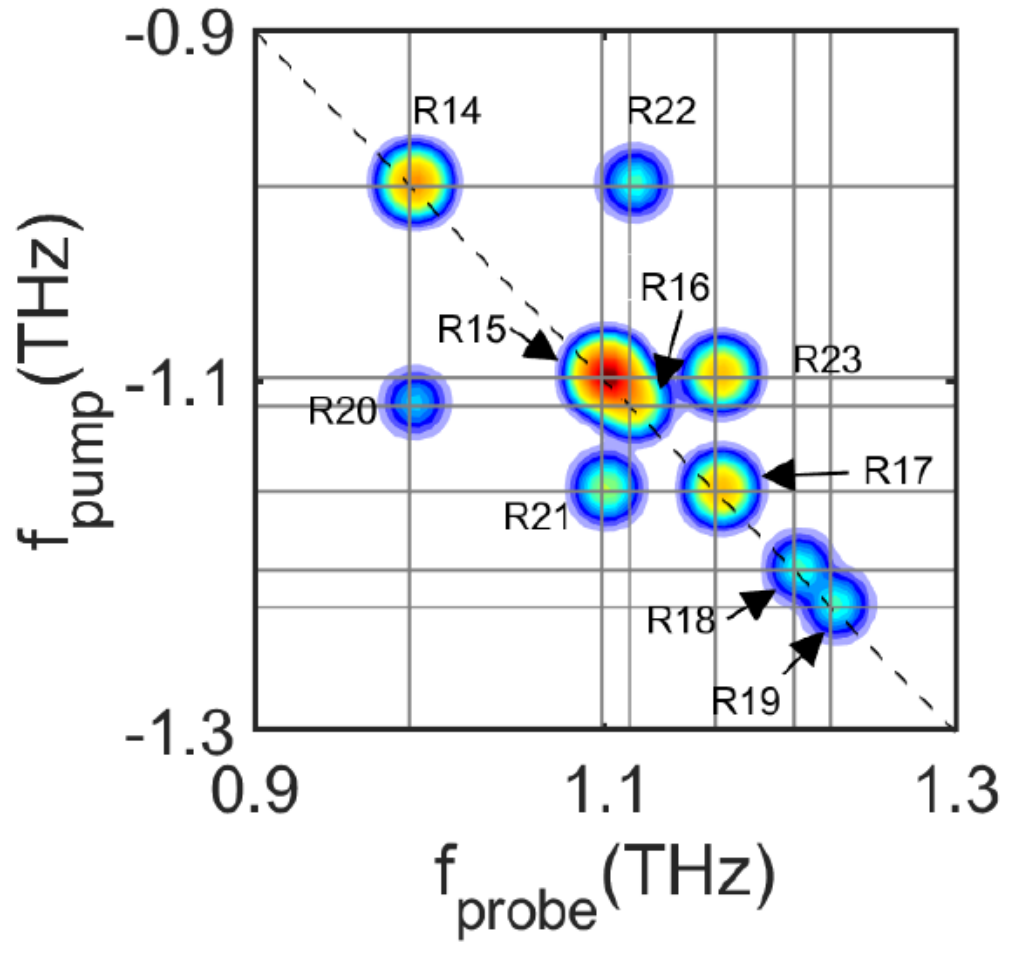

**Fig. S26. The complete simulated rephasing (R) spectrum. a**, R peaks and PP signals are shown and labelled. Some features are magnified by appropriate factors (indicated in red brackets), for clarity. For example, R6(×5) means a rephasing peak magnified by a factor of 5. Similar to weak peaks in NR spectra, weak peaks (WR1-5) are rephasing signals stemming from weak water transitions. **b**, Enlarged spectrum near 1.1 THz, which is indicated as a red dashed box in **a**. See also **Fig.2** for the measured spectrum.

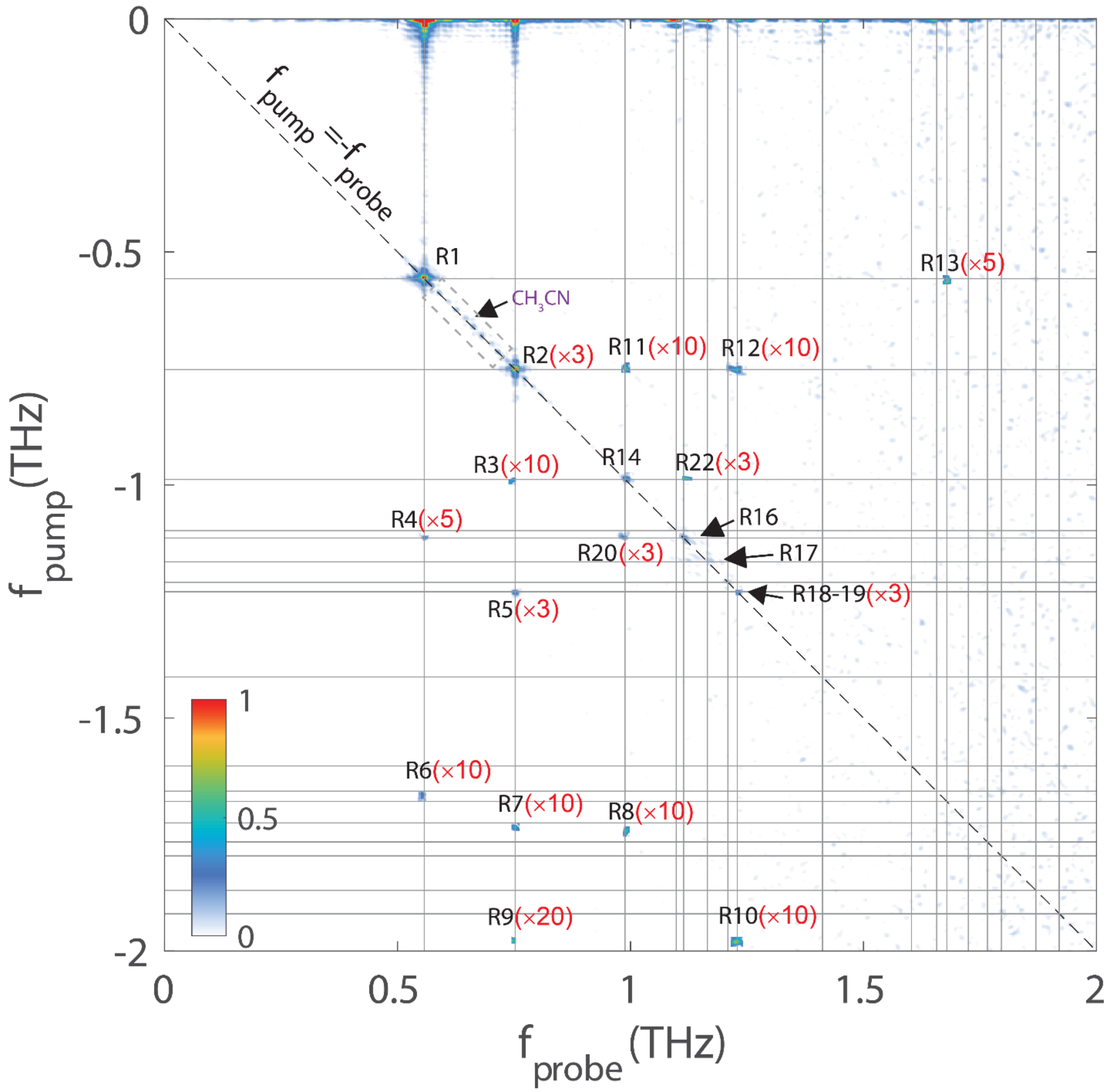

**Fig. S27. The complete experimental rephasing (R) spectrum.** Diagonal and off-diagonal R features are shown and labelled. Features from acetonitrile residue are indicated in a gray dashed box. Some features are magnified by appropriate factors (indicated in red brackets), for clarity. For example, R5(×3) means a rephasing peak magnified by a factor of 3. Peak labels follow those in the simulation (**Fig. S26**). See also **Fig.2** for separated spectra.

| Rotational constants | Values ( GHz) |
|---|---|
| $A$ | 835.83910 |
| $B$ | 435.347353 |
| $C$ | 278.139826 |
| $\Delta_J$ | $3.758280 \times 10^{-2}$ |
| $\Delta_{JK}$ | $-1.73361 \times 10^{-1}$ |
| $\Delta_K$ | $9.742470 \times 10^{-1}$ |
| $\delta_J$ | $1.521950 \times 10^{-2}$ |
| $\delta_K$ | $3.96461 \times 10^{-2}$ |
| $H$ | $1.6284 \times 10^{-5}$ |
| $H$ | $-5.2534 \times 10^{-5}$ |
| $H$ | $-5.42802 \times 10^{-4}$ |
| $H$ | $3.84436 \times 10^{-3}$ |
| $h_J$ | $8.1611 \times 10^{-6}$ |
| $h_{JK}$ | $-2.5607 \times 10^{-5}$ |
| $h_K$ | $9.4784 \times 10^{-4}$ |

**Table S1. Rotational constants of water molecules used in this work.** Rotational constants of water rotational states are listed for the implementation of Watson's reduced Hamiltonian. Parameter values are taken from references **(4, 5)**.

| Line number | Para/Ortho type | Observed freq (THz) | Simulated freq (THz) | Obs-Sim (THz) | Transition assignment |
|---|---|---|---|---|---|
| 1 | Ortho | 0.5574 | 0.5576 | -0.0002 | $1_{01} \longleftrightarrow 1_{10}$ |
| 2 | Para | 0.7525 | 0.7523 | 0.0002 | $2_{02} \longleftrightarrow 2_{11}$ |
| 3 | Para | 0.9883 | 0.9885 | -0.0002 | $1_{11} \longleftrightarrow 2_{02}$ |
| 4 | Ortho | 1.098 | 1.098 | 0 | $3_{03} \longleftrightarrow 3_{12}$ |
| 5 | Para | 1.114 | 1.114 | 0 | $0_{00} \longleftrightarrow 1_{11}$ |
| 6 | Ortho | 1.165 | 1.163 | 0.002 | $3_{12} \longleftrightarrow 3_{21}$ |
| 7 | Para | 1.209 | 1.208 | 0.001 | $4_{13} \longleftrightarrow 4_{22}$ |
| 8 | Para | 1.229 | 1.229 | 0 | $2_{11} \longleftrightarrow 2_{20}$ |
| 9 | Ortho | 1.412 | 1.411 | 0.001 | $5_{14} \longleftrightarrow 5_{23}$ |
| 10 | Para | 1.603 | 1.602 | 0.001 | $4_{04} \longleftrightarrow 4_{13}$ |
| 11 | Ortho | 1.657 | 1.660 | -0.003 | $2_{12} \longleftrightarrow 2_{21}$ |
| 12 | Ortho | 1.679 | 1.670 | 0.009 | $1_{01} \longleftrightarrow 2_{12}$ |
| 13 | Ortho | 1.725 | 1.717 | 0.008 | $2_{12} \longleftrightarrow 3_{03}$ |
| 14 | Para | 1.766 | 1.764 | 0.002 | $6_{24} \longleftrightarrow 6_{33}$ |
| 15 | Ortho<br>Para | 1.796 | 1.795 | 0.001 | $7_{25} \longleftrightarrow 7_{34}$<br>$6_{15} \longleftrightarrow 6_{24}$ |
| 16 | Ortho | 1.870 | 1.867 | 0.003 | $5_{23} \longleftrightarrow 5_{32}$ |
| 17 | Para | 1.920 | 1.920 | 0 | $3_{13} \longleftrightarrow 3_{22}$ |

**Table S2. Pure rotational transition lines of water vapor.** Both observed and simulated pure rotational transition lines from the present work are listed. The typical deviation between observation and simulation is approximately 1 GHz.

| Matrix elements | Results (given J, K, M) |
|---|---|
| $\langle JKM \vert J_x \vert J, K \pm 1, M\rangle$ | $\frac{1}{2}[J(J+1) - K(K \pm 1)]^{1/2}$ |
| $\langle JKM \vert J_y \vert J, K \pm 1, M\rangle$ | $\mp \frac{i}{2}[J(J+1) - K(K \pm 1)]^{1/2}$ |
| $\langle JKM \vert J_z \vert JKM\rangle$ | $K$ |
| $\langle JKM \vert \boldsymbol{J}^2 \vert JKM\rangle$ | $J(J+1)$ |
| $\langle JKM \vert J_x^2 \vert JKM\rangle$ | $\frac{1}{2}[J(J+1) - K^2]$ |
| $\langle JKM \vert J_y^2 \vert JKM\rangle$ | $\frac{1}{2}[J(J+1) - K^2]$ |
| $\langle JKM \vert J_z^2 \vert JKM\rangle$ | $K^2$ |
| $\langle JKM \vert J_-^2 \vert J, K \pm 2, M\rangle$ | $\frac{1}{2}[J(J+1) - K(K \pm 1)]^{1/2}[J(J+1) - (K \pm 1)(K \pm 2)]^{1/2}$ |
| $\langle JKM \vert \boldsymbol{J}^4 \vert JKM\rangle$ | $[J(J+1)]^2$ |
| $\langle JKM \vert \boldsymbol{J}^2 J_z^2 \vert JKM\rangle$ | $J(J+1)K^2$ |
| $\langle JKM \vert J_z^4 \vert JKM\rangle$ | $K^4$ |
| $\langle JKM \vert \boldsymbol{J}^2 J_-^2 \vert J, K \pm 2, M\rangle$ | $J(J+1) \cdot \frac{1}{2}[J(J+1) - K(K \pm 1)]^{1/2}[J(J+1) - (K \pm 1)(K \pm 2)]^{1/2}$ |
| $\langle JKM \vert J_z^2 J_-^2 \vert J, K \pm 2, M\rangle$ | $K^2 \cdot \frac{1}{2}[J(J+1) - K(K \pm 1)]^{1/2}[J(J+1) - (K \pm 1)(K \pm 2)]^{1/2}$ |
| $\langle JKM \vert J_-^2 J_z^2 \vert J, K \pm 2, M\rangle$ | $(K \pm 2)^2 \cdot \frac{1}{2}[J(J+1) - K(K \pm 1)]^{1/2}[J(J+1) - (K \pm 1)(K \pm 2)]^{1/2}$ |

**Table S3. Calculated matrix elements of the rotational Hamiltonian up to the 4$^{th}$ order.** Matrix elements of the angular momentum operator up to 4$^{th}$ order employed in this work are evaluated and listed here **(1)**.

| Matrix elements | Results (given J, K, M) |
| --- | --- |
| $\langle JKM \vert \boldsymbol{J}^6 \vert JKM\rangle$ | $[J(J+1)]^3$ |
| $\langle JKM \vert \boldsymbol{J}^4 J_z^2 \vert JKM\rangle$ | $[J(J+1)]^2 K^2$ |
| $\langle JKM \vert \boldsymbol{J}^2 J_z^4 \vert JKM\rangle$ | $[J(J+1)]K^4$ |
| $\langle JKM \vert J_z^6 \vert JKM\rangle$ | $K^6$ |
| $\langle JKM \vert \boldsymbol{J}^4 J_-^2 \vert J, K\pm 2, M\rangle$ | $[J(J+1)]^2 \cdot \frac{1}{2}[J(J+1)-K(K\pm 1)]^{1/2}[J(J+1)-(K\pm 1)(K\pm 2)]^{1/2}$ |
| $\langle JKM \vert \boldsymbol{J}^2 J_z^2 J_-^2 \vert J, K\pm 2, M\rangle$ | $J(J+1)K^2 \cdot \frac{1}{2}[J(J+1)-K(K\pm 1)]^{1/2}[J(J+1)-(K\pm 1)(K\pm 2)]^{1/2}$ |
| $\langle JKM \vert \boldsymbol{J}^2 J_-^2 J_z^2 \vert J, K\pm 2, M\rangle$ | $J(J+1)(K\pm 2)^2$ $\cdot \frac{1}{2}[J(J+1)-K(K\pm 1)]^{1/2}[J(J+1)-(K\pm 1)(K\pm 2)]^{1/2}$ |
| $\langle JKM \vert J_z^4 J_-^2 \vert J, K\pm 2, M\rangle$ | $K^4 \cdot \frac{1}{2}[J(J+1)-K(K\pm 1)]^{1/2}[J(J+1)-(K\pm 1)(K\pm 2)]^{1/2}$ |
| $\left\langle JKM \vert J_-^2 J_z^4 \vert J, K\pm 2, M\right\rangle$ | $(K\pm 2)^4 \cdot \frac{1}{2}[J(J+1)-K(K\pm 1)]^{1/2}[J(J+1)-(K\pm 1)(K\pm 2)]^{1/2}$ |
| | |

**Table S4. Calculated matrix elements of the rotational Hamiltonian of the 6th order.** Matrix elements of the angular momentum operator of 6th order are evaluated and listed **(1)**.

| Exp. # | T(°C) | Position ($f_{probe}$ (THz)) | 2Q diagonal peak (label (intensity)) | 2Q off-diagonal peak (label (intensity)) | THz field strength(kV/cm) |
|---|---|---|---|---|---|
| **1** | 21 | ~ 0.56 | a3(1.0) | $b3_1$(0.23); $b3_2$(0.16); $b3_3$(0.27) | E1=400, E2=400 |
| | | ~ 0.75 | a4(1.0) | $b4_1$(0.52); $b4_2$(0.46) | |
| **2** | 21 | ~ 0.56 | a3(1.0) | $b3_2$(0.20); $b3_3$(0.29); $b3_4$(0.28) | E1=300, E2=350 |
| | | ~ 0.75 | a2(1.0) | $b2_3$(0.43); $b2_4$(0.28); | |
| **3** | 21 | ~ 0.56 | a3(1.0) | $b3_2$(0.28); $b3_3$(0.44) | E1=300, E2=350 |
| | | ~ 0.75 | a4(1.0) | $b4_1$(0.35); $b4_2$(0.96) | |
| **4** | 60 | ~ 0.56 | a3(1.0) | $b1_3$(0.60); $b1_4$(0.73); $b1_5$(1.1); | E1=440, E2=290 |
| | | ~ 0.75 | a2(1.0) | $b2_2$(0.98); $b2_3$(1.3); $b2_4$(1.2) | |
| **5** | 60 | ~ 0.56 | a1(1.0) | $b1_4$(1.3); $b1_5$(1.8) | E1=300, E2=350 |
| | | ~ 0.75 | a2(1.0) | $b2_2$(2.6); $b2_3$(2.8) | |
| **6** | 60 | ~ 0.56 | a1(1.0) | $b1_3$(2.8); $b1_4$(2.5); $b1_5$(2.4); $b1_6$(2.2) | E1=300, E2=350 |
| | | ~ 0.75 | a2(1.0) | $b2_3$(1.0); $b2_4$(1.9) | |

**Table S5. Data for calculating the ratio of 2Q off-diagonal to 2Q diagonal peak intensities.** Data points from six experiments are used to calculate the ratio (Fig. 4d). Experiments (1 – 3) are data at room temperature, and experiments (4 – 6) are at 60°C. 2Q diagonal peak intensities are normalized to 1 (column 4). 2Q off-diagonal peak intensities are listed in column 5. The two THz field strengths (E1 and E2) used for each experiment are listed in the last column. Peak intensities (column 5) are normalized by E2 and the square of E1, and converted to the values when both THz strengths are 400 kV/cm. For example, the last data of the 2Q off-diagonal peak ($b2_4$) in column 5 is the product of its original value and the factor $1/(300^2\times350)\times(400^2\times400)$, where E1 = 300 kV/cm and E2=350 kV/cm are the THz fields used in the experiment. In addition, a gas-cell window transmission of 82.3% is taken into account to correct the THz field strength that reached the water molecules (because, for example, Exp. #2 and 3 measure the data from ambient air without the gas cell, whereas Exp. #5 and 6 are for water vapor in the gas cell). The averaged intensity ratio of 2Q off-diagonal peaks to 1Q diagonal peaks increases by a factor of 5.5 as the temperature rises from room temperature (21°C) to 60°C, while the water vapor pressure increases by a factor of 8.3.

| Peak labels | Positions ($f_{probe}$, $f_{pump}$) (THz) | Related rotational states ($J_{KaKc}$) and peak origins |
|---|---|---|
| NR1 | (0.558, 0.558) | $1_{01}$, $1_{10}$ |
| NR2 | (0.753, 0.753) | $2_{02}$, $2_{11}$ |
| NR3 | (1.41, 1.41) | $5_{14}$, $5_{23}$ |
| NR4 | (1.60, 1.60) | $4_{04}$, $4_{13}$ |
| NR5 | (1.67, 1.67) | $2_{12}$, $2_{21}$;<br>$1_{01}$, $2_{12}$ |
| NR6 | (0.753, 1.98) | $2_{02}$, $2_{11}$, $2_{20}$ |
| NR7 | (1.23, 1.98) | $2_{02}$, $2_{11}$, $2_{20}$ |
| NR8/3Q | near (0.558, 1.67) | Many-body interaction; Radiative coupling;<br>$1_{01}$, $1_{10}$, $2_{12}$ |
| NR9 | (0.753, 1.74) | $1_{11}$, $2_{02}$, $2_{11}$ |
| NR10 | (0.989, 1.74) | $1_{11}$, $2_{02}$, $2_{11}$ |
| NR11/2Q1 | near (0.558, 1.12) | Many-body interaction; Radiative coupling;<br>$1_{01}$, $1_{10}$, $2_{12}$ |
| NR12 | (0.753, 1.23) | $2_{02}$, $2_{11}$, $2_{20}$ |
| NR13 | (0.753, 0.989) | $1_{11}$, $2_{02}$, $2_{11}$ |
| NR14 | (0.989, 0.753) | $1_{11}$, $2_{02}$, $2_{11}$ |
| NR15 | (1.23, 0.753) | $2_{02}$, $2_{11}$, $2_{20}$ |
| NR16 | (1.67, 1.12) | $1_{01}$, $1_{10}$, $2_{12}$ |
| NR17 | (0.989, 0.989) | $1_{11}$, $2_{02}$ |
| NR18 | (1.10, 1.10) | $3_{03}$, $3_{12}$ |
| NR19 | (1.11, 1.11) | $0_{00}$, $1_{11}$ |
| NR20 | (1.16, 1.16) | $3_{12}$, $3_{21}$ |
| NR21 | (1.21, 1.21) | $4_{13}$, $4_{22}$ |
| NR22 | (1.23, 1.23) | $2_{11}$, $2_{20}$ |
| NR23 | (0.989, 1.12) | $0_{00}$, $1_{11}$, $2_{02}$ |
| NR24 | (1.10, 1.16) | $3_{03}$, $3_{12}$, $3_{21}$ |
| NR25 | (1.12, 0.989) | $0_{00}$, $1_{11}$, $2_{02}$ |
| NR26 | (1.16, 1.10) | $3_{03}$, $3_{12}$, $3_{21}$ |
| NR27<br>NR28<br>NR29<br>NR30<br>NR31<br>NR32 | (1.66, 1.66)<br>(~1.72, ~1.72)<br>(1.77, 1.77)<br>(1.80, 1.80)<br>(1.87, 1.87)<br>(1.92, 1.92) | $2_{12}$, $2_{21}$<br>$2_{12}$, $3_{03}$<br>$6_{24}$, $6_{33}$<br>$7_{25}$, $7_{34}$; $6_{15}$, $6_{24}$<br>$5_{23}$, $5_{32}$<br>$3_{13}$, $3_{22}$ |
| 2Q2 | near (0.753, 1.51) | Many-body interaction; Radiative coupling |
| PP1; PP2<br>PP3; PP4<br>PP5;<br>PP6 | (0.558, 0); (0.753, 0)<br>(0.989, 0); (1.10, 0)<br>(1.23, 0)<br>(1.67, 0) | $1_{01}$, $1_{10;}$ $2_{02}$, $2_{11}$<br>$1_{11}$, $2_{02;}$ $3_{03}$, $3_{12}$<br>$2_{11}$, $2_{20}$<br>$2_{12}$, $2_{21}$; $1_{01}$, $2_{12}$ |
| W1<br>W2<br>W3<br>W4<br>W5<br>W6<br>W7 | (0.185, 1.41)<br>(0.380, 1.54)<br>(0.753, 1.41)<br>(1.10, 1.54)<br>(1.16, 1.54)<br>(1.21, 1.53)<br>(1.23, 1.41) | $2_{11}$,$2_{20}$, $3_{13}$<br>$3_{12}$,$3_{21}$, $4_{14}$<br>$2_{02}$, $2_{11}$,$2_{20}$, $3_{13}$<br>$3_{03}$, $3_{12}$,$3_{21}$, $4_{14}$<br>$3_{12}$,$3_{21}$, $4_{14}$<br>$4_{13}$,$4_{22}$, $5_{15}$<br>$2_{11,}$ $2_{20,}$ $3_{13}$ |

**Table S6. Peak positions and related rotational states for NR spectra in Fig.S24&S25.**

| **Peak labels** | **Positions ($f_{probe}$, $f_{pump}$) (THz)** | **Related rotational states ($J_{KaKc}$) and peak origins** |
|---|---|---|
| R1 | (0.558, -0.558) | $1_{01}$, $1_{10}$ |
| R2 | (0.753, -0.753) | $2_{02}$, $2_{11}$ |
| R3 | (0.753, -0.989) | $1_{11}$, $2_{02}$, $2_{11}$ |
| R4 | (0.558, -1.11) | $1_{01}$, $1_{10}$, $2_{12}$ |
| R5 | (0.753, -1.23) | $2_{02}$, $2_{11}$, $2_{20}$ |
| R6 | (0.558, -1.67) | $1_{01}$, $1_{10}$, $2_{12}$ |
| R7 | (0.753, -1.74) | $1_{11}$, $2_{02}$, $2_{11}$ |
| R8 | (0.989, -1.74) | $1_{11}$, $2_{02}$, $2_{11}$ |
| R9 | (0.753, -1.98) | $1_{11}$, $2_{02}$, $2_{11}$ |
| R10 | (1.23, -1.98) | $1_{11}$, $2_{02}$, $2_{11}$ |
| R11 | (0.989, -0.753) | $1_{11}$, $2_{02}$, $2_{11}$ |
| R12 | (1.23, -0.753) | $2_{02}$, $2_{11}$, $2_{20}$ |
| R13 | (1.67, -0.558) | $1_{01}$, $1_{10}$, $2_{12}$ |
| R14 | (0.989, -0.989) | $1_{11}$, $2_{02}$ |
| R15 | (1.10, -1.10) | $3_{03}$, $3_{12}$ |
| R16 | (1.11, -1.11) | $0_{00}$, $1_{11}$ |
| R17 | (1.16, -1.16) | $3_{12}$, $3_{21}$ |
| R18 | (1.21, -1.21) | $4_{13}$, $4_{22}$ |
| R19 | (1.23, -1.23) | $2_{11}$, $2_{20}$ |
| R20 | (0.989, -1.11) | $0_{00}$, $1_{11}$, $2_{02}$ |
| R21 | (1.10, -1.16) | $3_{03}$, $3_{12}$, $3_{21}$ |
| R22 | (1.12, -0.989) | $0_{00}$, $1_{11}$, $2_{02}$ |
| R23 | (1.16, -1.10) | $3_{03}$, $3_{12}$, $3_{21}$ |
| WR1<br>WR2<br>WR3<br>WR4<br>WR5 | (0.185, -1.23)<br>(0.380, -1.16)<br>(1.16, -1.54)<br>(1.17, -0.380)<br>(1.23, -0.182) | $2_{11}$,$2_{20}$, $3_{13}$<br>$3_{12}$,$3_{21}$, $4_{14}$<br>$3_{12}$, $3_{21}$, $4_{14}$<br>$3_{12}$, $3_{21}$, $4_{14}$<br>$2_{11}$, $2_{20}$, $3_{13}$ |

**Table S7. Peak positions and related rotational states for R spectra in Fig.S26&S27.**

## Supplementary References